\documentclass[11pt]{article}

\usepackage{amsmath}
\usepackage{amssymb}
\usepackage{amsfonts}
\usepackage{tikz}
\usepackage{authblk}

\newcommand{\ket}[1]{\left|  #1 \right>  }
\newcommand{\bra}[1]{\left<  #1 \right|  }

\newcommand{\braket}[2]{ \left. \left<  #1 \right|  \hspace{-0.6mm} #2 \right>}

\begin{document}

\title{Entanglement scaling of operators: a conformal field theory approach, with a glimpse of simulability of long-time dynamics in 1+1d} 

\author{J. Dubail}
\affil{IJL, CNRS and Universit\'e de Lorraine UMR 7198, Boulevard des Aiguillettes F-54506 Vandoeuvre-l\`es-Nancy Cedex, France.}


\maketitle

\abstract{In one dimension, the area law and its implications for the approximability by Matrix Product States are the key to efficient numerical simulations
involving quantum states. Similarly, in simulations involving quantum operators, the approximability by Matrix Product Operators (in Hilbert-Schmidt norm) is tied to an {\it operator area law}, namely the fact that the Operator Space Entanglement Entropy (OSEE)---the natural analog of entanglement entropy for operators, investigated by Zanardi [Phys. Rev. A 63, 040304(R) (2001)] and by Prosen and Pi\v{z}orn [Phys. Rev. A 76, 032316 (2007)]---, is bounded. In the present paper, it is shown that the OSEE can be calculated in two-dimensional conformal field theory, in a number of situations that are relevant to questions of simulability of long-time dynamics in one spatial dimension. \phantom{\cite{zanardi2001entanglement,prosen2007operator}}\\
\indent It is argued that: (i) thermal density matrices $\rho \propto e^{-\beta H}$ and Generalized Gibbs Ensemble density matrices $\rho \propto e^{- H_{\rm GGE}}$ with local $H_{\rm GGE}$ generically obey the operator area law; (ii) after a global quench, the OSEE first grows linearly with time, then decreases back to its thermal or GGE saturation value, implying that, while the operator area law is satisfied both in the initial state and in the asymptotic stationary state at large time, it is strongly violated in the transient regime; (iii) the OSEE of the evolution operator $U(t) = e^{-i H t}$ increases linearly with $t$, unless the Hamiltonian is in a localized phase; (iv) local operators in Heisenberg picture, $\phi(t) = e^{i H t} \phi e^{-i H t}$, have an OSEE that grows sublinearly in time (perhaps logarithmically), however it is unclear whether this effect can be captured in a traditional CFT framework, as the free fermion case hints at an unexpected breakdown of conformal invariance.}

 \vspace{0.5cm}   {\it This is an invited contribution to the special issue of J. Phys. A "John Cardy's scale-invariant journey in low dimensions: a special issue for his 70th birthday'" edited by P. Calabrese, P. Fendley and U. T\"auber.}

\tableofcontents

\newpage

\section{Introduction}

In the past two decades, the way we think about quantum many-body systems, as well as the techniques we use to simulate them,
have been revolutionized by Tensor Networks; for reviews, see for instance Refs. \cite{verstraete2008matrix,orus2014practical}. In one spatial dimension, the success of numerical methods like DMRG \cite{white1992density,schollwock2011density} or TEBD \cite{vidal2003efficient,vidal2004efficient} is rooted in the fact that the physical states that are simulated have relatively low entanglement. In particular, the simulability of ground states of local Hamiltonians is intimately related (see Refs. \cite{verstraete2006matrix, schuch2008entropy} for precise statements) to the {\it area law} \cite{eisert2010colloquium}: for gapped Hamiltonians, the entanglement entropy of a subsystem $A$ is bounded \cite{hastings2007area}, while for gapless systems, the area law is only weakly violated---namely, {\it logarithmically} violated \cite{holzhey1994geometric, vidal2003entanglement, calabrese2004entanglement}---which still allows for efficient simulations based on Matrix Product States (MPSs) \cite{fannes1992finitely, perez2006matrix}. \vspace{0.3cm}

\noindent While MPSs approximate (or sometimes represent exactly) quantum states of one-dimensional systems, Matrix Product Operators (MPOs) are the natural candidates for approximating the operators acting on these states. In particular, algorithms that rely on MPOs to implement mixed states---{\it i.e.} density matrices---have been investigated since the early days of Tensor Networks \cite{verstraete2004matrix,zwolak2004mixed}. Nowadays, there is still a vivid interest in developing MPO algorithms, see {\it e.g.} Refs. \cite{banuls2009matrix, pirvu2010matrix, pivzorn2014real, cui2015variational, keller2015efficient, 2017arXiv170208894L}. \vspace{0.1cm}

\noindent Simulability/non-simulability by MPS methods is related to the area law and its violations (see Ref. \cite{schuch2008entropy} for a precise discussion), so, by analogy, it is natural to relate the question of simulability/non-simulability by MPO methods to an {\it operator area law}, namely to the boundedness/unboundedness of some analog of the entanglement entropy for operators. To the best of our knowledge, such an analog of entanglement entropy was first investigated in 2000 by Zanardi \cite{zanardi2001entanglement} and was later re-introduced and dubbed {\it Operator Space Entanglement Entropy} or {\it OSEE} by Prosen and Pi\v{z}orn \cite{prosen2007operator}. Adopting this terminology, the purpose of the present paper is to revisit this quantity from the point of view of 2d conformal field theory (CFT), in a number of situations that are relevant to the dynamics of one-dimensional systems. \vspace{0.3cm}

\noindent There is an important caveat in the line of thought consisting of treating operators in analogy with quantum states though, at least when the operator of interest is a density matrix\footnote{I am grateful to M.-C. Ba\~{n}uls and J.I. Cirac for emphasizing this point.} (as will sometimes happen in this paper): one tackles the approximability of operators {\it in Hilbert-Schmidt norm (or $L^2$-norm) only, as opposed to the $L^1$-norm}, which is the relevant one for density matrices. However, to the best of our knowledge there doesn't exist an analytically calculable quantity that would allow to tackle those questions for the $L^1$-norm. Thus, in the present paper, and in the absence of a better quantity on the market, we will focus on the OSEE, and discuss only approximability of operators in Hilbert-Schmidt norm.

\subsection{Brief review of entanglement entropy, the area law, and MPSs}

In a Hilbert space with a bipartition $\mathcal{H} = \mathcal{H}_A \otimes \mathcal{H}_B$, any state $\ket{\psi} \in \mathcal{H}$ has a Schmidt decomposition
\begin{equation*}
	\ket{\psi} \, = \, \sum_{i=1}^{r} \sqrt{\lambda_i}  \ket{\psi_{A,i}} \otimes \ket{\psi_{B,i}} 
\end{equation*}
where the $\lambda_i$ are positive real coefficients, ordered as $\lambda_1 \geq \lambda_2 \geq \dots \geq \lambda_r>0$. $r$ is the Schmidt rank, which cannot be larger than ${\rm min} ({\rm dim} \,\mathcal{H}_A, {\rm dim}\, \mathcal{H}_B)$. The $\ket{\psi_{A,i}}$'s (resp. $\ket{\psi_{B,i}}$) are an orthonormal set of states in $\mathcal{H}_A$ ($\mathcal{H}_B$). Although the Schmidt decomposition is not unique, the set of coefficients $\{ \lambda_i \}$ is. For a normalized state, $\braket{\psi}{\psi} = 1$, the {\it entanglement entropy} $S_\alpha ( \ket{\psi} )$ is defined as
\begin{equation}
	\label{eq:EE}
	S_\alpha (\ket{\psi}) \, =\, \frac{1}{1-\alpha} \log \left( \sum_{i=1}^r \lambda_i^\alpha \right) \, , \qquad \alpha >0 \, .
\end{equation}
[In this paper, we always refer to $S_\alpha$ as {\it the entanglement entropy}, without having $\alpha \rightarrow 1$ in mind; we do not give different names to the von Neumann and the Renyi entropies.] Because of the uniqueness of the $\lambda_i$'s, the entanglement entropy is basis independent: if $U_A$ and $U_B$ are two unitary transformations acting on $\mathcal{H}_A$ and $\mathcal{H}_B$ respectively, then $S_\alpha (U_A \otimes U_B \ket{\psi})= S_\alpha (\ket{\psi})$. For integer values of $\alpha \geq 2$, the entanglement entropy can be rewritten as the expectation value of an operator, dubbed {\it swap operator} by Zanardi, Zalka and Faoro \cite{zanardi2000entangling}, in a replicated system. One takes $\alpha$ identical replicas of the quantum state $\ket{\psi}$, thus creating a new state in a larger Hilbert space, $\ket{\psi}^{\otimes \alpha} = \ket{\psi} \otimes \ket{\psi} \otimes \dots \otimes \ket{\psi} \in \mathcal{H}^{\otimes \alpha}$. The swap operator is the operator that cyclically permutes the $\alpha$ replicas of the subsystem $A$, leaving $B$ untouched; we write it as $\mathcal{S}^A_{(1,2,3,\dots , \alpha)}$. The subscript $(1,2,3,\dots , \alpha)$ indicates the permutation of $\alpha$ elements, written in {\it cycle notation}; notice that here the permutation $\sigma$ has a single cycle $(1,\sigma (1) , \sigma\circ \sigma (1),etc.)$ of total length $\alpha$. The entanglement entropy is then
\begin{equation}
	\label{eq:EE_replicas}
	(\alpha \; {\rm integer} \; \geq 2) \qquad  \quad S_\alpha (\ket{\psi}) \, =\, \frac{1}{1-\alpha} \log \left(  \, \bra{\psi}^{\otimes \alpha} \;  \mathcal{S}^A_{(1,2,3,\dots,\alpha)}  \;  \ket{\psi}^{\otimes \alpha} \, \right) \, .
\end{equation}
The main advantage of the latter expression over Eq. (\ref{eq:EE}) is that it makes it clear that the quantity inside the logarithm is an observable \cite{zanardi2000entangling}; it allows to calculate the entanglement entropy in Monte-Carlo simulations \cite{hastings2010measuring}, it is the basic idea of various proposals to measure the entanglement entropy \cite{daley2012measuring,cardy2011measuring,abanin2012measuring}, and it is at heart of the recent groundbreaking experiment \cite{islam2015measuring} that managed to measure $S_2$ in a cold atom setup. It is also the key to field-theory calculations based on the replica trick \cite{calabrese2004entanglement}, although Eq. (\ref{eq:EE_replicas}) usually appears in a somewhat disguised form in that context, as those field theory calculations are typically formulated as path integrals on a replicated worldsheet that involve twist operators (we will come back to this point in section \ref{sec:CFT_thermal}). \vspace{0.3cm}

\noindent Now we turn to the case where $\mathcal{H} = (\mathbb{C}^d)^{\otimes L}$ is the Hilbert space of a spin chain; $d$ is the dimension of the local degree of freedom on each site, and $L$ is the total number of sites. The subsystem $A$ (resp. $B$) is made of the first $L_A$ sites ($L_B= L-L_A$ sites). Then one says that $\ket{\psi}$ obeys the {\it area law} (for a given $\alpha$) if $S_\alpha (\ket{\psi})$ remains bounded when $L_A, L_B \rightarrow \infty$. A famous theorem of Hastings \cite{hastings2007area} states that when $\ket{\psi}$ is the ground state of a gapped local Hamiltonian, it obeys the area law, with $\alpha =1$. Strictly speaking, the case $\alpha=1$ is usually not sufficient to ensure that a state can be approximated by an MPS, see Ref. \cite{schuch2008entropy}---although, for ground states of gapped Hamiltonians, Hastings \cite{hastings2007area} did prove the approximability as well---. As shown by Cirac and Verstraete \cite{verstraete2006matrix}, one sufficient condition for approximability, that is valid beyond ground states, is to obey the area law for $0<\alpha <1$. Then $\left| \psi \right>$ can be approximated as
\begin{equation*}
	\left|\psi \right> \, \simeq \,    \sum_{\{ \sigma_1,\dots,\sigma_L  \} } \sum_{\{ a_1,\dots ,a_{L-1} \}}  [M^{\sigma_1}]_{1 a_1} [M^{\sigma_2}]_{a_1 a_2}  [M^{\sigma_3}]_{a_2 a_3}  \dots [M^{\sigma_L}]_{a_{L-1} 1}   \left|   \sigma_1 , \dots ,\sigma_L   \right>
\end{equation*}
with finite matrices $M^{\sigma_i}$. The latter area law, for $0<\alpha<1$, can also be proved for the ground state of gapped Hamiltonians, see for instance the discussion in Ref. \cite{huang2015classical}. \vspace{0.3cm}

\noindent Notwithstanding these important aspects concerning the role of the (Renyi) parameter $\alpha$, it is fair to say that in most known situations, the behavior of $S_\alpha ( \ket{\psi} )$ does not appear to depend dramatically on the precise value of $\alpha$, and it is often the case in one-dimensional systems that the area law is either obeyed for all $\alpha >0$, or violated for all $\alpha >0$. Another way to say this is that, usually, there is no phase transition as one varies $\alpha$; notice that this is of course a requirement if one wants to use the replica trick. The analyticity in $\alpha$ is known to hold, in particular, for the (single interval) ground state entanglement entropy at quantum critical points described by CFT, where the area law is always violated logarithmically, no matter the precise value of $\alpha$. So in this paper, we will proceed as in the seminal work of Calabrese and Cardy \cite{calabrese2004entanglement}, and many subsequent papers: we will rely on the replica trick, calculate the OSEE for $\alpha$ integer $\geq 2$, and then make the assumption that the scaling one finds (as a function of system size, or of time) is similar for other values of $\alpha$.


\begin{figure}[ht]
	\begin{center}
	\includegraphics[width=0.75\textwidth]{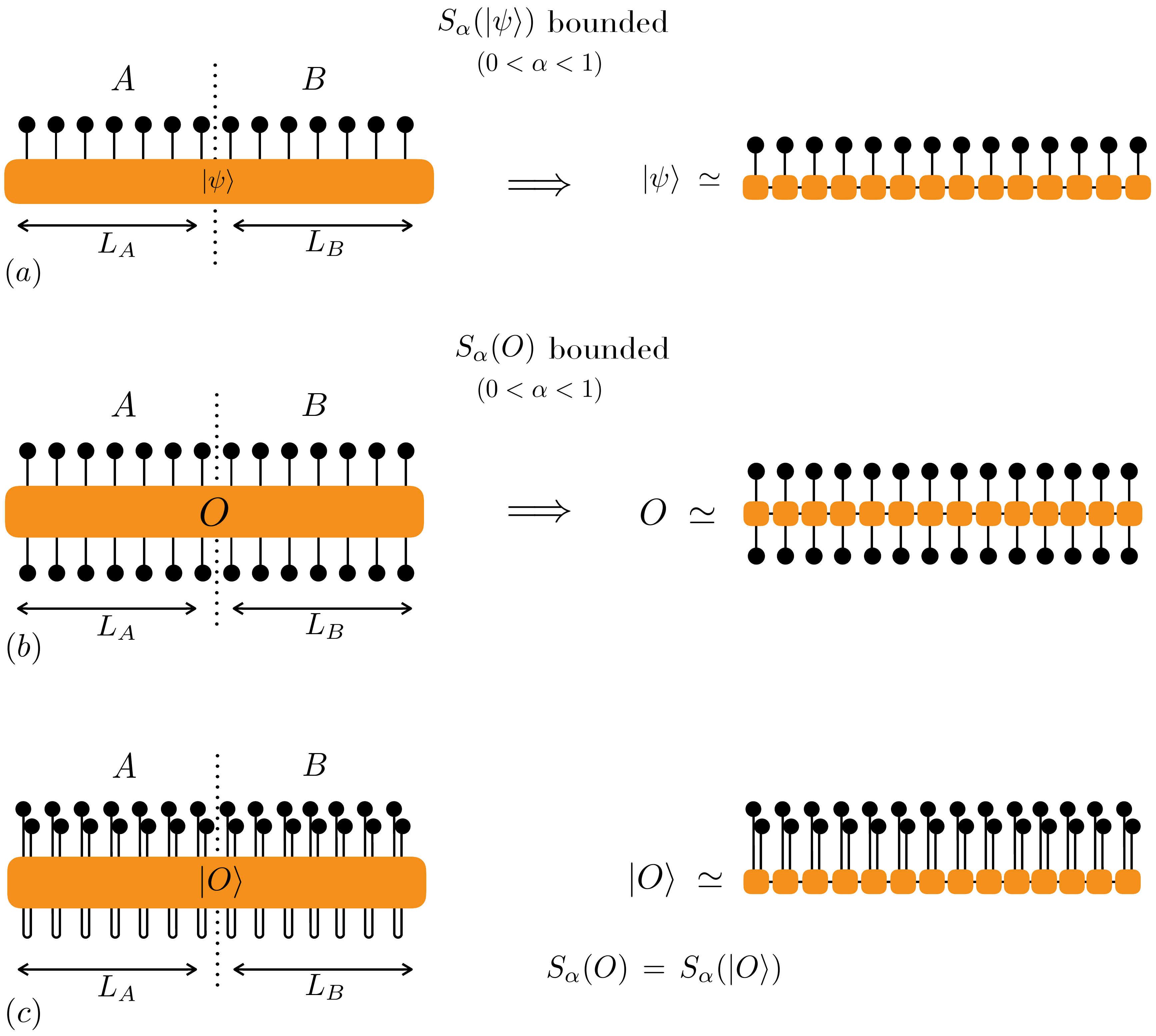}
	\end{center} \vspace{-0.6cm}
	\caption{Cartoon of the area law/operator area law and approximability by MPS/MPOs: (a) the states whose entanglement entropy remains bounded when $L_A,L_B \rightarrow \infty$ can be well approximated by MPSs with small (finite) bond dimension (b) the operators whose operator space entanglement entropy (OSEE) remains bounded can be well approximated (in Hilbert-Schmidt norm) by MPOs with small (finite) bond dimension. (c) Of course, by viewing $O \in {\rm End}(\mathcal{H})$ as a state $\ket{O} \in \mathcal{H} \otimes \overline{\mathcal{H}}$, the two things are exactly the same. This 'operator-folding' trick typically does not simplify analytic calculations, as it merely leads to rewritings of equivalent formulas. But it is sometimes helpful because it allows to use results that are well-established about the entanglement entropy to make analogous claims about the OSEE; it is also useful numerically, to turn MPS-algorithms into MPO-algorithms and vice versa.}
	\label{fig:cartoon_OSEE}
\end{figure}

\subsection{Operators, Operator Space Entanglement Entropy (OSEE), the operator area law, and MPOs}

Instead of approximating states by MPSs, we now want to approximate operators by MPOs. Mathematically, this is essentially the same problem though, since operators on $\mathcal{H}$ are nothing but elements of $\mathcal{H}' = {\rm End}(\mathcal{H})$, which is again a Hilbert space with the Hilbert-Schmidt inner product $\braket{O_1}{O_2}' = {\rm tr [O_1^\dagger O_2]}$, $O_1, O_2 \in {\rm End}(\mathcal{H})$. [To avoid mathematical difficulties, in this introduction we assume that we are dealing with a finite dimensional $\mathcal{H}$, so it is always true that ${\rm End}(\mathcal{H})$ is also a Hilbert space.] So one can simply replace $\mathcal{H}_A$ by $\mathcal{H}_A' = {\rm End}(\mathcal{H}_A)$ and $\mathcal{H}_B$ by $\mathcal{H}_B' = {\rm End}(\mathcal{H}_B)$ in the previous subsection; this straightforwardly leads to the notion of a Schmidt decomposition for the operator $O$, which we chose to write as
\begin{equation}
	\label{eq:schmidt_op}
	\frac{O}{\sqrt{{\rm tr } [O^\dagger O]} }  \, = \, \sum_{i=1}^{r} \sqrt{\lambda_i}  \, O_{A,i} \otimes O_{B,i} \, ,
\end{equation}
where the $\lambda_i$ are positive real coefficients, in decreasing order as above. The operators $O_{A,i} \in {\rm End}(\mathcal{H}_A)$ obey the orthonormality condition ${\rm tr} [O^\dagger_{A,i} O_{A,j}] = \delta_{i,j}$ (same for the $O_{B,i}$'s). In Eq. (\ref{eq:schmidt_op}) we have explicitly included a normalization factor in the left-hand side, because it should be emphasized that, while it is perfectly standard to work with quantum states that are properly normalized ($\braket{\psi}{\psi}=1$), it is unusual to ask that operators on $\mathcal{H}$ should obey a specific normalization condition. Again, the set of coefficients $\{ \lambda_i \}$ is unique, and is invariant under changes of basis of $\mathcal{H}_A$ and $\mathcal{H}_B$. Thus, we are naturally led to the notion of {\it Operator Space Entanglement Entropy} (OSEE), which, to the best of our knowledge, was investigated first by Zanardi in the context of Ref. \cite{zanardi2001entanglement}, then rediscovered by Prosen and Pi\v{z}orn \cite{prosen2007operator}, with subsequent studies in Refs. \cite{vznidarivc2008complexity,vznidarivc2008many,pivzorn2009operator},
\begin{equation}
	\label{eq:Renyi_alpha_op}
	S_\alpha (O) \, =\, \frac{1}{1-\alpha} \log \left( \sum_{i=1}^r \lambda_i^\alpha \right) \, .
\end{equation}
As in the previous subsection, when $\alpha$ is an integer $\geq 2$, one can rewrite $S_\alpha(O)$ in a way that involves the swap operator in a replicated system. The expression involves $\alpha$ replicas of $O$, $O^{\otimes \alpha} = O \otimes \dots \otimes O$, their complex conjugates, and the swap operator that cyclically permutes the replicas of subsystem $A$, $\mathcal{S}^A_{(1,2,3,\dots,\alpha)}$, as well as its inverse $\mathcal{S}^A_{(1,\alpha,\alpha-1,\dots,2)}$---recall that we use the cycle notation for the permutations $(1,\sigma(1), \sigma \circ \sigma(1),etc.)$---. The expression reads
\begin{equation}
	\label{eq:OSEE_replicas}
	(\alpha \; {\rm integer}\; \geq 2) \qquad \quad S_\alpha (O) \, = \, \frac{1}{1-\alpha} \log \left( \frac{{\rm tr} [ (O^\dagger)^{\otimes \alpha}  \cdot \mathcal{S}^A_{(1,\alpha,\alpha-1,\dots,2)} \cdot O^{\otimes \alpha} \cdot \mathcal{S}^A_{(1,2,3,\dots,\alpha)} ]}{  {\rm tr} \left[ (O^\dagger)^{\otimes \alpha}\cdot  O^{\otimes \alpha} \right] } \right)  .
\end{equation}
This is the key expression that we will use throughout this paper to calculate the OSEE; notice that the denominator inside the logarithm could also be written as $\left( {\rm tr} [O^\dagger O] \right)^\alpha$. Notice also that, if one is willing to relate the OSEE of, say, a density matrix $\rho$, to a physical observable, then Eq. (\ref{eq:OSEE_replicas}) does not quite do the job, because the traces involve expressions that are {\it quadratic} in the replicated density matrix $\rho^{\otimes \alpha}$, while expectation values must be {\it linear} in $\rho^{\otimes \alpha}$. It is not difficult to cure this though; but we defer this discussion to the next subsection. \vspace{0.3cm}

\noindent For now, let us focus on the relevance of the OSEE to the approximability by MPOs. Since the OSEE is exactly the same quantity as the 'usual' entanglement entropy once one has replaced the space of states $\mathcal{H}$ by the space of operators $\mathcal{H}'={\rm End}( \mathcal{H})$, one can make exactly the same statements as above. Namely, if we consider again the spin chain $\mathcal{H} = (\mathbb{C}^d)^{\otimes L}$ with subsystems $A$ and $B$ corresponding to the first $L_A$ sites and the last $L_B= L-L_A$ sites, then one says that $O$ obeys the {\it operator area law} (for a given $\alpha$) if $S_\alpha (O)$ remains bounded when $L_A, L_B, \rightarrow \infty$. Then the result of Cirac and Verstraete \cite{verstraete2006matrix} shows that if $O$ obeys the area law for $0<\alpha <1$, then it can be efficiently approximated (in Hilbert-Schmidt norm) by an MPO,
\begin{equation*}
	O \, \simeq \,    \sum_{\underset{\{ \sigma_1,\dots,\sigma_L  \}}{\{ \sigma'_1,\dots,\sigma'_L  \}} } \sum_{\{ a_1,\dots ,a_{L-1} \}}  [M^{\sigma_1'}_{\phantom{\sigma}\sigma_1}]_{1 a_1} [M^{\sigma_2'}_{\phantom{\sigma}\sigma_2}]_{a_1 a_2}  [M^{\sigma_3'}_{\phantom{\sigma}\sigma_3}]_{a_2 a_3} \dots  [M^{\sigma_L'}_{\phantom{\sigma}\sigma_L}]_{a_{L-1} 1}  \,   \left|   \sigma'_1 , \dots ,\sigma'_L   \right>  \left<   \sigma_1 , \dots ,\sigma_L   \right|,
\end{equation*}
with finite matrices $M^{\sigma_i'}_{\phantom{\sigma}\sigma_i}$. The discussion of Ref. \cite{schuch2008entropy} applies to other values of $\alpha$. \vspace{0.3cm}

\noindent Thus, in trying to approximate operators by MPOs, it is important to understand whether or not they violate the operator area law, and if they do violate it, whether this violation is logarithmic, or worse. Prosen and Pi\v{z}orn and collaborators investigated this question in a few concrete examples \cite{prosen2007operator,vznidarivc2008complexity,vznidarivc2008many,pivzorn2009operator}; in this paper, we revisit and extend their results, and in particular, we set up the framework for calculating the OSEE in CFT.

\subsection{Basic remarks about OSEE}

\paragraph{The OSEE of $ \ket{\psi}\bra{\psi}$ is exactly twice the entanglement entropy of $\ket{\psi}$.} Let us emphasize a simple but important point: the usual entanglement entropy is defined for states, while the OSEE is defined for operators. They do not probe the same class of objects, and it is therefore vain to look for a general relation between the two quantities. That being said, there is one case where they are obviously related: when the operator $O$ is the density matrix of a pure state, $\ket{\psi} \bra{\psi}$. In that case, the OSEE is exactly {\it twice} the entanglement entropy of $\ket{\psi}$,
\begin{equation}
	\label{eq:OSEE_pure}
	S_\alpha( \, \ket{\psi} \bra{\psi} \, ) \, = \, 2 \, S_\alpha( \,  \ket{\psi} \, ) .
\end{equation}
This simply follows from the fact that a Schmidt decomposition of the state $\ket{\psi}$ automatically induces a Schmidt decomposition for the operator $\ket{\psi} \bra{\psi}$. This simple relation will show up again in several places in this paper.

\paragraph{The OSEE of a density matrix $\rho$ is measurable.} For integer $\alpha \geq 2$ the OSEE of a density matrix $\rho$ is measurable. Indeed, using the hermiticity of $\rho$, elementary manipulations allow to rewrite Eq. (\ref{eq:OSEE_replicas}) in terms of expectation values in a system with $2\alpha$ replicas---instead of $\alpha$ replicas in Eq. (\ref{eq:OSEE_replicas})---,
\begin{equation*}
	 S_\alpha (\rho) \, = \, \frac{1}{1-\alpha}\log \left( \frac{{\rm tr} \left[ \rho^{\otimes 2\alpha}  \cdot \mathcal{S}^{A\cup B}_{(1,\alpha+1)(2,\alpha+2)\dots(\alpha,2\alpha)} \cdot \mathcal{S}^{A}_{(1,2,3,\dots,\alpha)(\alpha+1,2\alpha,2\alpha-1,\dots,\alpha+2)}  \right]}{  {\rm tr} \left[ \rho^{\otimes 2 \alpha} \cdot  \mathcal{S}^{A\cup B}_{(1,\alpha+1)(2,\alpha+2)\dots (\alpha,2\alpha)}  \right] } \right) \, .
\end{equation*}
Notice that the different swap operators involved act on different subsystems: $\mathcal{S}^A$ acts on the replicas of subsystem $A$ only, while $\mathcal{S}^{A \cup B}$ acts on the full system; notice also that the permutations are different. This formula shows that $S_\alpha (\rho)$ is {\it measurable}, at least in principle, since both the numerator and the denominator in the logarithm are expectation values of observables. Notice also the identity
\begin{eqnarray*}
	&& \mathcal{S}^{A \cup B}_{(1,\alpha+1)(2,\alpha+2)\dots(\alpha,2\alpha)} \mathcal{S}^{A}_{(1,2,3,\dots,\alpha)(\alpha+1,2\alpha,2\alpha-1,\dots,\alpha+2)} \\
	&& \qquad = \, \mathcal{S}^{A}_{(1,\alpha+2)(2,\alpha+3)\dots (\alpha-1,2\alpha) (\alpha,\alpha+1)} \mathcal{S}^{B}_{(1,\alpha+1)(2,\alpha+2)\dots(\alpha,2\alpha)} \, ,
\end{eqnarray*}
which follows from the composition of the two permutations in part $A$, and which leads to a more symmetric expression for the OSEE,
\begin{equation}
	\label{eq:OSEE_replicas_rho}
	  S_\alpha (\rho) \, = \, \frac{1}{1-\alpha} \log \left( \frac{{\rm tr} \left[ \rho^{\otimes 2\alpha}  \cdot \mathcal{S}^{A}_{(1,\alpha+2)(2,\alpha+3)\dots (\alpha-1,2\alpha) (\alpha,\alpha+1)}  \cdot \mathcal{S}^{B}_{(1,\alpha+1)(2,\alpha+2)\dots(\alpha,2\alpha)}  \right]}{  {\rm tr} \left[ \rho^{\otimes 2 \alpha} \cdot  \mathcal{S}^{A\cup B}_{(1,\alpha+1)(2,\alpha+2)\dots (\alpha,2\alpha)}  \right] } \right)  .
\end{equation}

\paragraph{The OSEE is not a measure of entanglement in a mixed state.} Finally, another point, already discussed in Ref. \cite{prosen2007operator}, that deserves to be emphasized is that, even when the operator of interest is a density matrix, the OSEE is not a measure of entanglement in a mixed state. The 'counterexample' pointed out by Prosen and Pi\v{z}orn is a state of the form $\frac{1}{2} \left( \rho_{A,1} \otimes \rho_{B,1}+ \rho_{A,2} \otimes \rho_{B,2}\right)$, which possesses a non-zero OSEE, and yet can be prepared by a protocol that is entirely classical. There is also no obvious relation between the OSEE and other quantities that are usually studied in mixed states, such as {\it e.g.} the mutual information or the negativity; see for instance Ref. \cite{nielsen2003quantum} for a common discussion of all these quantities.

\subsection{Organization of the paper}

In the next section, we illustrate how the OSEE can be calculated in practical situations, and we introduce the CFT tools that are needed later; in particular, we discuss the case of thermal density matrices. In section 3, we discuss the OSEE of the reduced density matrix after a global quench, the approach to the thermal density matrix or Generalized Gibbs Ensemble, and the consequences for numerical simulations of the evolution of one-dimensional quantum systems at large times. In section 4, we turn to the evolution operator $e^{- i H t}$, and argue that its OSEE typically grows linearly in time, which is reminiscent of the ballistic spreading of entanglement. In section 5, we report an attempt at understanding  the OSEE of local operators in Heisenberg picture, which is relevant to Heisenberg-picture DMRG. We conclude in section 6. More details about free fermion calculations, permutations and twist operators in replicated CFT, and about the reported attempt of section 5, are included in three appendices.

\newpage

\section{Calculation of OSEE in two illustrative examples}

To gain some further intuition of what exactly is measured by the OSEE, and of how to calculate it in practice, we start by exploring two simple examples: the translation operator---one of the most basic examples of an MPO---, and the thermal density matrix---which is not an MPO, but it obeys the operator area law, and can therefore be approximated by an MPO---. The thermal density matrix is the occasion for us to set up the framework for CFT calculations; with these tools, we are able to recover the results of Ref. \cite{vznidarivc2008complexity}, where the OSEE of the thermal density matrix was studied with a combination of numerical and exact free-fermion calculations.

\subsection{OSEE of the translation operator}

Consider the operator $T$ that translates all sites by one (modulo $L$) to the right: it acts on the basis states of $\mathcal{H}=(\mathbb{C}^d)^{\otimes L}$ as $T \ket{i_1,i_2,\dots ,i_L} = \ket{i_L,i_1,\dots, i_{L-1}}$. Equivalently, the components of $T$ are $[T]_{\phantom{i} i_1 i_2 \dots i_L}^{i_1' i_2' \dots i_L'} = \delta_{i_L, i_1'} \delta_{i_1, i_2'} \dots \delta_{i_{L-1},i_{L}'}$. It is easy to see that this operator is an MPO with bond dimension $d$: for each pair $(i,i')$ define the $d \times d$ matrix $M^{i'}_{\phantom{a}i}$ that has elements $[M^{i'}_{\phantom{a}i}]_{a b} = \delta_{a,i} \delta_{b,i'}$; then
\begin{equation*}
	[T]_{\phantom{i} i_1 i_2 \dots i_L}^{i_1' i_2' \dots i_L'} \, = \, {\rm tr} \left[ M^{i_1'}_{\phantom{i} i_1}  M^{i_2'}_{\phantom{i} i_2} \dots  M^{i_L'}_{\phantom{i} i_L} \right]  .
\end{equation*}
Now take $\mathcal{H}_A$ (resp. $\mathcal{H}_B$) as the first $L_A$ sites (resp. $L_B=L-L_A$ last sites) of the chain, and let us calculate the OSEE. We need to introduce the 'partial translators' $T_{A, a ,b}$, defined as
\begin{equation*}
	[T_{A,a,b}]_{\phantom{i} i_1 i_2 \dots i_{L_A}}^{i_1' i_2' \dots i_{L_A}'} \, = \,  \delta_{a, i_1'} \delta_{i_1, i_2'} \dots \delta_{i_{L_A},b} \, ,
\end{equation*}
as well as partial translators on part $B$, $T_{B,b,a}$, with a similar definition. We have
\begin{equation*}
	T \, = \, \sum_{a=1}^d \sum_{b=1}^d T_{A,a,b} \otimes T_{B,b,a} \, .
\end{equation*}
This is close to the Schmidt decomposition (\ref{eq:schmidt_op}), however we still need to check that the different term are orthonormal.
First, notice that in the left-hand side, $T$ has norm ${\rm tr} [T^\dagger T] = d^L$, which follows from the fact that $T^\dagger T$ is the identity on $\mathcal{H}$. Second, the inner product between two partial translators in the right-hand side is
\begin{eqnarray*}
	{\rm tr} \left[ T_{A,a,b}^\dagger T_{A,a',b'}  \right] &=& \sum_{\{i\},\{ i'\} } [T^\dagger_{A,a,b}]_{\phantom{i} i_1' i_2' \dots i_{L_A}'}^{i_1 i_2 \dots i_{L_A}} \times  [T_{A,a',b'}]_{\phantom{i} i_1 i_2 \dots i_{L_A}}^{i_1' i_2' \dots i_{L_A}'} \\
	&=&  \sum_{\{i\},\{ i'\} }    \delta_{a,i_1'} \delta_{i_1, i_2'} \dots \delta_{i_{L_A},b}   \times   \delta_{a',i_1'} \delta_{i_1,i_2'} \dots \delta_{i_{L_A},b'} \\
	&=& d^{L_A-1} \, \delta_{a,a'} \,  \delta_{b,b'} \, ,
\end{eqnarray*}
so the partial translators are orthogonal, with some fixed normalization factor. Thus, the correct way to write the Schmidt decomposition of $T$, with properly normalized terms, is
\begin{equation*}
	\frac{1}{d^{L/2}} T \, = \,   \sum_{a=1}^d \sum_{b=1}^d \frac{1}{d} \left( \frac{1}{d^{(L_A-1)/2}} T_{A,a,b} \right) \otimes \left( \frac{1}{d^{(L_B-1)/2}}  T_{B,b,a} \right) .
\end{equation*}
There are exactly $d^2$ non-zero Schmidt coefficients, all equal to $1/d$. It follows that
\begin{equation}
	S_\alpha( T ) \, = \, \frac{1}{1-\alpha} \log \left(  \sum_{a,b} (1/d^2)^\alpha \right) \, = \, 2 \,\log  d  ,
\end{equation}
independently of $\alpha$. Thus, in this simple example, the OSEE is simply the log of the bond dimension of the MPO, counted twice because there are two boundary points that separate $A$ and $B$.\vspace{0.3cm}

\noindent Alternatlvely, instead of writing the Schmidt decomposition explicitly, we could have used the formula with the replicas (\ref{eq:OSEE_replicas}), leading to the following amusing calculation. Let us represent the operator $T$ graphically, as
$$
\begin{tikzpicture}
	\draw (-1.5,0.5) node{$T \, =$};
	\draw[line width=2pt] (-0.45,0.78) arc (90:270:0.3cm and 0.12cm) arc (-90:0:0.55cm);
	\draw[line width=2pt] (0,0) arc (180:90:0.55cm) arc (-90:0:0.55cm);
	\draw[line width=2pt] (1,0) arc (180:90:0.55cm) arc (-90:0:0.55cm);
	\draw[line width=2pt] (2,0) arc (180:90:0.55cm) arc (-90:0:0.55cm);
	\draw[line width=2pt] (3,0) arc (180:90:0.55cm) arc (-90:0:0.55cm);
	\draw[line width=2pt] (4,0) arc (180:90:0.55cm) arc (-90:0:0.55cm);
	\draw[line width=2pt] (5,0) arc (180:90:0.55cm) arc (-90:0:0.55cm);
	\draw[line width=2pt] (6,0) arc (180:90:0.55cm) arc (-90:0:0.55cm);
	\draw[line width=2pt] (7,0) arc (180:90:0.55cm) arc (-90:90:0.3cm and 0.12cm);
	\draw[line width=1.8pt,dashed] (-0.4,0.78) -- (7.6,0.78);	
	\draw (0,-0.3) node{\small $1$};
	\draw (1,-0.3) node{\small $2$};
	\draw (7,-0.3) node{\small $L$};
\end{tikzpicture} \vspace{-0.3cm}
$$
where each line corresponds to a Kronecker delta $\delta_{i, i'}$. Then, graphically, one just has to count the number
of closed loops; each loop gives a factor $d$. For instance, the squared norm of $T$ is given by
$$
\begin{tikzpicture}
	\draw (-2,1.2) node{${\rm tr} [T^\dagger T] \, =$};
	\draw[line width=1.8pt,gray] (0,-0.02) arc (-110:110:0.2cm and 1.15cm);
	\draw[line width=1.8pt,gray] (1,-0.02) arc (-110:110:0.2cm and 1.15cm);
	\draw[line width=1.8pt,gray] (2,-0.02) arc (-110:110:0.2cm and 1.15cm);
	\draw[line width=1.8pt,gray] (3,-0.02) arc (-110:110:0.2cm and 1.15cm);
	\draw[line width=1.8pt,gray] (4,-0.02) arc (-110:110:0.2cm and 1.15cm);
	\draw[line width=1.8pt,gray] (5,-0.02) arc (-110:110:0.2cm and 1.15cm);
	\draw[line width=1.8pt,gray] (6,-0.02) arc (-110:110:0.2cm and 1.15cm);
	\draw[line width=1.8pt,gray] (7,-0.02) arc (-110:110:0.2cm and 1.15cm);
	\begin{scope}
		\draw[line width=2pt] (-0.45,0.78) arc (90:270:0.3cm and 0.12cm) arc (-90:0:0.55cm);
		\draw[line width=2pt] (0,0) arc (180:90:0.55cm) arc (-90:0:0.55cm);
		\draw[line width=2pt] (1,0) arc (180:90:0.55cm) arc (-90:0:0.55cm);
		\draw[line width=2pt] (2,0) arc (180:90:0.55cm) arc (-90:0:0.55cm);
		\draw[line width=2pt] (3,0) arc (180:90:0.55cm) arc (-90:0:0.55cm);
		\draw[line width=2pt] (4,0) arc (180:90:0.55cm) arc (-90:0:0.55cm);
		\draw[line width=2pt] (5,0) arc (180:90:0.55cm) arc (-90:0:0.55cm);
		\draw[line width=2pt] (6,0) arc (180:90:0.55cm) arc (-90:0:0.55cm);
		\draw[line width=2pt] (7,0) arc (180:90:0.55cm) arc (-90:90:0.3cm and 0.12cm);
		\draw[line width=1.8pt,dashed] (-0.4,0.78) -- (7.6,0.78);	
	\end{scope}
	\begin{scope}[yshift=2.15cm]
		\draw[line width=2pt] (-0.45,-0.3) arc (90:270:0.3cm and 0.12cm) arc (90:0:0.55cm);
		\draw[line width=2pt] (0,0) arc (180:270:0.55cm) arc (90:0:0.55cm);
		\draw[line width=2pt] (1,0) arc (180:270:0.55cm) arc (90:0:0.55cm);
		\draw[line width=2pt] (2,0) arc (180:270:0.55cm) arc (90:0:0.55cm);
		\draw[line width=2pt] (3,0) arc (180:270:0.55cm) arc (90:0:0.55cm);
		\draw[line width=2pt] (4,0) arc (180:270:0.55cm) arc (90:0:0.55cm);
		\draw[line width=2pt] (5,0) arc (180:270:0.55cm) arc (90:0:0.55cm);
		\draw[line width=2pt] (6,0) arc (180:270:0.55cm) arc (90:0:0.55cm);
		\draw[line width=2pt] (7,0) arc (180:270:0.55cm) arc (-90:90:0.3cm and 0.12cm);
		\draw[line width=1.8pt,dashed] (-0.4,-0.3) -- (7.6,-0.3);
	\end{scope}
	\draw (0,-0.3) node{\small $1$};
	\draw (1,-0.3) node{\small $2$};
	\draw (7,-0.3) node{\small $L$};
	\draw (8.5,1.2) node{$ = \; d^{L} .$};
\end{tikzpicture} \vspace{-0.3cm}
$$
Now let us evaluate ${\rm tr} [ (T^\dagger)^{\otimes \alpha}  \cdot \mathcal{S}_{(1,\alpha,\alpha-1,\dots,2)} \cdot T^{\otimes \alpha} \cdot \mathcal{S}_{(1,2,3,\dots,\alpha)} ]$. For simplicity, we draw the case $\alpha=2$ only. We represent the replicated $T$ as
$$
\begin{tikzpicture}
	\draw (-1.5,0.5) node{$T^{\otimes 2} \, =$};
	\begin{scope}
		\draw[line width=2pt] (-0.45,0.78) arc (90:270:0.3cm and 0.12cm) arc (-90:0:0.55cm);
		\draw[line width=2pt] (0,0) arc (180:90:0.55cm) arc (-90:0:0.55cm);
		\draw[line width=2pt] (1,0) arc (180:90:0.55cm) arc (-90:0:0.55cm);
		\draw[line width=2pt] (2,0) arc (180:90:0.55cm) arc (-90:0:0.55cm);
		\draw[line width=2pt] (3,0) arc (180:90:0.55cm) arc (-90:0:0.55cm);
		\draw[line width=2pt] (4,0) arc (180:90:0.55cm) arc (-90:0:0.55cm);
		\draw[line width=2pt] (5,0) arc (180:90:0.55cm) arc (-90:0:0.55cm);
		\draw[line width=2pt] (6,0) arc (180:90:0.55cm) arc (-90:0:0.55cm);
		\draw[line width=2pt] (7,0) arc (180:90:0.55cm) arc (-90:90:0.3cm and 0.12cm);
		\draw[line width=1.8pt,dashed] (-0.4,0.78) -- (7.6,0.78);	
		\draw (0,-0.3) node{\small $1$};
		\draw (1,-0.3) node{\small $2$};
		\draw (7,-0.3) node{\small $L$};
	\end{scope}
	\begin{scope}[xshift=0.1cm,yshift=-0.1cm]
		\draw[line width=2pt] (-0.45,0.78) arc (90:270:0.3cm and 0.12cm) arc (-90:0:0.55cm);
		\draw[line width=2pt] (0,0) arc (180:90:0.55cm) arc (-90:0:0.55cm);
		\draw[line width=2pt] (1,0) arc (180:90:0.55cm) arc (-90:0:0.55cm);
		\draw[line width=2pt] (2,0) arc (180:90:0.55cm) arc (-90:0:0.55cm);
		\draw[line width=2pt] (3,0) arc (180:90:0.55cm) arc (-90:0:0.55cm);
		\draw[line width=2pt] (4,0) arc (180:90:0.55cm) arc (-90:0:0.55cm);
		\draw[line width=2pt] (5,0) arc (180:90:0.55cm) arc (-90:0:0.55cm);
		\draw[line width=2pt] (6,0) arc (180:90:0.55cm) arc (-90:0:0.55cm);
		\draw[line width=2pt] (7,0) arc (180:90:0.55cm) arc (-90:90:0.3cm and 0.12cm);
		\draw[line width=1.8pt,dashed] (-0.4,0.78) -- (7.6,0.78);	
	\end{scope}
\end{tikzpicture} \vspace{-0.3cm}
$$
and the swap operator that permutes the two replicas of part $A$ as
$$
\begin{tikzpicture}
	\draw (-1.2,0.5) node{$\mathcal{S}^A_{(1,2)} \; =$};
	\begin{scope}
		\draw[line width=1.8pt] (4,0) -- ++(0,1);
		\draw[line width=1.8pt] (5,0) -- ++(0,1);
		\draw[line width=1.8pt] (6,0) -- ++(0,1);
		\draw[line width=1.8pt] (7,0) -- ++(0,1);
		\draw[line width=1.8pt] (4+0.1,-0.1) -- ++(0,1);
		\draw[line width=1.8pt] (5+0.1,-0.1) -- ++(0,1);
		\draw[line width=1.8pt] (6+0.1,-0.1) -- ++(0,1);
		\draw[line width=1.8pt] (7+0.1,-0.1) -- ++(0,1);
		\draw[line width=1.8pt] (0,0) -- ++(0.1,0.9);
		\draw[line width=1.8pt] (0,0) ++(0.1,-0.1) -- ++(-0.1,1.1);
		\draw[line width=1.8pt] (1,0) -- ++(0.1,0.9);
		\draw[line width=1.8pt] (1,0) ++(0.1,-0.1) -- ++(-0.1,1.1);
		\draw[line width=1.8pt] (2,0) -- ++(0.1,0.9);
		\draw[line width=1.8pt] (2,0) ++(0.1,-0.1) -- ++(-0.1,1.1);
		\draw[line width=1.8pt] (3,0) -- ++(0.1,0.9);
		\draw[line width=1.8pt] (3,0) ++(0.1,-0.1) -- ++(-0.1,1.1);
	\end{scope}
	\draw (0,-0.3) node{\small $1$};
	\draw (1,-0.3) node{\small $2$};
	\draw (3,-0.3) node{\small $L_A$};
	\draw (4.3,-0.3) node{\small $L_A+1$};
	\draw (7,-0.3) node{\small $L$};
	\draw (7.5,0.2) node{.};
\end{tikzpicture}   \vspace{-0.3cm}
$$
Then we find, counting the number of closed loops on the diagram, that
$$
\begin{tikzpicture}
	\draw (-4.1,2) node{${\rm tr} [ (T^\dagger)^{\otimes 2}  \cdot \mathcal{S}_{(1,2)} \cdot T^{\otimes 2} \cdot \mathcal{S}_{(1,2)} ]$};
	\draw (-1.3,2) node{$=$};
	\draw (9,2) node{$= \; d^{2L-2}$.};
	\draw[line width=1.6pt,gray] (0,-0.02) arc (-120:120:0.2cm and 2.45cm);
	\draw[line width=1.6pt,gray] (0.1,-0.12) arc (-120:120:0.2cm and 2.45cm);
	\draw[line width=1.6pt,gray] (1,-0.02) arc (-120:120:0.2cm and 2.45cm);
	\draw[line width=1.6pt,gray] (1.1,-0.12) arc (-120:120:0.2cm and 2.45cm);
	\draw[line width=1.6pt,gray] (2,-0.02) arc (-120:120:0.2cm and 2.45cm);
	\draw[line width=1.6pt,gray] (2.1,-0.12) arc (-120:120:0.2cm and 2.45cm);
	\draw[line width=1.6pt,gray] (3,-0.02) arc (-120:120:0.2cm and 2.45cm);
	\draw[line width=1.6pt,gray] (3.1,-0.12) arc (-120:120:0.2cm and 2.45cm);
	\draw[line width=1.6pt,gray] (4,-0.02) arc (-120:120:0.2cm and 2.45cm);
	\draw[line width=1.6pt,gray] (4.1,-0.12) arc (-120:120:0.2cm and 2.45cm);
	\draw[line width=1.6pt,gray] (5,-0.02) arc (-120:120:0.2cm and 2.45cm);
	\draw[line width=1.6pt,gray] (5.1,-0.12) arc (-120:120:0.2cm and 2.45cm);
	\draw[line width=1.6pt,gray] (6,-0.02) arc (-120:120:0.2cm and 2.45cm);
	\draw[line width=1.6pt,gray] (6.1,-0.12) arc (-120:120:0.2cm and 2.45cm);
	\draw[line width=1.6pt,gray] (7,-0.02) arc (-120:120:0.2cm and 2.45cm);
	\draw[line width=1.6pt,gray] (7.1,-0.12) arc (-120:120:0.2cm and 2.45cm);
	\begin{scope}
		\draw[line width=2pt] (-0.45,0.78) arc (90:270:0.3cm and 0.12cm) arc (-90:0:0.55cm);
		\draw[line width=2pt] (0,0) arc (180:90:0.55cm) arc (-90:0:0.55cm);
		\draw[line width=2pt] (1,0) arc (180:90:0.55cm) arc (-90:0:0.55cm);
		\draw[line width=2pt] (2,0) arc (180:90:0.55cm) arc (-90:0:0.55cm);
		\draw[line width=2pt] (3,0) arc (180:90:0.55cm) arc (-90:0:0.55cm);
		\draw[line width=2pt] (4,0) arc (180:90:0.55cm) arc (-90:0:0.55cm);
		\draw[line width=2pt] (5,0) arc (180:90:0.55cm) arc (-90:0:0.55cm);
		\draw[line width=2pt] (6,0) arc (180:90:0.55cm) arc (-90:0:0.55cm);
		\draw[line width=2pt] (7,0) arc (180:90:0.55cm) arc (-90:90:0.3cm and 0.12cm);
		\draw[line width=1.8pt,dashed] (-0.4,0.78) -- (7.6,0.78);	
	\end{scope}
	\begin{scope}[xshift=0.1cm,yshift=-0.1cm]
		\draw[line width=2pt] (-0.45,0.78) arc (90:270:0.3cm and 0.12cm) arc (-90:0:0.55cm);
		\draw[line width=2pt] (0,0) arc (180:90:0.55cm) arc (-90:0:0.55cm);
		\draw[line width=2pt] (1,0) arc (180:90:0.55cm) arc (-90:0:0.55cm);
		\draw[line width=2pt] (2,0) arc (180:90:0.55cm) arc (-90:0:0.55cm);
		\draw[line width=2pt] (3,0) arc (180:90:0.55cm) arc (-90:0:0.55cm);
		\draw[line width=2pt] (4,0) arc (180:90:0.55cm) arc (-90:0:0.55cm);
		\draw[line width=2pt] (5,0) arc (180:90:0.55cm) arc (-90:0:0.55cm);
		\draw[line width=2pt] (6,0) arc (180:90:0.55cm) arc (-90:0:0.55cm);
		\draw[line width=2pt] (7,0) arc (180:90:0.55cm) arc (-90:90:0.3cm and 0.12cm);
		\draw[line width=1.8pt,dashed] (-0.4,0.78) -- (7.6,0.78);	
	\end{scope}
	\begin{scope}[yshift=1.1cm,xshift=0.1cm]
		\draw[line width=1.8pt] (4,0) -- ++(0,1);
		\draw[line width=1.8pt] (5,0) -- ++(0,1);
		\draw[line width=1.8pt] (6,0) -- ++(0,1);
		\draw[line width=1.8pt] (7,0) -- ++(0,1);
		\draw[line width=1.8pt] (4+0.1,-0.1) -- ++(0,1);
		\draw[line width=1.8pt] (5+0.1,-0.1) -- ++(0,1);
		\draw[line width=1.8pt] (6+0.1,-0.1) -- ++(0,1);
		\draw[line width=1.8pt] (7+0.1,-0.1) -- ++(0,1);
		\draw[line width=1.8pt] (0,0) -- ++(0.1,0.9);
		\draw[line width=1.8pt] (0,0) ++(0.1,-0.1) -- ++(-0.1,1.1);
		\draw[line width=1.8pt] (1,0) -- ++(0.1,0.9);
		\draw[line width=1.8pt] (1,0) ++(0.1,-0.1) -- ++(-0.1,1.1);
		\draw[line width=1.8pt] (2,0) -- ++(0.1,0.9);
		\draw[line width=1.8pt] (2,0) ++(0.1,-0.1) -- ++(-0.1,1.1);
		\draw[line width=1.8pt] (3,0) -- ++(0.1,0.9);
		\draw[line width=1.8pt] (3,0) ++(0.1,-0.1) -- ++(-0.1,1.1);
	\end{scope}
	\begin{scope}[yshift=3.2cm]
		\draw[line width=2pt] (-0.45,-0.3) arc (90:270:0.3cm and 0.12cm) arc (90:0:0.55cm);
		\draw[line width=2pt] (0,0) arc (180:270:0.55cm) arc (90:0:0.55cm);
		\draw[line width=2pt] (1,0) arc (180:270:0.55cm) arc (90:0:0.55cm);
		\draw[line width=2pt] (2,0) arc (180:270:0.55cm) arc (90:0:0.55cm);
		\draw[line width=2pt] (3,0) arc (180:270:0.55cm) arc (90:0:0.55cm);
		\draw[line width=2pt] (4,0) arc (180:270:0.55cm) arc (90:0:0.55cm);
		\draw[line width=2pt] (5,0) arc (180:270:0.55cm) arc (90:0:0.55cm);
		\draw[line width=2pt] (6,0) arc (180:270:0.55cm) arc (90:0:0.55cm);
		\draw[line width=2pt] (7,0) arc (180:270:0.55cm) arc (-90:90:0.3cm and 0.12cm);
		\draw[line width=1.8pt,dashed] (-0.4,-0.3) -- (7.6,-0.3);
	\end{scope}
	\begin{scope}[yshift=3.1cm,xshift=0.1cm]
		\draw[line width=2pt] (-0.45,-0.3) arc (90:270:0.3cm and 0.12cm) arc (90:0:0.55cm);
		\draw[line width=2pt] (0,0) arc (180:270:0.55cm) arc (90:0:0.55cm);
		\draw[line width=2pt] (1,0) arc (180:270:0.55cm) arc (90:0:0.55cm);
		\draw[line width=2pt] (2,0) arc (180:270:0.55cm) arc (90:0:0.55cm);
		\draw[line width=2pt] (3,0) arc (180:270:0.55cm) arc (90:0:0.55cm);
		\draw[line width=2pt] (4,0) arc (180:270:0.55cm) arc (90:0:0.55cm);
		\draw[line width=2pt] (5,0) arc (180:270:0.55cm) arc (90:0:0.55cm);
		\draw[line width=2pt] (6,0) arc (180:270:0.55cm) arc (90:0:0.55cm);
		\draw[line width=2pt] (7,0) arc (180:270:0.55cm) arc (-90:90:0.3cm and 0.12cm);
		\draw[line width=1.8pt,dashed] (-0.4,-0.3) -- (7.6,-0.3);
	\end{scope}
	\begin{scope}[yshift=3.2cm]
		\draw[line width=1.8pt] (4,0) -- ++(0,1);
		\draw[line width=1.8pt] (5,0) -- ++(0,1);
		\draw[line width=1.8pt] (6,0) -- ++(0,1);
		\draw[line width=1.8pt] (7,0) -- ++(0,1);
		\draw[line width=1.8pt] (4+0.1,-0.1) -- ++(0,1);
		\draw[line width=1.8pt] (5+0.1,-0.1) -- ++(0,1);
		\draw[line width=1.8pt] (6+0.1,-0.1) -- ++(0,1);
		\draw[line width=1.8pt] (7+0.1,-0.1) -- ++(0,1);
		\draw[line width=1.8pt] (0,0) -- ++(0.1,0.9);
		\draw[line width=1.8pt] (0,0) ++(0.1,-0.1) -- ++(-0.1,1.1);
		\draw[line width=1.8pt] (1,0) -- ++(0.1,0.9);
		\draw[line width=1.8pt] (1,0) ++(0.1,-0.1) -- ++(-0.1,1.1);
		\draw[line width=1.8pt] (2,0) -- ++(0.1,0.9);
		\draw[line width=1.8pt] (2,0) ++(0.1,-0.1) -- ++(-0.1,1.1);
		\draw[line width=1.8pt] (3,0) -- ++(0.1,0.9);
		\draw[line width=1.8pt] (3,0) ++(0.1,-0.1) -- ++(-0.1,1.1);
	\end{scope}
	\draw (0,-0.7) node{\small $1$};
	\draw (1,-0.7) node{\small $2$};
	\draw (3,-0.7) node{\small $L_A$};
	\draw (4.3,-0.7) node{\small $L_A+1$};
	\draw (7,-0.7) node{\small $L$};
\end{tikzpicture} \vspace{-0.3cm}
$$
More generally, for $\alpha \geq 2$, one finds ${\rm tr} [ (T^\dagger)^{\otimes \alpha}  \cdot \mathcal{S}_{(1,\alpha,\alpha-1,\dots,2)} \cdot T^{\otimes \alpha} \cdot \mathcal{S}_{(1,2,3,\dots,\alpha)} ]= d^{\alpha (L-1)}$. Plugging this into Eq. (\ref{eq:OSEE_replicas}), we recover the fact that $S_\alpha(T) = 2\, \log d$, independently of $\alpha$.

\begin{figure}[ht]
	\begin{center}
		\includegraphics[width=0.8\textwidth]{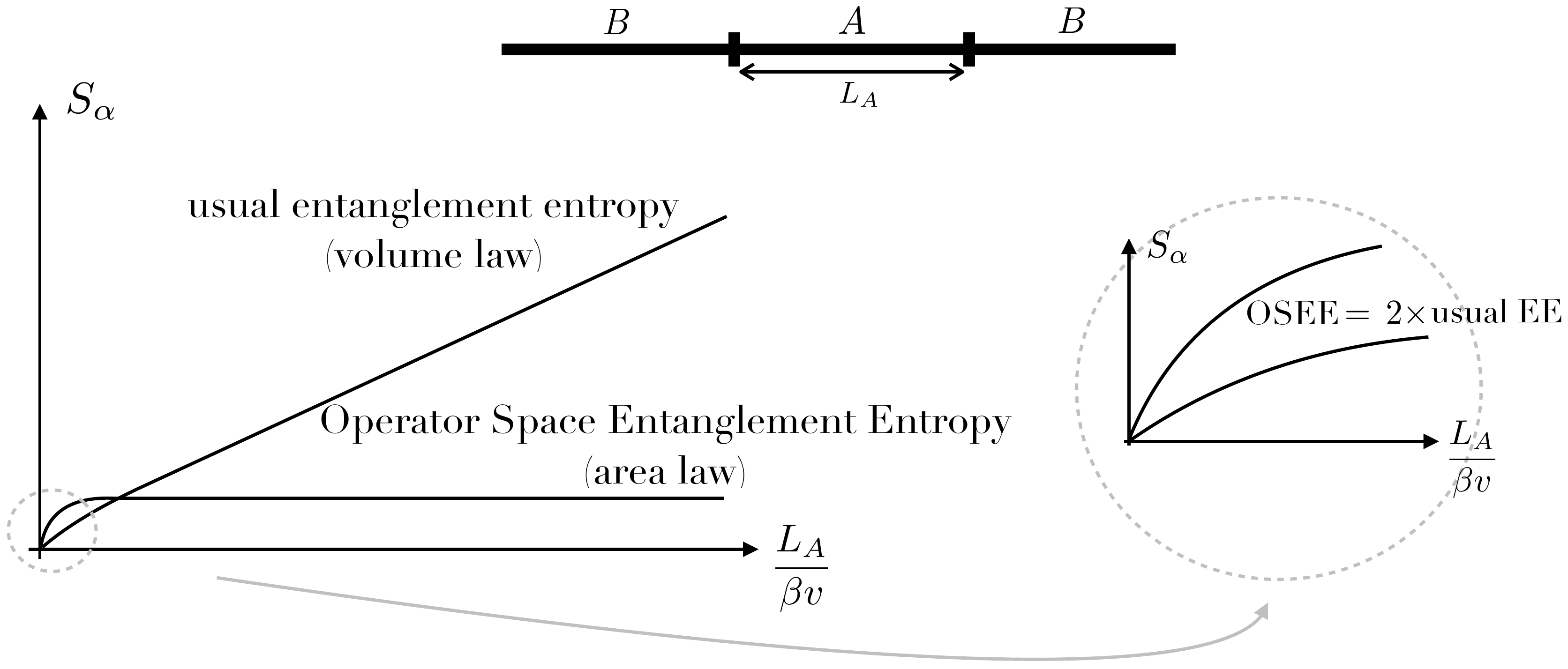}
	\end{center}
	\caption{Cartoon of an infinitely long one-dimensional quantum system at finite temperature $1/\beta$, bipartitioned as $A \cup B$ where $A$ is a finite interval. The entanglement entropy famously obeys a volume law, while the OSEE of the thermal density matrix $\rho_\beta \, \propto \, e^{-\beta H}$ saturates as $L_A \gg \beta v$: the thermal density matrix obeys an {\it operator area law}. In a CFT, $v$ is the velocity of gapless excitations; in more general systems, $v$ would be some natural velocity, for instance a Lieb-Robinson velocity. In the opposite limit of low temperatures, $\beta  v \gg L_A$, the thermal density matrix is close to the projector onto the ground-state, so relation (\ref{eq:OSEE_pure}) holds.}
	\label{fig:thermal}
\end{figure}

\subsection{OSEE from CFT: the case of the thermal (Gibbs) density matrix}
\label{sec:CFT_thermal}

Next, consider an infinitely long critical system described by conformal field theory, with a Hamiltonian $H$ and gapless excitations that propagate at velocity $v$. The system is at inverse temperature $\beta$. If we think of this setup as being the effective, large-scale, description of some microscopic model, then we have to assume that $\beta v \gg a_0$ where $a_0$ is some characteristic UV length scale ({\it e.g.} lattice spacing or Fermi wavelength). \vspace{0.4cm}

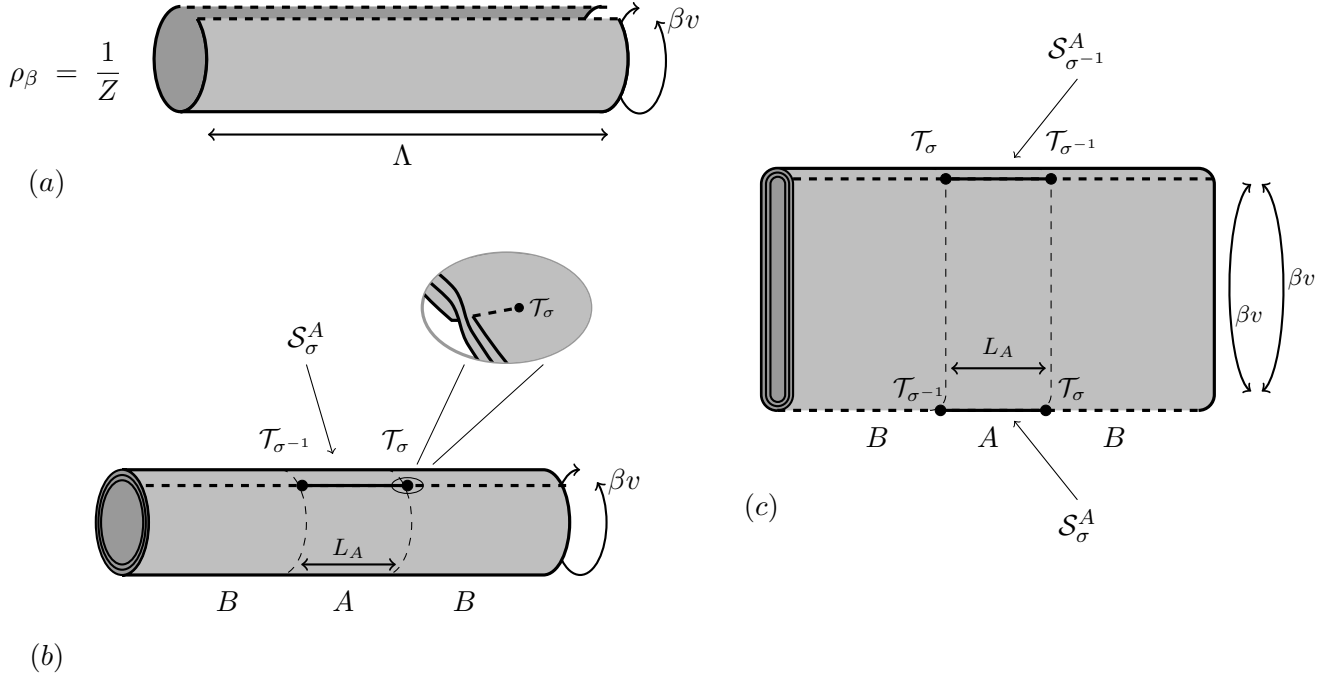
\begin{figure}[ht]
	\begin{tikzpicture}
	\begin{scope}[scale=0.7]
		\draw[thick,<->] (8.4,0.2) arc (-270:50:0.5cm and 1cm) node[right]{$\beta v$};
		\filldraw[very thick,draw=none,fill=gray!80] (0,0) arc (50:-270:0.5cm and 1cm) -- ++(8,0) arc (-270:50:0.5cm and 1cm) -- ++(-8,0);
		\draw[very thick] (0,0) arc (50:-270:0.5cm and 1cm) ++(8,0) arc (-270:50:0.5cm and 1cm) ++(-8,0);
		\filldraw[very thick,draw=none,fill=gray!50] (0,0) arc (50:-90:0.5cm and 1cm) -- ++(8,0) arc (-90:50:0.5cm and 1cm) -- ++(-8,0);
		\draw[very thick,dashed] (-0.2,0.22) -- ++(8,0);
		\draw[very thick,dashed] (0,0) -- ++(8,0);
		\draw[very thick] (0,0) arc (50:-90:0.5cm and 1cm) -- ++(8,0) arc (-90:50:0.5cm and 1cm);
		\draw[thick,<->] (0.2,-2.2) -- (7.8,-2.2);
		\draw (3.9,-2.6) node{$\Lambda$};
		\draw (-2.5,-1) node{$\rho_\beta \; = \; \displaystyle \frac{1}{Z}$};
	\end{scope}
	\draw (-2,-2.2) node{$(a)$};
	
	\begin{scope}[xshift=-1cm,yshift=-6cm,scale=0.7]
		\draw[thick,<->] (8.7,0) arc (-270:50:0.5cm and 1cm) node[right]{$\beta v$};
		\filldraw[very thick,draw=black,fill=gray!50] (0,0) arc (90:-90:0.5cm and 1cm) -- ++(8,0) arc (-90:90:0.5cm and 1cm) -- ++(-8,0);
		\filldraw[thick,draw=black,fill=gray!80] (0,-1) ellipse (0.5cm and 1cm);
		\filldraw[thick,draw=black,fill=gray!80] (0,-1) ellipse (0.45cm and 0.9cm);
		\filldraw[thick,draw=black,fill=gray!80] (0,-1) ellipse (0.4cm and 0.8cm);
		\draw[very thick, dashed] (0.45,-0.3) -- ++(8,0);
		\draw[very thick] (3.42,-0.3) -- ++(2,0);
		\filldraw (3.42,-0.3) circle (1mm);
		\filldraw (5.42,-0.3) circle (1mm);
		\draw[dashed] (3,0) arc (90:-90:0.5cm and 1cm);
		\draw[dashed] (5,0) arc (90:-90:0.5cm and 1cm);
		\draw (4.2,-2.5) node{$A$};
		\draw (2,-2.5) node{$B$};
		\draw (6.5,-2.5) node{$B$};
		\draw[thick,<->] (3.4,-1.8) -- ++(1.8,0);
		\draw (4.3,-1.5) node{\footnotesize $L_A$};
		\draw (3.1,0.55) node{\small $\mathcal{T}_{\sigma^{-1}}$};
		\draw (5.2,0.55) node{\small $\mathcal{T}_\sigma$};

		\draw[very thin] (5.6,0.1) -- (6.5,2);
		\draw[very thin] (5.9,0.1) -- (8,2);
		\draw[black] (5.42,-0.3) ellipse (0.3cm and 0.15cm);		
		
		\begin{scope}[xshift=6.5cm,yshift=3cm,scale=0.8]
			\draw[very thick,gray!80] (1,0.1) ellipse (2cm and 1.3cm);		
			\clip (1,0.1) ellipse (2cm and 1.3cm);		
			\filldraw [draw=black,fill=gray!50,very thick] plot [smooth] coordinates {(1.8,1.4) (-1.3,1.1) (-0.3,-0.2)} -- ++(3,0) -- ++(0,1);
			\filldraw [draw=black,fill=gray!50,very thick] plot [smooth cycle] coordinates {(-1.5,1.7) (-0.3,0.2) (0.1,-0.8) (1.3,-2.3) (2.9,-2.3) (2.9,1.7)};
			\filldraw [draw=black,fill=gray!50,very thick] plot [smooth cycle] coordinates {(-1.4,2) (-0.2,0.5) (0.2,-0.5) (1.4,-2) (3,-2) (3,2)};
			\filldraw [draw=black,fill=gray!50,very thick] plot [smooth] coordinates {(0.2,-0.1) (1.5,-1.7) (3.1,-1.7) (3.1,1.7)};
			\draw[very thick, dashed] (0.2,-0.1) -- (1.3,0.1);
			\filldraw (1.3,0.1) circle (1mm) node[right]{\footnotesize$\mathcal{T}_{\sigma}$};
		\end{scope}
		\draw[->] (3.5,2) node[above]{$\mathcal{S}^A_{\sigma}$} -- (4,0.3);
	\end{scope}
	\draw (-2,-8.5) node{$(b)$};

	\begin{scope}[xshift=7.5cm,yshift=-5cm,scale=0.7]
		\draw[thick,<->] (9.5,4) arc (80:-80:0.5cm and 2cm);
		\draw[thick,<->] (9.3,4) arc (100:260:0.5cm and 2cm);
		\draw (9.8,2.2) node[right]{\footnotesize$\beta v$};
		\draw (8.8,1.5) node[right]{\footnotesize$\beta v$};
		\filldraw[very thick,draw=none,fill=gray!50] (0,0) arc (-180:-90:0.3cm) -- ++(8,0) arc (-90:0:0.3cm) -- ++(0,4) arc (0:90:0.3cm) -- ++(-8,0) arc (90:180:0.3cm);
		\draw[very thick] (0,0) arc (-180:-90:0.3cm) ++(8,0) arc (-90:0:0.3cm) -- ++(0,4) arc (0:90:0.3cm) -- ++(-8,0) arc (90:180:0.3cm) -- (0,0);		
		\filldraw[thick,fill=gray!80] (0,0) arc (-180:0:0.3cm) -- ++(0,4) arc (0:180:0.3cm) -- cycle;		
		\filldraw[thick,fill=gray!80] (0.085,0) arc (-180:0:0.22cm) -- ++(0,4) arc (0:180:0.22cm) -- cycle;		
		\filldraw[thick,fill=gray!80] (0.17,0) arc (-180:0:0.14cm) -- ++(0,4) arc (0:180:0.14cm) -- cycle;		
		\draw[very thick,dashed] (0.3,-0.3) -- ++(8,0);		
		\draw[very thick,dashed] (0.6,4.1) -- ++(8,0);
		\draw[dashed] (3.2,-0.3) arc (-90:0:0.3cm) -- ++(0,4) arc (0:90:0.3cm);
		\draw[dashed] (5.2,-0.3) arc (-90:0:0.3cm) -- ++(0,4) arc (0:90:0.3cm);
		\draw (4.3,-0.8) node{$A$};
		\draw (2.2,-0.8) node{$B$};
		\draw (6.7,-0.8) node{$B$};
		\draw[thick,<->] (3.6,0.5) -- ++(1.8,0);
		\draw (4.5,0.9) node{\footnotesize $L_A$};
		\draw[very thick] (3.4,-0.3) -- (5.4,-0.3);
		\filldraw (3.4,-0.3) circle (1mm) ++(-0.4,0.4) node{\small $\mathcal{T}_{\sigma^{-1}}$};
		\filldraw (5.4,-0.3) circle (1mm) ++(0.5,0.4) node{\small $\mathcal{T}_\sigma$};
		\draw[very thick] (3.5,4.1) -- (5.5,4.1);
		\filldraw (3.5,4.1) circle (1mm) ++(-0.4,0.7) node{\small $\mathcal{T}_\sigma$};
		\filldraw (5.5,4.1) circle (1mm) ++(0.4,0.7) node{\small $\mathcal{T}_{\sigma^{-1}}$};
		\draw[->] (6,-2) node[below]{$\mathcal{S}^A_{\sigma}$} -- (4.8,-0.5);
		\draw[->] (6,6) node[above]{$\mathcal{S}^A_{\sigma^{-1}}$} -- (4.8,4.5);
	\end{scope}
	\draw (7.5,-6.5) node{$(c)$};

	\end{tikzpicture}
	\caption{(a) Drawing of the thermal density matrix; taking the trace corresponds to identifying the two dashed lines; this gives the partition function $Z = {\rm tr} [e^{-\beta H}]$ of the CFT on a cyclinder of circumference $\beta v$. In order for $Z$ to be finite, one needs an IR cutoff $\Lambda$, namely some truncation of the system at the two ends. (b) The replicated surface that is used to calculate the entanglement entropy, in the setup of Cardy and Calabrese \cite{calabrese2004entanglement}. The swap operator $\mathcal{S}^A_\sigma$, with a cyclic permutation $\sigma = (1,2\dots, \alpha)$, can be equivalently represented by two twist field $\mathcal{T}_{\sigma}$ and $\mathcal{T}_{\sigma^{-1}}$ located at the two ends of the interval $A$. (c) For the OSEE, there are two swap operators $\mathcal{S}^A_{\sigma}$ and $\mathcal{S}^A_{\sigma^{-1}}$, that are equivalent to a four-point function of twist fields.}
	\label{fig:CFT_thermal}
\end{figure}

\noindent The thermal density matrix is
\begin{equation}
	\rho_\beta \, = \, \frac{1}{Z} e^{-\beta H} \, ,
\end{equation}
where the normalization factor $Z = {\rm tr} [e^{-\beta H}]$ is the partition function of the CFT on an infinitely long
cylinder of circumference $\beta v$, see Fig. \ref{fig:CFT_thermal}.(a). Strictly speaking, this partition function is IR divergent; it is convenient
to introduce an IR cutoff by cutting the cylinder at its two extremities, leaving a finite cylinder of length
$\Lambda \gg \beta v$. It is then a standard result that the partition function goes as
\begin{equation}
	\label{eq:Zcylinder}
	Z  \, \simeq \, e^{ \frac{\pi c}{6} \frac{\Lambda}{\beta v} } \, ,
\end{equation}
where $c$ is the central charge. Now, let us briefly review the calculation of the 'usual' entanglement entropy. This will provide an opportunity to emphasize the well-known but crucial fact that the 'usual' entanglement entropy obeys a {\it volume} law, as opposed to the {\it area law} obeyed by the OSEE; it will also be a convenient way to introduce twist fields, which we will need later on. \vspace{0.4cm}

\paragraph{Preliminary: the volume law for the 'usual' entanglement entropy.} Take $A = [0, L_A]$ and $B = (-\infty, 0) \cup ( L_A, +\infty )$, and consider $S^{(\rm EE)}_\alpha  =  \frac{1}{1-\alpha} \log \left(  {\rm tr} [\rho_A^\alpha] \right)$, for $\alpha$ integer. Here we use the superscript (EE) to emphasize that this quantity is {\it not} the OSEE $S_\alpha(\rho_\beta)$. As briefly reviewed in the introduction, $S_\alpha^{({\rm EE})}$ may also been written as the expectation value of the swap operator,
\begin{eqnarray*}
	S^{({\rm EE})}_\alpha  \, = \, \frac{1}{1-\alpha} \log \left(  {\rm tr} [ \mathcal{S}^A_{(1,2,\dots,\alpha)}   \cdot  (\rho_\beta)^{\otimes \alpha}  ] \right) \, = \, \frac{1}{1-\alpha} \log \left(  \frac{  {\rm tr} [ \mathcal{S}^A_{(1,2,\dots,\alpha)}   \cdot (e^{- \beta H})^{\otimes \alpha} ]  }{ Z^\alpha  } \right) .
\end{eqnarray*}
The key point from Ref. \cite{calabrese2004entanglement} is that $Z^{\rm twist}_\alpha \, \equiv \,  {\rm tr} [ \mathcal{S}^A_{(1,2,\dots,\alpha)}   \cdot (e^{- \beta H})^{\otimes \alpha}]$ is again the partition function of the CFT, but this time on a more complicated surface, see Fig. \ref{fig:CFT_thermal}.(b): it is a cyclinder with $\alpha$ identical sheets, where one can go continuously from one sheet to the next by turning around the two points $(x,y) = (0,0)$ and $(0,L_A)$---namely, the two end points of the interval $A$---. It is possible to calculate this partition function $Z^{\rm twist}_\alpha$ directly, but it is more efficient to view the ratio of partition functions $Z^{\rm twist}_\alpha / Z^\alpha$ as a two-point correlator of {\it twist fields} $\mathcal{T}_{(1,\alpha,\alpha-1,\dots,2)}$ and $\mathcal{T}_{(1,2,\dots, \alpha)}$ located at $(0,0)$ and $(0,L_A)$ respectively (by convention, the subscript $\sigma$ in $\mathcal{T}_\sigma$ is the permutation of the replicas one picks when one turns {\it clockwise} around the point; see also the appendix for a discussion of twist fields). The twist fields are primary operators with scaling dimension $\Delta_\alpha = \frac{c}{12} \left( \alpha - 1/\alpha\right)$. The two-point function of primary operators on the cylinder is of course entirely fixed by conformal invariance,
\begin{equation*}
	\frac{Z^{\rm twist}_\alpha}{Z^\alpha} \, = \, \left< \mathcal{T}_{(1,\alpha,\dots,2)}(0,0) \mathcal{T}_{(1,2,\dots,\alpha)}(L_A,0) \right> \, = \,  \frac{1}{[ \frac{ \beta v}{ \pi }  \sinh ( \frac{\pi L_A}{\beta v} ) ]^{2 \Delta_\alpha}} \, ,
\end{equation*}
which finally leads to
\begin{eqnarray}
	\label{eq:thermal_EE}
	S^{(\rm EE)}_\alpha & =  & \frac{c}{6} \left( 1 + \frac{1}{\alpha} \right) \log \left(   \frac{ \beta v}{\pi }  \sinh ( \frac{\pi L_A}{ \beta v} )    \right)  \;  \underset{ L_A \gg \beta v }{\simeq} \;  \left( 1 + \frac{1}{\alpha} \right)  \frac{\pi c}{6 \beta v} L_A  \, .
\end{eqnarray}
It is extensive: the entanglement entropy satisfies a volume law. However, most of this entropy is just thermal entropy. Indeed, if we calculate the (Renyi version of the) classical thermodynamic entropy at temperature $\beta$, we find:
\begin{eqnarray*}
\nonumber	S_\alpha^{\rm (class.)}   &=& \frac{1}{1-\alpha} \log \left( {\rm tr} [ (\rho_\beta)^\alpha] \right) \\
\nonumber	&=&  \frac{1}{1-\alpha} \log \left( \frac{  {\rm tr} [ e^{ - \alpha \beta H}  ]}{ ({\rm tr} [ e^{ - \beta H}  ] )^\alpha}  \right) \\
	& \underset{\Lambda \gg \beta v}{\simeq} &   \frac{1}{1-\alpha} \log \left(  \frac{ e^{\frac{\pi c}{6} \frac{\Lambda}{\alpha \beta v} } }{ e^{\frac{\pi c}{6} \frac{\Lambda}{\beta v} \alpha }  } \right)  \,  = \,   \left( 1+\frac{1}{\alpha}  \right)   \frac{\pi c }{6 \beta v} \Lambda  \, ,
\end{eqnarray*}
where we used Eq. (\ref{eq:Zcylinder}) in the last line. Thus, the entropy per unit length is $\left( 1+\frac{1}{\alpha}  \right)   \frac{\pi c }{6 \beta v}$, which precisely coincides with the prefactor of Eq. (\ref{eq:thermal_EE}), demonstrating that the extensive part of the 'usual' entanglement entropy at finite temperature entirely comes from thermal entropy, and has nothing to do with quantum entanglement.

\paragraph{OSEE from CFT, and an operator area law for $\rho_\beta$.} Now, instead of the 'usual' entanglement entropy, consider the OSEE. To calculate the latter in CFT, we use again the replica trick, in the form
\begin{equation*}
	S_\alpha (\rho_\beta) \, =\, \frac{1}{1-\alpha} \log \left( \frac{  {\rm tr} [\rho_\beta^{\otimes \alpha} \cdot \mathcal{S}^A_{(1,\alpha,\alpha-1,\dots,2)} \cdot  \rho_\beta^{\otimes \alpha} \cdot  \mathcal{S}^A_{(1,2,\dots,\alpha)}]   }{ ( {\rm tr} [ (\rho_\beta)^2  ] )^{\alpha} }  \right) .
\end{equation*}
Exactly like in the calculation of the entanglement entropy which we just reviewed, the two traces inside the logarithm can be regarded as CFT partition functions on certain multi-sheeted surfaces. Again, it is useful to view the ratio of these two partition functions as a correlation function of twist operators, living on an infinitely long cylinder of circumference $2 \beta v$, as illustrated in Fig. \ref{fig:CFT_thermal}.(c). If we parametrize the cylinder by the complex coordinate $x+i y$ with $(x,y) \in \mathbb{R} \times [0,2\beta v]$, then the four twist operators are located at the points $0$, $L_A$, $i \beta v$ and $L_A + i \beta v$, and the four-point function we need to calculate is (see Fig. \ref{fig:CFT_thermal})
\begin{equation*}
		 \frac{  {\rm tr} [\rho^{\otimes \alpha} \cdot \mathcal{S}^A_{\sigma^{-1}} \cdot  \rho^{\otimes \alpha} \cdot  \mathcal{S}_\sigma ]   }{ ( {\rm tr} [ \rho^2  ] )^{\alpha} }  \, =\,  \left< \mathcal{T}_{\sigma^{-1}} (0) \mathcal{T}_\sigma (i\beta v) \mathcal{T}_{\sigma} (L_A) \mathcal{T}_{\sigma^{-1}} (i\beta v+L_A)  \right> ,
\end{equation*}
with the cyclic permutation $\sigma = (1, 2, \dots , \alpha)$. Since this is a four-point function, it is not entirely fixed by conformal invariance. In fact, the four-point function of twist operators is an object that has attracted a lot of attention in recent years in the context of the two-interval entanglement entropy\cite{furukawa2009mutual,calabrese2009entanglement,calabrese2011entanglement_e}, and it is now well-established that is depends on the full spectrum of the theory one is working with. Therefore, the calculation of the four-point function is usually complicated, and we want to avoid such complications here. So we stick to the information that can be most easily extracted, without relying on any specific details of the CFT. This concerns the two regimes $L_A \ll \beta v$ and $L_A \gg \beta v$, where the four-point function factorizes into a product of two two-point functions; the latter are easily calculated. This gives
\begin{eqnarray*}
	L_A\ll \beta v: && \left< \mathcal{T}_{\sigma^{-1}} (0) \mathcal{T}_\sigma (i\beta v) \mathcal{T}_{\sigma} (L_A) \mathcal{T}_{\sigma^{-1}} (i\beta v+L_A)  \right>  \\
	&& \quad \simeq \;   \left< \mathcal{T}_{\sigma^{-1}} (0)\mathcal{T}_\sigma (L_A) \right> \left< \mathcal{T}_{\sigma^{-1}} (i\beta v+L_A)   \mathcal{T}_\sigma (i\beta v)  \right>
	\, = \, \frac{1}{[  \frac{ 2\beta v}{\pi }  \sinh ( \frac{ \pi L_A}{ 2\beta v} )   ]^{2 \Delta_\alpha}}  \frac{1}{[  \frac{ 2\beta v}{ \pi }  \sinh ( \frac{ \pi L_A}{ 2\beta v} )   ]^{2 \Delta_\alpha}} \,  \\
	L_A \gg \beta v: && \left< \mathcal{T}_{\sigma^{-1}} (0) \mathcal{T}_\sigma (i\beta v) \mathcal{T}_{\sigma} (L_A) \mathcal{T}_{\sigma^{-1}} (i\beta v+L_A)  \right>   \\
	&& \quad \simeq \;   \left< \mathcal{T}_{\sigma^{-1}} (0)\mathcal{T}_\sigma (i \beta v) \right> \left< \mathcal{T}_{\sigma^{-1}} (i\beta v+L_A)   \mathcal{T}_\sigma (L_A)  \right> 
	\, = \, \frac{1}{[  \frac{ 2\beta v}{\pi }  \sin ( \frac{ \pi \beta v}{ 2\beta v} )   ]^{2 \Delta_\alpha}}  \frac{1}{[  \frac{ 2\beta v}{ \pi }  \sin ( \frac{ \pi \beta v}{ 2\beta v} )   ]^{2 \Delta_\alpha}} \, ,
\end{eqnarray*}
which leads to the following asymptotic behavior for the OSEE,
\begin{subequations}
\begin{eqnarray}
	\label{eq:CFT_twice_EE}
		L_A \ll \beta v : &\qquad & S_\alpha(\rho_\beta) \, \simeq  \,  \frac{c}{3}\left( 1+ \frac{1}{\alpha}\right)  \log \left( \frac{L_A}{a_0}  \right) \\
	\label{eq:thermal_area_law}
		L_A \gg \beta v : &\qquad & S_\alpha(\rho_\beta) \, \simeq  \,  \frac{c}{3}\left( 1+ \frac{1}{\alpha}\right) \log \left(   \frac{ \beta v}{a_0 }  \right) \, ,
\end{eqnarray}
\end{subequations}
where we have reinstated the UV cutoff $a_0$ in order to get dimensionless expressions. Of course, strictly speaking, the results hold only for $\alpha$ integer, however we see now that the final expressions are analytic in $\alpha$, and so it is very plausible that they hold for any real positive $\alpha$. Such reasoning is standard in the case of the 'usual' entanglement entropy; we are simply repeating these well-known arguments here in the case of the OSEE; moreover, the analyticity in $\alpha$ will be supported by numerical evidence below. \vspace{0.3cm}

The physical meaning of Eq. (\ref{eq:CFT_twice_EE}) is clear: at low temperatures, the reduced density matrix $\rho_A$ is very close to the one of the ground state $\ket{\psi_0}$, whose entanglement entropy is $S_\alpha(\ket{\psi_0}) = \frac{c}{6} (1+1/\alpha) \log (L_A/a_0)$. Thus, Eq. (\ref{eq:CFT_twice_EE}) is just a particular occurrence of 
the relation (\ref{eq:OSEE_pure}). Eq. (\ref{eq:thermal_area_law}) is more exciting: {\it it is the operator area law} for the thermal density matrix $\rho_\beta$. It shows that the OSEE is {\it bounded} at finite temperature; moreover, it increases logarithmically with inverse temperature $\beta$. As discussed in the introduction, this has an interesting practical consequence, which is that $\rho_\beta$ may be efficiently represented by an MPO with finite bond dimension. In addition, we see from Eq. (\ref{eq:thermal_area_law}) that, when one approaches zero temperature, the bond dimension grows polynomially with inverse temperature $\beta$. We thus recover, within the framework of CFT, some of the conclusions obtained in Ref. \cite{vznidarivc2008complexity}. Let us also emphasize that the question of approximability of thermal density matrices by MPOs or their higher dimensional versions has been tackled directly, with rigorous statements and proofs, in Refs. \cite{hastings2006solving, molnar2015approximating}.

\begin{figure}[h]
	\begin{center}
		\includegraphics[width=0.6\textwidth]{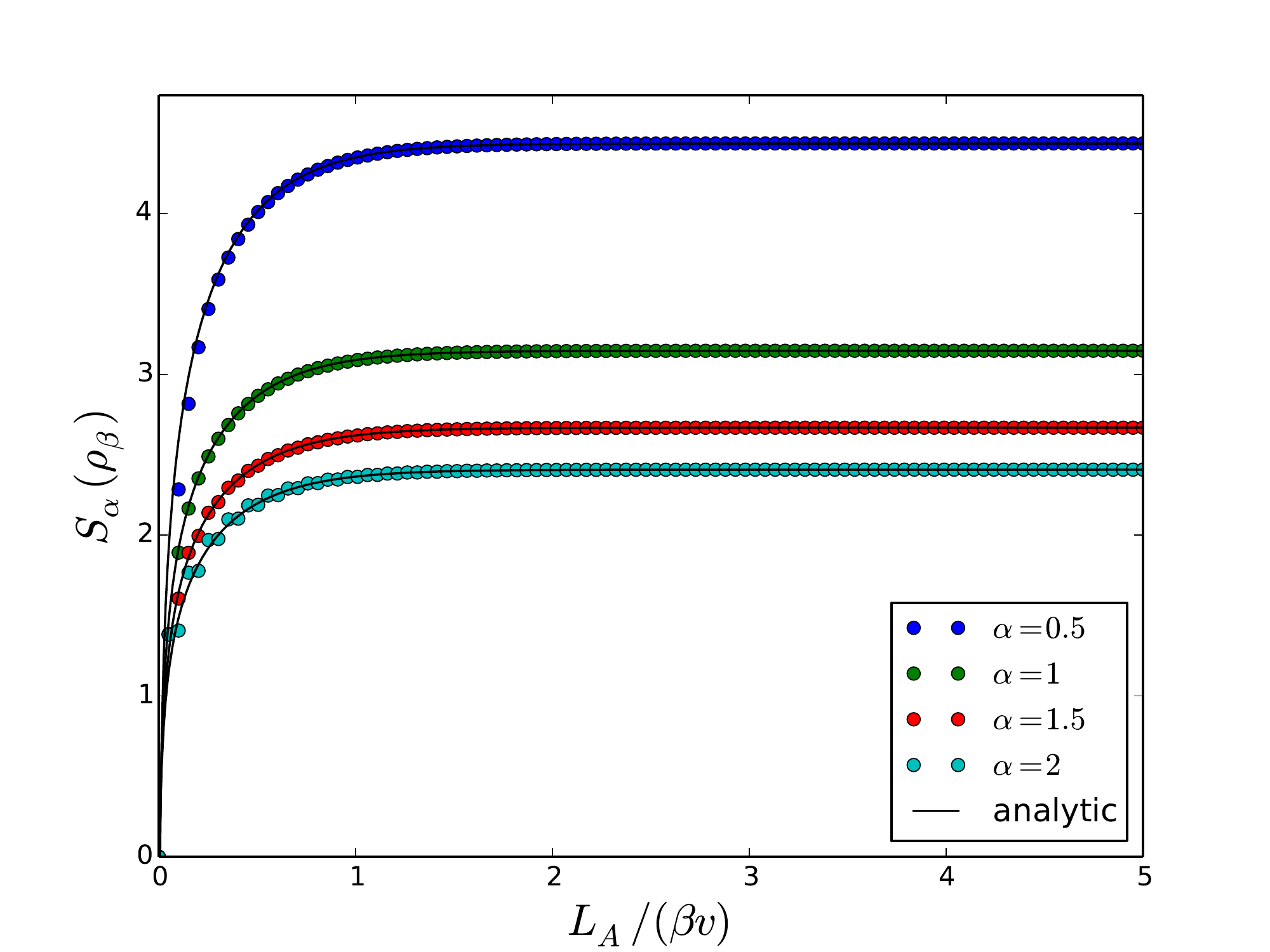}
	\end{center}
	\caption{Numerical check of formula (\ref{eq:exact_thermal_free}) for free fermions on the lattice with dispersion $\varepsilon(k) = -\cos k$ ({\it a.k.a} the XX chain after a Jordan-Wigner transformation); the Fermi velocity is $v=1$, and the constant $a_0$ in (\ref{eq:exact_thermal_free}) is obtained by fitting. We take a chain of $1000$ sites at inverse temperature $\beta = 20$, and cut an interval of length $L_A= 1,2,\dots ,100$ in the middle. Notice that formula (\ref{eq:exact_thermal_free}) works for non-integer $\alpha$, and, in particular, it works for $\alpha < 1$.}
\end{figure}

\paragraph{The free fermion case.} Contrary to most CFTs, for the free Dirac fermion ($c=1$), the four-point function of twist operators is known (more generally, all correlations of twist operators are known in that case, and they are given in Ref. \cite{calabrese2004entanglement}). Thus, for completeness, we now give the complete formula for $S_\alpha (\rho_\beta)$ in the free fermion case. The four-point function on the cylinder of circumference $2\beta v$ is
\begin{eqnarray*}
	({\rm free \; fermions})&& \left< \mathcal{T}^\dagger_\alpha (0) \mathcal{T}_\alpha (i\beta v) \mathcal{T}_\alpha (L_A) \mathcal{T}_\alpha^\dagger (i\beta v+L_A)  \right>   \\
	&& \quad = \, \left[ \frac{    \frac{ 2\beta v}{\pi }  \sinh ( \frac{ \pi (L_A+i\beta v)}{ 2\beta v} )     \frac{ 2\beta v}{\pi }  \sinh ( \frac{ \pi (L_A-i \beta v)}{ 2\beta v} )   }{   \left(  \frac{ 2\beta v}{\pi }  \sin ( \frac{ \pi \beta v}{ 2\beta v} )    \right)^2   \left( \frac{ 2\beta v}{\pi }  \sinh ( \frac{ \pi L_A}{ 2\beta v} ) \right)^2   }    \right]^{2 \Delta_\alpha} 
	\,=\, \left[  \frac{  1 }{  \frac{ 2\beta v}{\pi }  \tanh ( \frac{ \pi L_A}{ 2\beta v} )    }    \right]^{4 \Delta_\alpha}  ,
\end{eqnarray*}
which gives the OSEE:
\begin{equation}
	\label{eq:exact_thermal_free}
	({\rm free \; fermions}) \qquad S_\alpha(\rho_\beta) \, = \, \frac{1}{3} \left( 1+ \frac{1}{\alpha}\right)  \log \left(   \frac{ 2\beta v}{\pi a_0}  \tanh ( \frac{ \pi L_A}{ 2\beta v} )  \right) .
\end{equation}


\section{Time-evolution of the reduced density matrix after a global quench, and consequences for simulability}
\label{sec:global_quenches}

\subsection{An operator area law for Generalized Gibbs Ensembles?}
\label{sec:GGE}
In the past years, there has been a lot of work on the problem of equilibration in closed quantum systems, and on Generalized Gibbs Ensembles (GGE), motivated by Refs. \cite{rigol2007relaxation,rigol2008thermalization}. In particular, many recent works (see, for instance, \cite{wouters2014quenching, pozsgay2014correlations, pereira2014exactly, ilievski2015quasilocal,ilievski2015complete, doyon2015thermalization,essler2015generalized,fagotti2016local,ilievski2016quasilocal,essler2016truncated,piroli2016quasi,bastianello2016quasi}) have dealt with the problem of what charges exactly should be used as a basis set in the construction of the GGE: should they be {\it local?}, {\it non-local?}, {\it quasi-local?}, {\it pseudo-local?}, {\it localized?}, {\it semi-local?}, {\it ultra-local?} and so on (these terms are all properly defined in the previous list of references). These discussions are very advanced and concern subtle points about the thermodynamic Bethe Ansatz solution of specific models such as the XXZ chain or integrable field theories (for instance: what exactly is the minimal set of conserved charges whose expectation values are needed to uniquely  specify a distribution of Bethe roots in the thermodynamic limit?). Such questions are far beyond the scope of the present paper. These recent works do inspire, however, the following more pedestrian question, which
is directly related to our topic. \vspace{0.3cm}

\noindent Assume, as usual in discussions about GGE, that the reduced density matrix of a subsystem $A$ of size $L_A$ ($L_A$ small compared to the total size of the system) converges to some
\begin{equation}
	\label{eq:H_GGE}
	\rho_{\rm GGE} \, = \, e^{- H_{\rm GGE}}
\end{equation}
at large times (assume for instance that this convergence holds in $L^1$-norm, which then also implies convergence in $L^2$-norm). As usual, one needs to be careful about possible revivals of the system in the definition of the convergence, for instance by averaging over time (possibly by excluding some time-window where the revival occurs); such details are well discussed in the standard treatments of GGE, and we will skip these subtle points here. We assume that $\rho_{\rm GGE}$ is properly defined, and we
would like to get some idea of the locality of $H_{\rm GGE}$ not by focusing directly on the basis set for the space of conserved charges that should be used to write $H_{\rm GGE}$, but by asking instead whether it would make sense to try to approximate $\rho_{\rm GGE} \, = \, e^{-H_{\rm GGE}}$ by an MPO. We thus ask:
\begin{center}
{\it (Q1) \qquad Does the GGE density matrix $\rho_{\rm GGE}$ satisfy the operator area law?}
\end{center}
In other words, does $\rho_{\rm GGE}$ behave morally like a thermal (Gibbs) density matrix, or is it spectacularly different from a Gibbs state, for instance with a logarithmic violation of the area law? [We note that the seemingly related question of whether $\rho_{\rm GGE}$ satisfies an area law for the mutual information has been addressed in Ref. \cite{eisler2014area}; there, a violation of the area law was found for a specific quench protocol. The question of the area law for the mutual information has also been investigated in open quantum systems, see for instance the recent Refs. \cite{cubitt2015stability, brandao2015area, mahajan2016entanglement,zanoci2016entanglement}.]

\vspace{0.2cm}

\noindent Since the construction of the GGE is in itself a hard task for interacting systems \cite{caux2012constructing}, it seems that a direct evaluation of the OSEE of $\rho_{\rm GGE}$ is out of reach at this point. However, it is of course easy to make progress in free fermion systems, which are already quite instructive for our purposes. Indeed, the free fermion case is arguably the worst when it comes to locality, in the sense that the conserved quantities that must be included in the construction of the GGE are the mode occupation themselves, which are less local than the ones that show up, say, in the XXZ chain ({\it e.g.} the 'quasi-local' ones \cite{ilievski2015complete}). So, we expect that, if a simple calculation in the free fermion case shows that $\rho_{\rm GGE}$ satisfies the operator area law, then it will be a good indication that the same
will be true in other integrable systems. With this reasoning in mind, we now turn to a simple free-fermion calculation. \vspace{0.3cm}

 
\noindent In this simple example, we will reach the conclusion that, for a global (translation invariant) quench to a critical point, as long as the initial state is the ground state of a gapped Hamiltonian, $\rho_{\rm GGE}$ satisfies the operator area law. This conclusion holds even though, perhaps surprisingly, $H_{\rm GGE}$ itself is not local\footnote{I thank Viktor Eisler for pointing this out to me.} (in the sense of exponentially decaying terms). A similar observation was made previously for the mutual information in the transverse-field Ising chain: in that case, $H_{\rm GGE}$ is a sum of algebraically decaying terms \cite{fagotti2013reduced}, and yet the mutual information satisfies an area law according to the discussion in the conclusion of Ref. \cite{eisler2014area}.

\paragraph{A free-fermion example.} For simplicity, we would like to treat an example of a quantum quench to the XX chain. For the initial state, a simple
choice would be the N\'eel state as in Fig. \ref{fig:blowup} below, however since all occupation numbers in the N\'eel state are equal, one finds that $\rho_{\rm GGE} = 1$ in that case. So we can indeed 
conclude that $H_{\rm GGE}=0$ is local and that $\rho_{\rm GGE}=1$ satisfies the area law in that example; however it is too simplistic.

So, instead, we turn to a slightly less trivial example. We consider the following Hamiltonian for an infinitely long one-dimensional superconductor on a lattice,
\begin{equation}
	H_{\delta,h} \, =\, \frac{1}{2} \int_{-\pi}^\pi \frac{dk}{2\pi} \left( \begin{array}{cc} c^\dagger_k & c_{-k} \end{array} \right)  \left(  \begin{array}{cc}  -\cos k - h  & i \delta \sin k \\ -i \delta \sin k &  \cos k +h  \end{array}  \right) \left(  \begin{array}{c} c_k \\ c^\dagger_{-k} \end{array} \right) , 
\end{equation}
which is nothing but the XY chain in transverse field after a Jordan-Wigner transformation, see for instance Refs. \cite{mukherjee2007quenching, fagotti2008evolution}. We focus on the following quench protocol: the system is prepared in the ground state of $H_{\delta, h}$, and at time $t=0$, we switch off the parameters $\delta$ and $h$,
\begin{equation}
	H_{\delta, h} \quad {\rm at \;} t<0 \quad \longrightarrow \quad  H_{0,0} \quad {\rm at \;} t>0 \, .
\end{equation}
Notice that $H_{0,0}$ is the Hamiltonian of the XX chain. The ground state of $H_{\delta, h}$ is
\begin{equation*}
	\ket{\psi_{\delta,h}} \, = \, \prod_{k>0} \left(  \cos (\theta_k/2) + i \sin  (\theta_k/2)  c^\dagger_{k} c^\dagger_{-k}  \right) \ket{0} \, ,
\end{equation*}
where
\begin{equation}
	\label{eq:theta_k}
	e^{i \theta_k} \, = \, \frac{h+\cos k + i  \delta \sin k }{\left|  h+\cos k + i \delta \sin k  \right|} \, .
\end{equation}
The conserved charges of $H_{0,0}$ are the mode occupations at each wave vector $k$; when evaluated in the initial state, they give
\begin{equation}
	n_k \, = \, \bra{\psi_{\delta,h}} c^\dagger_k c_k \ket{\psi_{\delta,h}} \, = \, \sin (\theta_k/2)^2 \, .
\end{equation}
The crucial property on which we are going to rely shortly is the (real-) analyticity of $n_k$ as a function of $k$ in the Brillouin zone. Notice that, assuming $\delta \neq 0$, $n_k$ is always an analytic function of $k$, provided $|h| \neq 1$. This is simply because
$n_k$ is an analytic function of $\theta_k$, and $\theta_k$ itself is real-analytic in $k$ as long as $h + \cos k$ is non-zero at $k = 0$ and $k=\pi$, as can be seen from Eq. (\ref{eq:theta_k}).

\vspace{0.3cm}
\noindent The GGE density matrix is
\begin{equation}
	\rho_{\rm GGE} \, \propto \, e^{-\sum_k  \lambda_k c^\dagger_k c_k}
\end{equation}
where $e^{-\lambda_k} = \frac{n_k}{1-n_k}$, such that the expectation values of the occupation modes are the same as in the initial state, ${\rm tr} [ c^\dagger_k c_k\,  \rho_{\rm GGE} ]\, = \,  \bra{\psi_{\delta,h}} c^\dagger_k c_k \ket{\psi_{\delta,h}}$. Here the density matrix is the one of an infinite system; to obtain the one of a finite subsytem $A$, one can of course trace over all the degrees of freedom outside $A$. \vspace{0.3cm}

\noindent Thus, we have $H_{\rm GGE}   =  -\sum_k  \lambda_k c^\dagger_k c_k$, up to some irrelevant additional constant. The question of the locality of $H_{\rm GGE}$ follows from elementary results in Fourier analysis: $H_{\rm GGE}$ has exponentially decaying terms if $\lambda_k$ is real-analytic in $k$ in the Brillouin zone.
But we see that, if there is a point $k$ where $n_k \rightarrow 0$ or $n_k \rightarrow 1$, then $\lambda_k$ is not analytic, and so $H_{\rm GGE}$ cannot be local in the sense of exponentially decaying terms. From Eq. (\ref{eq:theta_k}), we see that this is what happens when $k \rightarrow 0$ and $k \rightarrow \pi$. So $H_{\rm GGE}$ cannot be local. \vspace{0.3cm}

\noindent Let us now see why, despite the fact that $H_{\rm GGE}$ is not local (in the sense of exponentially decaying terms), $\rho_{\rm GGE}$ still satisfies the
operator area law. To do that, one can use the trick that $\rho_{\rm GGE}$, which is an operator acting on a Hilbert space $\mathcal{H}$, can be viewed as a state
in the space $\mathcal{H} \otimes \overline{\mathcal{H}}$ (see also Fig. \ref{fig:cartoon_OSEE} and the appendix for details on how to implement this in the free fermion case). We write the state $\ket{ \rho_{\rm GGE} }$ in this larger space as
\begin{equation}
	\label{eq:GGE_op_state}
	\rho_{\rm GGE} \, \propto \, \prod_k \left[ 1 + (e^{-\lambda_k}-1) c^\dagger_k c_k \right] \quad \longrightarrow \quad  \ket{\rho_{\rm GGE}}  \propto \prod_k \left[ 1 + (e^{-\lambda_k}-1) c^\dagger_k \tilde{c}^\dagger_k \right]  \ket{0} ,
\end{equation}
where the $\tilde{c}^\dagger_k$'s are new creation operators that anti-commute with all the $c_k$'s, and $\ket{0}$ is the state annihilated by all the $c_k$'s and $\tilde{c}_k$'s.
The OSEE of $\rho_{\rm GGE}$ is, by definition, the 'usual' entanglement entropy of the state $\ket{\rho_{\rm GGE}}$ (see Fig. \ref{fig:cartoon_OSEE}). Thus, we need to argue that the entanglement entropy of 
$\ket{\rho_{\rm GGE}}$ satisfies the area law. Since we are dealing with translation-invariant free fermions, this can be done by expressing the entanglement entropy in terms 
of a certain (block-) Toeplitz determinant, whose asymptotics depend on whether or not the symbol of the Toeplitz matrix possesses singularities \cite{jin2004quantum,its2005entanglement}. Since the symbol is related to $n_k$, it is possible to argue that the analyticity of $n_k$ implies that the symbol itself is analytic, such that the area law then follows from the strong Szeg\"o limit theorem. This line of arguments can presumably be made mathematically rigorous. However, for conciseness, we now take a slightly different route, which also sheds some light on the role played by the analyticity of $n_k$. Namely, we exhibit a local, gapped Hamiltonian acting on $\mathcal{H}\otimes \overline{\mathcal{H}}$, of which $\ket{\rho_{\rm GGE}}$ is the ground state. And we then rely on the standard fact that it is sufficient to imply the area law. \vspace{0.3cm}

\noindent Such a Hamiltonian can be cooked up as follows. Consider the operator
\begin{equation}
	\label{eq:super_ham}
	\sum_k  \left( \begin{array}{cc}  c^\dagger_k & \tilde{c}_k  \end{array} \right)   \left(\begin{array}{cc}  \cos \alpha_k  &  -\sin \alpha_k \\  -\sin \alpha_k   & -\cos \alpha_k  \end{array} \right)  \left( \begin{array}{c}  c_k \\ \tilde{c}^\dagger_k  \end{array} \right),
\end{equation}
which is a flat-band Hamiltonian, because the $2 \times 2$ matrix in the middle has eigenvalues $+1, -1$, independently of $\alpha_k$. The ground state of this
operator is $\prod_k  \left[  1 +  {\rm tan \,} (\alpha_k/2) c^\dagger_k  \tilde{c}^\dagger_k \right] \ket{0}$. Comparing this to Eq. (\ref{eq:GGE_op_state}), we see that $\ket{\rho_{\rm GGE}}$ is the ground state of this flat-band hamiltonian if
\begin{equation}
	 {\rm tan \,} (\alpha_k/2)  \, = \, e^{-\lambda_k}-1 \, = \, \frac{1}{1-n_k} -2\, .
\end{equation}
This defines a function $n_k \mapsto \alpha_k (n_k)$ which is analytic in $n_k$. [One way of seeing that it is analytic is to write it as $\alpha_k (n_k ) = 2\, {\rm Im \,} [ \log ( 1-i - (1-2i)n_k) ]$, where the branch cut of the $\log(.)$ is along the negative real axis; then observe that, for $n_k$ real, $1-i - (1-2i)n_k$ is never real and negative.] As a consequence, if $n_k$ is an analytic function of $k$, so is $\alpha_k$. This is precisely the property we need in order to ensure that the operator (\ref{eq:super_ham}) is local in real space (again, in the sense of exponentially decaying terms). So, in summary, provided that $n_k$ is a real-analytic function of $k$ in the Brillouin zone, $\ket{\rho_{\rm GGE}}$ is the ground state of a gapped, local Hamiltonian. As such, it must satisfy the area law. In contrast, in the critical case $|h|=1$, the occupation number $n_k$ is {\it not} analytic as a function of $k$, therefore the operator (\ref{eq:super_ham}) is {\it not} local in real space, leaving the possibility open that $\ket{\rho_{\rm GGE}}$ violates the area law.


\vspace{0.3cm}
\noindent So we have reached the conclusion that, in this simple example, as long as the initial XY Hamiltonian is gapped, $\rho_{\rm GGE}$ satisfies the operator area law;  this is true even though $H_{\rm GGE}$ itself is usually not local (in the sense of exponentially decaying terms). The only case where $\rho_{\rm GGE}$ can possibly violate the operator area law is when the initial XY Hamiltonian is gapless, when $n_k$ is no longer an analytic function of $k$.


\subsection{Can one overcome the exponential growth of the bond dimension by trading MPSs for MPOs?}

We now focus on a quench from a translationally invariant initial state that is the ground state of some gapped hamiltonian, or {\it global quench} in the (now standard) terminology of Cardy and Calabrese \cite{calabrese2005evolution,calabrese2006time,calabrese2016quantum}. One of the most important results they obtained about global quenches is the fact that the entanglement entropy of a subsystem $A$ increases linearly in time, until it saturates to a value proportional to the length $L_A$. They interpreted this result in terms of pairs of quasi-particles being emitted from the initial state and generating a ballistic spreading of the entanglement (for a recent discussion, see also Ref. \cite{alba2016entanglement}). This result has dramatic consequences for numerical simulations, as it implies that any attempt at approximating the state at time $t>0$, $\ket{\psi(t)}$, with an MPS, will quickly face the exponential growth of the bond dimension. This unfortunately induces very strong limitations on MPS-based algorithms like time-dependent DMRG or TEBD. \vspace{0.3cm}

\noindent However, in view of the discussion in the previous section, we could think of the following alternative way of simulating the dynamics after a global quench. If the system {\it is not integrable}, the reduced density matrix $\rho_A$ should converge to the thermal one, and thus should be approximable by an MPO. If the system {\it is integrable}, then $\rho_A$ is expected to converge to some $\rho_{\rm GGE}$, and we have given arguments for the approximability of this object by MPOs. So, as far as approximability is concerned, it makes no difference whether the dynamics is integrable or not: either way, the reduced density matrix at large times can be well approximated by an MPO. At very short times also, it is obviously approximable by an MPO (because the initial state itself, being the ground state of a gapped local Hamiltonian, is approximable by an MPS). So it seems that, both at short and large times, an MPO-based algorithm that would approximate the reduced density matrix $\rho_A(t)$---as opposed to an MPS approximation of the pure state $\ket{\psi(t)}$---would perform well. This raises the question\footnote{I'm grateful to Frank Pollmann for a very stimulating discussion about this question.}:
\begin{center}
{\it (Q1-a) \qquad Is it possible to overcome the exponential growth of the bond dimension after a global quench simply by trading the MPS for an MPO?} \vspace{0.3cm}
\end{center}

\begin{figure}[ht]
	\begin{center}
	\includegraphics[width=0.8\textwidth]{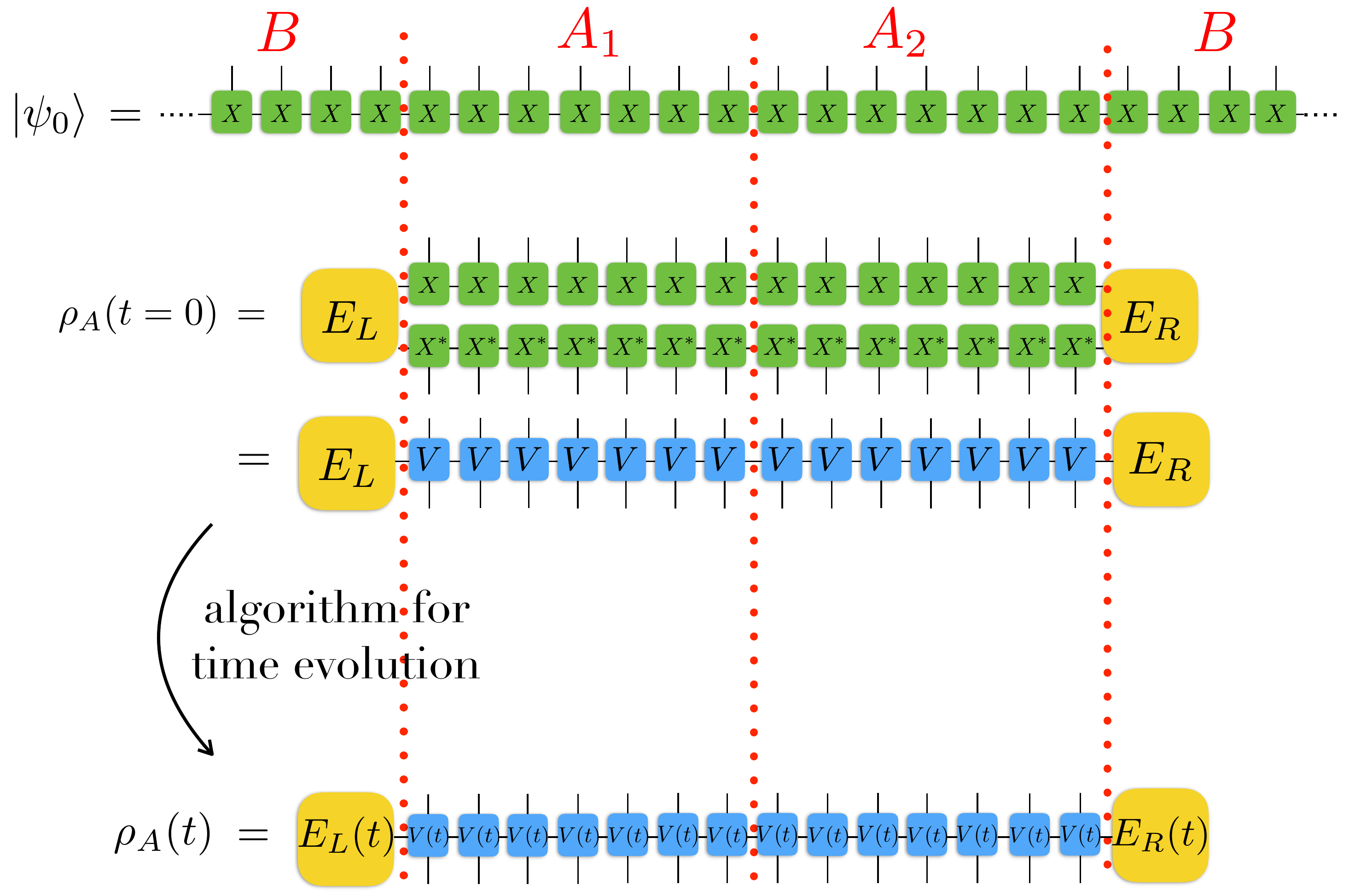}
	\caption{Cartoon of an algorithm that would try to exploit the fact that $\rho_A$ can be approximated by an MPO at short and large times. Top: starting from an initial state written as an MPS, one would construct the reduced density matrix $\rho_A$ in the form of an MPO, with left and right 'environments' $E_L$ and $E_R$. One would then design a protocol for updating both the matrices $V$ and the two vectors $E_L$ and $E_R$ that encode the environment, at each time step. Designing such a protocol for the update is of course the non-trivial part. However, what we argue in this section is that, no matter how smart this protocol is, such an algorithm will face an exponential growth of the bond dimension at intermediate times $t \sim L_A/{2v}$.}
	\label{fig:MPO_rhoA}
	\end{center}
\end{figure}

\noindent  A naive sketch of an algorithm that would exploit this idea is the following (Fig. \ref{fig:MPO_rhoA}). First, start from an MPS approximation of the initial state $\ket{\psi_0}$, say with bond dimension $\chi$, and use it to write the reduced density matrix $\rho_A$ as an MPO with bond dimension $\chi^2$. Tracing over the degrees of freedom in part $B$ naturally gives two 'environments' $E_L$, $E_R$, that are two vectors, each of dimension $\chi^2$. Then one needs to design a protocol for updating both the bulk matrices (the $V$'s in Fig. \ref{fig:MPO_rhoA}) and the environments $E_{L,R}$ that define the MPO for $\rho_A(t)$. This could be some version of TEBD for the update of the $V$'s; then, if one thinks of the simplest case where all the $V$'s are equal at every time, $V_j(t) = V(t)$, one could for instance use the left- and right-eigenvectors (with maximal eigenvalue) of $V(t)$ as updates for $E_{L,R}(t)$.

More elaborate versions of such algorithms have been investigated much more thoroughly in the literature, for instance in Ref. \cite{banuls2009matrix}.

\subsection{The barrier of the transient regime}

\begin{figure}[ht]
\begin{center}
	\includegraphics[width=0.95\textwidth]{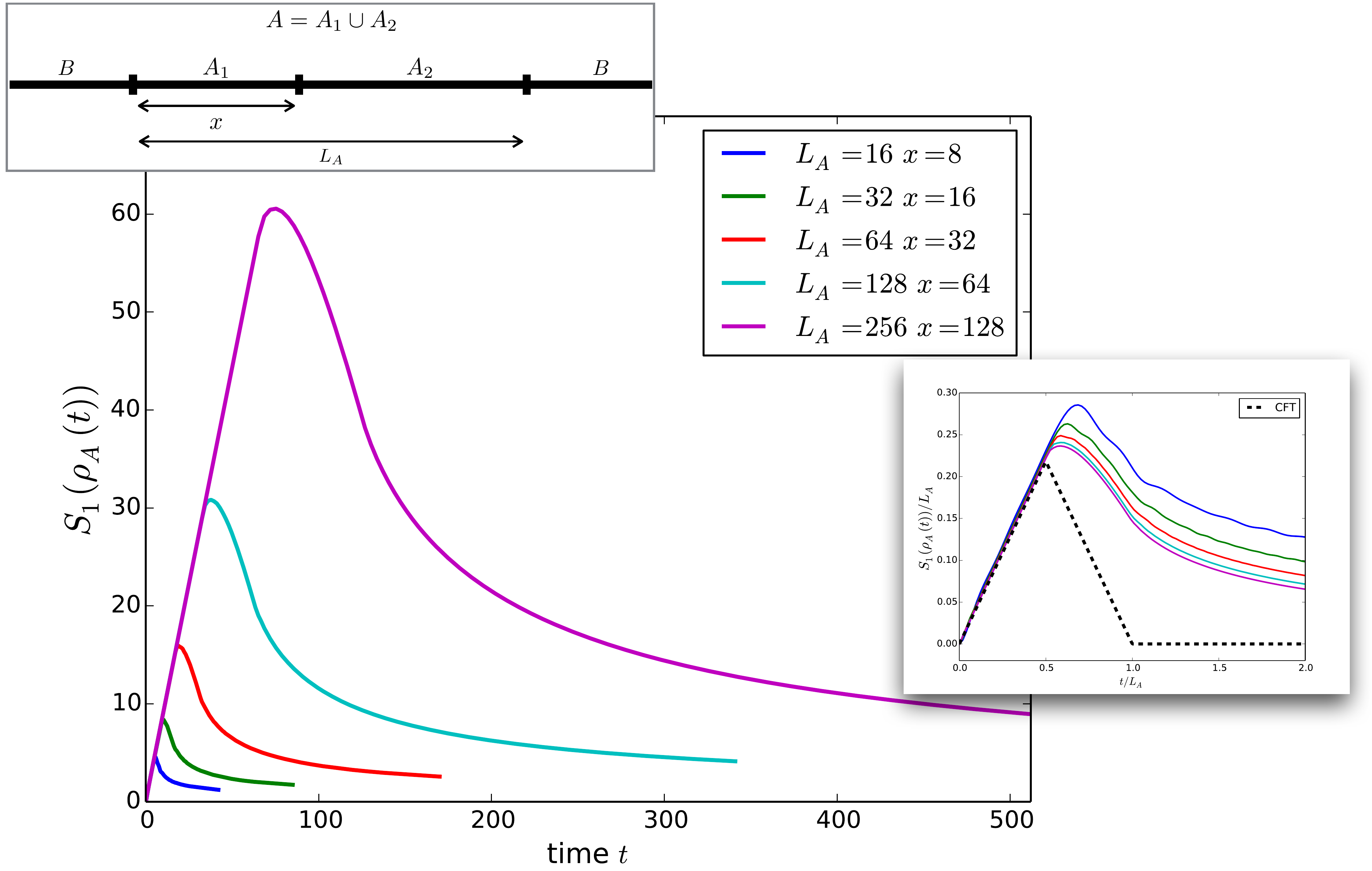}
	\caption{OSEE of the reduced density matrix $\rho_A$, for a bipartition $A=A_1 \cup A_2$, where $A_1$ and $A_2$ are of lengths $x$ and $L_A-x$ respectively (here with $x= L_A/2$). The numerical results are obtained for a quench in the XX chain from the N\'eel state (the calculation is performed in a finite periodic chain of total size $8 \times L_A$; at larger times the system exhibits revivals that are not shown here). The inset shows the rescaled OSEE, compared to the result of the CFT calculation. The deviations from the CFT result are quite large (possible explanations for those are given in the main text), however the blow-up of the OSEE, which is the main message of this section, is well captured by the CFT argument.}
	\label{fig:blowup}
\end{center}
\end{figure}

Sadly enough, the answer to {\it (Q1-a)} is negative; the reason is the following. If one calculates the OSEE of the reduced density matrix $\rho_A$, for a bipartition of the subsystem $A$ in the form of two consecutive intervals $A = A_1 \cup A_2$ of lengths $x$ and $L_A-x$ respectively (Fig. \ref{fig:blowup}), then, as expected, the OSEE is small both at short and at large times. This supports the fact that both
the short-time and large-time behaviors can be captured by an MPO with small bond dimension. However, the OSEE blows up linearly in the transient regime $t \lesssim \frac{L_A}{2 v}$, implying that the bond dimension that is needed to approximate $\rho_A(t)$ in that time window does blow up exponentially. This is illustrated in Fig. \ref{fig:blowup}, where numerical results for a quench in the XX chain from a N\'eel initial state are displayed. In fact, the linear growth of the OSEE at times $t \leq \frac{x}{v}$ can be traced back to Eq. (\ref{eq:OSEE_pure}): at short times, the reduced density matrix is still very close to the one of the pure state, and one does not (yet) gain much by approximating $\rho_A (t)$ instead of $\ket{\psi(t)}$; this trick becomes efficient only at later times.

\subsection{CFT calculation of the OSEE after a global quench}
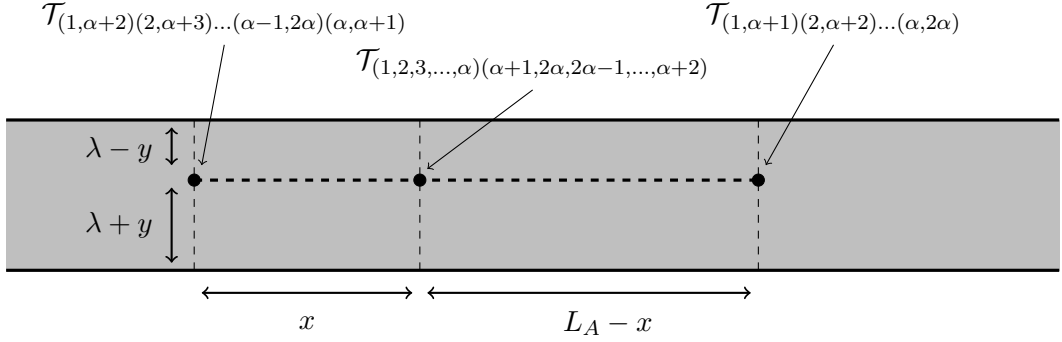
\begin{figure}[ht]
	\begin{center}
	\begin{tikzpicture}
	\begin{scope}[scale=1.]
		\filldraw[very thick,draw=none,fill=gray!50] (0,0) rectangle (14,2);
		\draw[very thick,dashed] (2.5,1.2) -- (10,1.2);
		\draw[very thick] (0,0) -- ++(14,0);
		\draw[very thick] (0,2) -- ++(14,0);
		\draw[dashed] (2.5,0) -- ++(0,2);
		\draw[dashed] (5.5,0) -- ++(0,2);
		\draw[dashed] (10,0) -- ++(0,2);
		\draw[thick,<->] (2.6,-0.3) -- (5.4,-0.3);
		\draw[thick,<->] (5.6,-0.3) -- (9.9,-0.3);
		\draw[thick,<->] (2.2,0.1) -- (2.2,1.1);
		\draw[thick,<->] (2.2,1.4) -- (2.2,1.9);
		\draw (4,-0.7) node{$x$};
		\draw (8,-0.7) node{$L_A-x$};
		\draw (1.5,0.6) node{$\lambda+y$};
		\draw (1.5,1.6) node{$\lambda-y$};
		\filldraw (2.5,1.2) circle (0.8mm);
		\draw[<-] (2.6,1.4) -- (2.9,3) node[above]{$\mathcal{T}_{(1,\alpha+2)(2,\alpha+3)\dots (\alpha-1,2\alpha)(\alpha,\alpha+1)}$};	
		\filldraw (5.5,1.2) circle (0.8mm);	
		\draw[<-] (5.6,1.35) -- (7,2.4) node[above]{$\mathcal{T}_{(1,2,3,\dots,\alpha)(\alpha+1,2\alpha,2\alpha-1,\dots,\alpha+2)}$};
		\filldraw (10,1.2) circle (0.8mm);	
		\draw[<-] (10.1,1.4) -- (11,3) node[above]{$\mathcal{T}_{(1,\alpha+1)(2,\alpha+2)\dots (\alpha,2\alpha)}$};	
	\end{scope}
	\end{tikzpicture}
	\caption{The CFT calculation of the OSEE of $\rho_A$ after a global quench involves a three-point function on an infinite strip with conformal boundary conditions on both sides.}
	\end{center}
	\label{fig:CFT_global_quench}
\end{figure}
Finally, we turn to the CFT calculation. As in previous sections, we start by rewriting $S_\alpha (\rho_A(t))$ in terms of swap operators for $\alpha$ integer. From Eq. (\ref{eq:OSEE_replicas_rho}), we see that $S_\alpha (\rho_A(t)) = \frac{1}{1-\alpha} \log r(t)$, where $r(t)$ is the ratio of two expectation values in a system with $2\alpha$ replicas,
\begin{equation}
	r(t) \, \equiv \, \frac{   \bra{\psi(t)}^{\otimes 2 \alpha}  \;  \mathcal{S}^{A_1}_{(1,\alpha+2)(2,\alpha+3)\dots (\alpha-1,2\alpha) (\alpha,\alpha+1)}  \mathcal{S}^{A_2}_{(1,\alpha+1)(2,\alpha+2)\dots(\alpha,2\alpha)}\; \ket{\psi(t)}^{\otimes 2 \alpha}  }{   \bra{\psi(t)}^{\otimes 2 \alpha} \, \mathcal{S}^A_{(1,\alpha+1)(2,\alpha+2)\dots (\alpha,2\alpha)} \,  \ket{\psi(t)}^{\otimes 2 \alpha}     }\, .
\end{equation}
This expression is very similar to the one appearing in the calculation of the negativity in Ref. \cite{coser2014entanglement}; the only difference is that the permutations involved are not the same. It is then straightforward to adapt the calculation of Ref. \cite{coser2014entanglement} to the OSEE. We start by making a crucial assumption about the initial state $\ket{\psi_0}$: we assume that it is of the form
\begin{equation}
	\label{eq:CC_initial}
	\ket{\psi(t)} \, = \, e^{- \frac{\lambda}{v} H} \ket{b} ,
\end{equation}
where $\ket{b}$ is a conformally invariant boundary state, $H$ is the Hamiltonian of the CFT, and $\lambda >0$ is some UV cutoff, known as the {\it extrapolation length} in the field theory approach to surface critical phenomena \cite{lubensky1975critical}. One simple way to understand the appearance of this length is as follows (this argument appeared in Refs. \cite{dubail2012edge,stephan2013logarithmic} and was later adapted by Cardy \cite{cardy2015quantum} to quantum quenches): since the initial state $\ket{\psi_0}$ is assumed to have short-range correlations, it must flow (under the RG) to a conformal boundary state; therefore $\ket{\psi_0}$ itself can be written as the RG fixed point $\ket{b}$ perturbed by some irrelevant operators $\phi_p$, $\ket{\psi_0} \propto e^{- \int d x  \sum_p \lambda_p \phi_p(x)} \ket{b}$. One then needs to discuss what irrelevant operators are allowed in this expression. The most generic one, which is also often the least irrelevant one, is the stress-tensor itself, which comes with a coefficient $\lambda$ homogeneous to a length. But the effect of perturbing the boundary condition by the stress-tensor is precisely to shift the position of the boundary. Therefore, taking $\ket{\psi_0}$ in the form (\ref{eq:CC_initial}) is equivalent to neglecting all perturbations but the one due to the stress-tensor. Of course, in principle, one should take any possible perturbations into account (as sketched in Ref. \cite{dubail2012edge} in a different context, and by Cardy in the context of quantum quenches in Ref. \cite{cardy2015quantum}), however this is much more difficult in practice, so we will omit all other perturbations and proceed from Eq. (\ref{eq:CC_initial}), as in most references dealing with global quenches in CFT, including \cite{coser2014entanglement}. \vspace{0.3cm}

\noindent The state at time $t$ is then
\begin{equation*}
	\ket{\psi(t)} \, = \, e^{- i H t} e^{- \frac{\lambda}{v} H} \ket{b} ,
\end{equation*}
and, as usual in this type of calculations, it is useful to switch to imaginary time: we replace $i v t$ by $y$, or equivalently
\begin{equation}
	\ket{\psi(t)} \, \rightarrow \, e^{- \frac{\lambda+y}{v} H} \ket{b} \qquad  {\rm and} \qquad  \bra{\psi(t)} \, \rightarrow \,  \bra{b} e^{- \frac{\lambda-y}{v} H} \, .
\end{equation}
Then, for $-\lambda <y<\lambda$, the ratio above becomes a ratio of correlation functions of three twist operators in an infinite strip (Fig. \ref{fig:CFT_global_quench}) of width $2\lambda$, parametrized by $(x,y) \in \mathbb{R} \times [- \lambda, \lambda]$,
\begin{equation*}
r(t) \, \rightarrow \, \frac{  ( \bra{b} e^{- \frac{\lambda-y}{v} H})^{\otimes 2 \alpha}  \;  \mathcal{S}^{A_1}_{(1,\alpha+2)(2,\alpha+3)\dots (\alpha-1,2\alpha) (\alpha,\alpha+1)} \mathcal{S}^{A_2}_{(1,\alpha+1)(2,\alpha+2)\dots(\alpha,2\alpha)}\; (e^{- \frac{\lambda+y}{v} H} \ket{b})^{\otimes 2 \alpha}  }{   ( \bra{b} e^{- \frac{\lambda-y}{v} H})^{\otimes 2 \alpha}\, \mathcal{S}^A_{(1,\alpha+1)(2,\alpha+2)\dots (\alpha,2\alpha)} \,   (e^{- \frac{\lambda+y}{v} H} \ket{b})^{\otimes 2 \alpha}   } \, = \, \frac{G_3}{G_2} \\
\end{equation*}
with
\begin{eqnarray*}
	 G_2 &\equiv& \left< \mathcal{T}_{(1,\alpha+1)(2+\alpha+2)\dots (\alpha,2\alpha)} ( i y ) \, \mathcal{T}_{(1,\alpha+1)(2+\alpha+2)\dots (\alpha,2\alpha)} ( L_A+ i y ) \right> \\
	 G_3 &\equiv& \left<  \mathcal{T}_{(1,\alpha+2)(2,\alpha+3)\dots (\alpha-1,2\alpha) (\alpha,\alpha+1)}(i y)  \,  \mathcal{T}_{(1,2,3,\dots,\alpha)(\alpha+1,2\alpha,2\alpha-1,\dots,\alpha+2)}(x+i y)   \, \mathcal{T}_{(1,\alpha+1)(2,\alpha+2)\dots(\alpha,2\alpha)}(L_A+i y)  \right>  .
\end{eqnarray*}
The strip is conformally mapped to the upper half-plane by $x+i y \mapsto \zeta = i e^{ \frac{\pi}{2 \lambda} (x+ iy)}$; it is convenient to work with the coordinates $\zeta_1$, $\zeta_2$, $\zeta_3$ in the upper half-plane, and their cross-ratios,
\begin{equation}
	\left\{ \begin{array}{rcl}  \zeta_1 &=&  i e^{\frac{\pi}{2 \lambda} (i y) }  \\
	\zeta_2 &=&  i e^{\frac{\pi}{2 \lambda} (L_A+ i y) } \\
	\zeta_3 &=&  i e^{\frac{\pi}{2 \lambda} (x+ i y) }    \end{array}  \right. \, ;  \qquad \eta_{ij} \, = \, \frac{(\zeta_i - \zeta_j)(\overline{\zeta}_i-\overline{\zeta}_j)}{(\zeta_i- \overline{\zeta}_j) (\overline{\zeta}_i- \zeta_j)  }\, .
\end{equation}
Then we have
\begin{eqnarray*}
	 G_2 &=&  \left| \frac{d \zeta_1}{ d z_1} \right|^{\alpha \Delta_2}   \left| \frac{d \zeta_2}{ d z_2} \right|^{\alpha \Delta_2}  \frac{1}{| \zeta_1-\overline{\zeta}_1|^{\alpha \Delta_2} |\zeta_2-\overline{\zeta}_2|^{\alpha \Delta_2}}  \frac{1}{\eta_{12}^{\alpha \Delta_2}}  \; f_2( \eta_{12}, \overline{\eta}_{12} ) ,   \\
	 G_3 &=& \left| \frac{d \zeta_1}{ d z_1} \right|^{\alpha \Delta_2}   \left| \frac{d \zeta_2}{ d z_2} \right|^{\alpha \Delta_2}  \left| \frac{d \zeta_3}{ d z_3} \right|^{2 \Delta_\alpha}  \frac{1}{|\zeta_1-\overline{\zeta}_1|^{\alpha \Delta_2} |\zeta_2-\overline{\zeta}_2|^{\alpha \Delta_2}  |\zeta_3-\overline{\zeta}_3|^{2 \Delta_\alpha}} \left( \frac{  \eta_{12}^{2 \Delta_\alpha - 2 \alpha \Delta_2} }{\eta_{13}^{2 \Delta_\alpha} \eta_{23}^{2 \Delta_\alpha} } \right)^{\frac{1}{2}} \; f_3( \{ \eta_{ij} \} , \{ \overline{\eta}_{ij} \}  )  \, .
\end{eqnarray*}
The Jacobians in front of both expressions come from the conformal mapping $z\mapsto \zeta$, while the rest is the correlation function in the upper half-plane. At the points $\zeta_1$ and $\zeta_2$, the twist operators have scaling dimension $\alpha \Delta_2$, since they correspond to permutation with $\alpha$ cycles of length $2$. At the point $\zeta_3$, the permutation has two cycles of length $\alpha$, therefore the scaling dimension is $2 \Delta_\alpha$.
There is of course not a unique way to write formulas for $G_2$ and $G_3$, since there is some arbitrariness in the choice of the functions of the anharmonic ratios $f_2$ and $f_3$. As in Ref. \cite{coser2014entanglement}, we have chosen our expressions such that $f_{2,3}$ always go to a finite constant whenever one of the anharmonic ratios $\eta_{ij}$ approaches $0$ or $1$, as can be seen from the fusion rules of the twist operators. The ratio of interest is then
\begin{eqnarray*}
	 \frac{G_3}{G_2} &=& \left( \left| \frac{d \zeta_3}{ d z_3} \right|   \frac{1}{| \zeta_3-\overline{\zeta}_3 |}\,  \left( \frac{  \eta_{12} }{\eta_{13}\eta_{23} }\right)^{\frac{1}{2}}  \right)^{2 \Delta_\alpha}    \frac{  f_3( \eta_{12}, \eta_{23} ,\eta_{13}, \overline{\eta}_{12}, \overline{\eta}_{23} ,\overline{\eta}_{13} )     }{  f_2( \eta_{12}, \overline{\eta}_{12} ) }  \\
	 & =& \left( \frac{\pi}{4 \lambda}  \frac{1}{ \cos \frac{\pi y}{2 \lambda} }\,  \left( \frac{  \eta_{12} }{\eta_{13}\eta_{23} }\right)^{\frac{1}{2}}  \right)^{2\Delta_\alpha}    \frac{  f_3( \eta_{12}, \eta_{23} ,\eta_{13}, \overline{\eta}_{12}, \overline{\eta}_{23} ,\overline{\eta}_{13} )     }{  f_2( \eta_{12}, \overline{\eta}_{12} ) }  \, .
\end{eqnarray*}
The anharmonic ratios are equal to
\begin{equation}
	\eta_{ij} = \frac{2 \sinh^2 \left( \frac{\pi}{4 \lambda} (x_i-x_j) \right)}{ \cos \left(\frac{\pi y}{\lambda} \right) + \cosh \left( \frac{\pi}{2\lambda} (x_i-x_j) \right) }
\end{equation}
and when we Wick-rotate back $y\rightarrow i vt$, they become
\begin{eqnarray*}
	\eta_{ij} &=& \frac{2 \sinh^2 \left( \frac{\pi}{4 \lambda} (x_i-x_j) \right)}{ \cosh \left(\frac{\pi v t}{\lambda} \right) + \cosh \left( \frac{\pi}{2\lambda} (x_i-x_j) \right) } \\
	& \underset{x_i-x_j, vt \gg \lambda}{\simeq} & \frac{ 1 }{1+ e^{\frac{\pi }{\lambda} (vt- \frac{1}{2} | x_i-x_j | )}  } \\
	& \sim & \left\{  \begin{array}{ccl}  1  & {\rm if} &  \frac{1}{2} |x_i - x_j| - vt \gg \lambda >0\\ e^{-\frac{\pi }{2 \lambda} (2vt-  | x_i-x_j | )}  &{\rm if} & vt - \frac{1}{2} |x_i - x_j|  \gg \lambda >0   \, .   \end{array} \right.	
\end{eqnarray*}
To study the regime where $\left| vt - \frac{1}{2} |x_i - x_j| \right|$ is of order $\lambda$, we would need the full dependence on the aspect ratios, which means that we would need to know the functions $f_2$ and $f_3$. In general, this is difficult. But there would be little to gain anyway, since we are most interested in results in the thermodynamic limit, where all length scales involved are much larger than $\lambda$. Under this assumption, $f_2$ and $f_3$ can be replaced by constants, and we see that
\begin{equation}
	\frac{G_3}{G_2} \, \simeq \, \left\{ \begin{array}{ccl}   \left( \frac{\pi}{2 \lambda}  e^{-\frac{\pi v t}{2 \lambda}}\,  \left( \frac{ 1 }{1 \times 1 }\right)^{\frac{1}{2}}  \right)^{2\Delta_\alpha}  \times   {\rm const.} \quad &{\rm if} &  vt <  \frac{1}{2} {\rm min} ( x, L_A -x) \\
	 \left( \frac{\pi}{2 \lambda}  e^{-\frac{\pi v t}{2 \lambda}}\,  \left( \frac{  1}{  e^{-\frac{\pi }{2\lambda} (2vt-  {\rm min} ( x, L_A -x)  )}  }\right)^{\frac{1}{2}}  \right)^{2\Delta_\alpha}  \times   {\rm const.} \quad &{\rm if} &  \frac{1}{2} {\rm min} ( x, L_A -x) < vt < \frac{1}{2} {\rm max} ( x, L_A -x) \\
	 \left( \frac{\pi}{2 \lambda}  e^{-\frac{\pi v t}{2 \lambda}}\,  \left( \frac{  1}{  e^{-\frac{\pi }{2\lambda} (4vt- L_A )}  }\right)^{\frac{1}{2}}  \right)^{2\Delta_\alpha}  \times   {\rm const.} \quad &{\rm if} &  \frac{1}{2} {\rm max} ( x, L_A -x)  < v t < L_A \\
	  \left( \frac{\pi}{2 \lambda}  e^{-\frac{\pi v t}{2 \lambda}}\,  \left( \frac{  e^{-\frac{\pi }{2\lambda} (2vt-  L_A )}   }{  e^{-\frac{\pi }{2\lambda} (4vt-  L_A )}  }\right)^{\frac{1}{2}}  \right)^{2 \Delta_\alpha}  \times   {\rm const.} \quad &{\rm if} &  L_A < v t \, .
	\end{array} \right.
\end{equation}
We are finally ready to write the CFT result for the OSEE, obtained by taking the logarithm of $G_3/G_2$, with the prefactor $1/(1-\alpha)$,
\begin{equation}
	\label{eq:OSEE_global_quench}
	S_\alpha(\rho_A(t)) \, \simeq \,  \frac{c}{6} \left( 1+\frac{1}{\alpha}\right)  \times  \left\{ \begin{array}{ccl}  \frac{\pi v t}{2 \lambda}  +   {\rm const.} \quad &{\rm if} &  vt < \frac{1}{2}{\rm min} ( x, L_A -x) \\
	 \frac{\pi }{4\lambda}   {\rm min} ( x, L_A -x )  +   {\rm const.} \quad &{\rm if} &  \frac{1}{2} {\rm min} ( x, L_A -x) < vt < \frac{1}{2} {\rm max} ( x, L_A -x) \\
	  \frac{\pi L_A}{4\lambda}  -   \frac{\pi v t}{2\lambda}    +   {\rm const.} \quad &{\rm if} & \frac{1}{2}  {\rm max} ( x, L_A -x)  < v t < L_A \\
	 0+{\rm const.} \quad &{\rm if} &  L_A < v t \, .
	\end{array} \right.
\end{equation}
The additive constant needs not be the same in the four lines, and the possible shift involved reflects the role of $f_3$ and $f_2$; however, this is always a subleading effect.
 A comparison between this formula and numerical results in the XX chain is displayed in Fig. \ref{fig:blowup} for the case $x = L_A/2$. The agreement is qualitatively good. At $t< L_A/(2v)$, the numerical result is clearly compatible with the linear growth we just found. However, for $t> L_A/(2v)$ the discrepancy between the numerics and the analytical prediction is larger. There are at least two possible explanations for that. One explanation could be that the initial state (\ref{eq:CC_initial}) we considered is not a good CFT representation of the N\'eel state for the XX chain, and that other perturbations should be included, as discussed above. Another explanation could be that the subleading effects due to the functions $f_2$ and $f_3$ are still quite large for the system sizes we probe numerically. These questions lie beyond the scope of the present paper and will be investigated more thoroughly elsewhere.

 \noindent In conclusion, our CFT calculation---which focuses on the leading behavior only and could possibly be improved to grasp also subleading effects---confirms that the OSEE initially grows linearly in time, and therefore blows up in the transient regime, even though it decays at later times. This is in perfect agreement with the discussion above.


\section{OSEE of the evolution operator, and localized phases}
\label{sec:evolution_op}

In this section, we focus directly on the OSEE of the evolution operator $U(t) = e^{-i H t}$. The results from this section will not be surprising to the expert reader, as they closely parallel famous results for the behavior of the entanglement entropy after a global quench. However, to the best of our knowledge, there seems to have been no attempts at a direct characterization of the approximability of $U(t)$ itself\footnote{But, just before this draft was submitted to the arXiv, another preprint by Zhou and Luitz \cite{2016arXiv161207327Z} appeared, that has a very large overlap with this section.} (as opposed to $U(t) \ket{\psi_0}$, in problems of quenches from some initial state $\ket{\psi_0}$), with the exception of a recent preprint by Zhang {\it et al.} \cite{zhang2016density} which focused on the same question in the context of Floquet time-evolution operators. So, although not very surprising, the results of this section are interesting because they concern the evolution operator $U(t)$ itself, as opposed to particular quench protocols; in that sense, they give a more direct characterization of the features of a given Hamiltonian $H$.

Of course, if $U(t)$ could be efficiently approximated by MPOs, then any time-evolution problem would be easy to simulate. Since this is not the case, we obviously expect the OSEE of $U(t)$ to grow quickly with $t$. This is actually easy to show in CFT.


\subsection{OSEE of the evolution operator in CFT: linear growth}

\begin{figure}[ht]
	\begin{center}
	\begin{tikzpicture}
	\begin{scope}[xshift=7.5cm,yshift=-5cm,scale=0.7]
		\draw[thick,<->] (9.5,4) arc (80:-80:0.5cm and 2cm);
		\draw[thick,<->] (9.3,4) arc (100:260:0.5cm and 2cm);
		\draw[thick,->] (-1,-0.1) arc (-70:250:0.4cm and 2.2cm) node[left]{$y$};
		\draw[thick,->] (1.5,-0.9) -- ++(6,0) node[below]{$x$};
		\draw[thick] (4.5,-0.75) -- ++(0,-0.3) node[below]{$0$};
		\draw[thick] (-1.13,4) -- ++(0,0.28) ++(0,0.3) node{\small $y=0$};
		\draw (9.8,2.2) node[right]{\footnotesize$\frac{\beta v}{2} $};
		\draw (8.8,1.5) node[right]{\footnotesize$\frac{\beta v}{2}$};
		\filldraw[very thick,draw=none,fill=gray!50] (0,0) arc (-180:-90:0.3cm) -- ++(8,0) arc (-90:0:0.3cm) -- ++(0,4) arc (0:90:0.3cm) -- ++(-8,0) arc (90:180:0.3cm);
		\draw[very thick] (0,0) arc (-180:-90:0.3cm) ++(8,0) arc (-90:0:0.3cm) -- ++(0,4) arc (0:90:0.3cm) -- ++(-8,0) arc (90:180:0.3cm) -- (0,0);		
		\filldraw[thick,fill=gray!80] (0,0) arc (-180:0:0.3cm) -- ++(0,4) arc (0:180:0.3cm) -- cycle;		
		\filldraw[thick,fill=gray!80] (0.085,0) arc (-180:0:0.22cm) -- ++(0,4) arc (0:180:0.22cm) -- cycle;		
		\filldraw[thick,fill=gray!80] (0.17,0) arc (-180:0:0.14cm) -- ++(0,4) arc (0:180:0.14cm) -- cycle;		
		\draw[dashed] (4.2,-0.3) arc (-90:0:0.3cm) -- ++(0,4) arc (0:90:0.3cm);
		\draw (2.3,2) node{$A$};
		\draw (6.5,2) node{$B$};
		\filldraw (4.4,-0.3) circle (1mm) ++(-0.3,0.5) node{\small $\mathcal{T}_\sigma (-i \beta v/2)$};
		\draw[very thick,dashed] (0.4,4.1) -- (8.4,4.1);
		\draw[very thick] (0.6,3.1) -- (4.5,3.1);
		\draw[very thick] (0.4,-0.3) -- (4.4,-0.3);
		\draw[very thick,dashed] (4.4,-0.3) -- (8.3,-0.3);
		\filldraw (4.5,3.1) circle (1mm) ++(-0.3,0.4) node{\small $\mathcal{T}_{\sigma^{-1}} (i y)$};
		
		\draw[<-] (2.5,-0.6) -- (1,-2.5) node[below]{swap $\mathcal{S}_{(1,2,\dots,\alpha)}$};
		\draw[<-] (2.2,3.3) -- (1,6) node[above]{swap $\mathcal{S}_{(1,\alpha,\dots,2)}$};
	\end{scope}
	\end{tikzpicture}
	\end{center}
	\caption{The $\alpha$-sheeted cylinder that is used to calculate the OSEE of $U(t)$ in imaginary time: the result is given by the two-point function $\left< \mathcal{T}_{\sigma^{-1}} (-i \beta v /2) \mathcal{T}_{\sigma} (i y)\right>$, with $\sigma = (1,2,\dots, \alpha)$, on an infinite cylinder of circumference $\beta v$.}
	\label{fig:CFT_evolution_operator}
\end{figure}
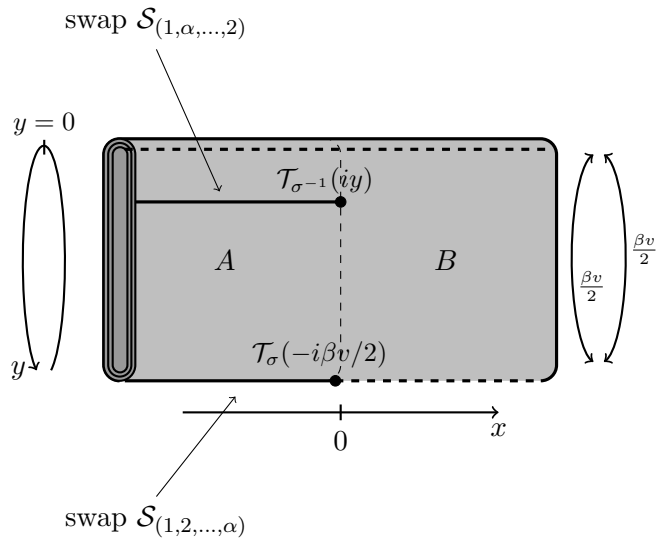
We consider once again an infinitely long system described by a CFT with a Hamiltonian $H$, with the corresponding evolution operator
\begin{equation}
	U(t)\, = \, e^{- i H t} \, .
\end{equation}
We want to calculate the OSEE for the bipartition $A \cup B = (-\infty, 0] \cup [ 0, +\infty )$. At time $t=0$, $U(t)$ is the identity operator, 
whose OSEE is obviously zero; at later times, $U(t)$ is a non-trivial operator with non-zero OSEE. The question is: {\it how fast does it increase as a function of time?} \vspace{0.3cm}

\noindent In CFT, this question is easily answered as follows. We rely again on Eq. (\ref{eq:OSEE_replicas}), assuming $\alpha$ integer $\geq 2$. Here, Eq. (\ref{eq:OSEE_replicas}) reads
\begin{eqnarray*}
	S_\alpha (U(t)) &=& \frac{1}{1-\alpha} \log \left( \frac{  {\rm tr} [(e^{i H t})^{\otimes \alpha} \cdot \mathcal{S}_{(1,\alpha,\alpha-1,\dots,2)} \cdot  (e^{-i H t})^{\otimes \alpha} \cdot  \mathcal{S}_{(1,2,\dots,\alpha)}]   }{ ( {\rm tr} [ e^{i H t} e^{-i H t} ] )^{\alpha} }  \right) .
\end{eqnarray*}
Notice that the denominator inside the logarithm is trivially equal to one; the numerator cannot be evaluated that easily though. As in the previous, it is convenient to switch to imaginary time, and to start by calculating the quantity
\begin{eqnarray*}
	\frac{  {\rm tr} [(e^{-(\frac{\beta}{2}-\frac{y}{v}) H} )^{\otimes \alpha} \cdot \mathcal{S}_{(1,\alpha,\alpha-1,\dots,2)} \cdot  (e^{-(\frac{\beta}{2}+\frac{y}{v}) H})^{\otimes \alpha} \cdot  \mathcal{S}_{(1,2,\dots,\alpha)}]   }{ ( {\rm tr} [ e^{- \frac{\beta}{2} H} e^{- \frac{\beta}{2} H} ] )^{\alpha} } ,
\end{eqnarray*}
and at the end of the calculation, do the Wick rotation $y \rightarrow i v t$ and take the limit $\beta \rightarrow 0^+$. We will keep $\beta$ finite though, as it conveniently plays the role of a UV cutoff. The big advantage of looking at this imaginary-time version of the ratio is that, as long as $|y| < \beta v/2$, it is again (exactly like in section \ref{sec:CFT_thermal}) a ratio of two partition functions of the CFT on a cylinder with $\alpha$ sheets, see Fig. \ref{fig:CFT_evolution_operator}, that boils down to a two-point function of twist operators. Indeed, if we parametrize the cylinder by the complex coordinates $x+i y$, $(x,y) \in \mathbb{R} \times [-\beta v/2 , \beta v/2]$, then the two swap operator $\mathcal{S}_\alpha$ can be replaced by a pair of twist operators at $- i \beta v/2$ and $ i y$ respectively. There are also two operators at the left end of the cylinder (at infinity), but those two can be fused together to the identity; so they don't affect the result in the end. We are left with the two-point function
\begin{eqnarray*}
	&& \frac{  {\rm tr} [(e^{-(\frac{\beta}{2}-\frac{y}{v}) H} )^{\otimes \alpha} \cdot \mathcal{S}_{(1,\alpha,\dots,2)} \cdot  (e^{-(\frac{\beta}{2}+\frac{y}{v}) H})^{\otimes \alpha} \cdot  \mathcal{S}_{(1,2,\dots, \alpha)} ]   }{ ( {\rm tr} [  e^{- \beta H} ] )^{\alpha} } \\
	& = & \left< \mathcal{T}_{(1,\alpha,\dots,2)} (-i \beta v/2) \mathcal{T}_{(1,2,\dots,\alpha)} (i y)\right>  \, \propto \, \frac{1}{[  \frac{ \beta v }{\pi }  \sin ( \frac{ \pi   (\frac{\beta v}{2}+y)}{ \beta v} )   ]^{2 \Delta_\alpha}} \, ,
\end{eqnarray*}
up to some non-universal constant factor that is homogeneous to a length to the power $2 \Delta_\alpha$, because the ratio of traces must be dimensionless. Then, Wick-rotating back to $y \rightarrow i v t$, and adjusting the constant factor such that $S_\alpha (U(t))$ vanishes at $t=0$, we find
\begin{eqnarray*}
	S_\alpha (U_t) & = & \frac{c}{6}\left( 1+ \frac{1}{\alpha}\right)  \log \left(  \cosh \left( \frac{ \pi   t }{ \beta} \right) \right) \\
		& \underset{ t \gg \tau }{\simeq} &   \frac{c}{6}\left( 1+ \frac{1}{\alpha}\right)  \frac{ \pi   t }{ \beta}  \, .
\end{eqnarray*}
The OSEE increases linearly in time. In other words, if one wants to approximate $U(t)$ by an MPO, then one needs a bond dimension that blows up exponentially with time.
This conclusion, of course, in not surprising, and it is intimately tied to the famous linear growth of the entanglement entropy after a global quench, first pointed out by Calabrese and Cardy \cite{calabrese2005evolution}. \vspace{0.3cm}

\noindent In fact, there are two good reasons why the OSEE of the evolution operator must be closely related to the entanglement entropy after a global quench from an initial state with short-range correlations $\ket{\psi_0}$. First, it is because the entanglement of $U(t) \ket{\psi_0}$ obviously does not come from $\ket{\psi_0}$ itself, so it must come from $U(t)$; in other words, since $\ket{\psi_0}$ is well approximated by an MPS with small bond dimension, the only explanation for the fact that the entanglement entropy of $U(t) \ket{\psi_0}$ blows up must be that the bond dimension of $U(t)$ itself blows up. Second, upon using the 'operator-folding' trick shown in Fig. \ref{fig:cartoon_OSEE}.(c), namely replacing the operator $U(t) \in {\rm End}(\mathcal{H})$ by the state $\ket{U(t)} \in \mathcal{H} \otimes \overline{\mathcal{H}}$, on sees that $U(t)$ itself can be viewed as resulting from a global quench,
\begin{equation}
	\ket{U(t)}  \, = \,  e^{-i (H \otimes 1 - 1 \otimes H) t}  \ket{{\rm Id}}  .
\end{equation}
This implies that the linear growth of the OSEE of $U(t)$ is valid far beyond the range of applicability of CFT: all systems with ballistic spreading of correlations are known to
have an entanglement entropy that grows linearly after a global quench (see {\it e.g.} the recent discussion \cite{alba2016entanglement}), so they must have an evolution operator $U(t)$ whose OSEE increases linearly in time. Again, this should not be surprising to most readers, however the point we want to emphasize here is that this is really a property of the evolution operator (and therefore of the Hamiltonian itself), which does not have anything to do with any particular quench protocol.

\subsection{The evolution operator in Anderson and many-body localized phases} 

For the same reason, in phases where there is no ballistic spreading of correlations, namely in {\it localized} phases,
the OSEE of the evolution must mimic the generic behavior of the entanglement entropy after a global quench. In
free fermion systems that are (Anderson) localized, the OSEE of $U(t)$ must be bounded, while in
many-body localized phases, it must grow logarithmically \cite{vznidarivc2008many,bardarson2012unbounded,serbyn2013universal}. In the (Anderson-localized) free-fermion case, this is simply because the Hamiltonian is diagonalized in a basis of orbitals that are exponentially localized. For instance, if we consider the following simple model
\begin{equation}
	H \, = \,  - \frac{1}{2} \sum_x  [c^\dagger_x c_{x+1} \, + \,h.c.] \, + \, \sum_x V(x) c^\dagger_x c_x ,
\end{equation}
for some random onsite potential $V(x)$, then it is always true that it can be put in diagonal form
\begin{equation}
	H \, = \, \sum_x \varepsilon_x a^\dagger_x a_x ,
\end{equation}
with some random energies $\varepsilon_x$, and where the new creation/annihilation modes are exponentially localized around the original ones,
\begin{equation}
	u_{x,x'} \equiv \bra{0} a(x)  c^\dagger(x') \ket{0} \, \sim \, e^{-|x-x'|/\xi} .
\end{equation}
$\xi > 0$ is the localization length, it depends on the strength of the random potential. The amplitudes $u_{x,x'}$ satisfy the normalization condition $\sum_{x''} u_{x,x''} u^*_{x',x''} = \delta_{x,x'}$. The Hilbert space $\mathcal{H}$ is the Fock space generated by the $c^\dagger$'s; it is useful to introduce a copy of this Fock space, noted $\mathcal{H}_{f^\dagger}$, generated by other fermion modes $f^\dagger$'s. The $c^\dagger$'s and $f^\dagger$'s do not act on the same space, but it is nevertheless convenient to write expressions that involve both kinds of operators; we decide, by convention, that the $c^\dagger$'s and the $f^\dagger$'s always anticommute: $\{ c^\dagger_x , f_{x'} \} =\{ c^\dagger_x , f^\dagger_{x'} \}  =0$. Then we can write the following map $W: \mathcal{H} \rightarrow \mathcal{H}_{f^\dagger}$,
\begin{equation}
 	W \, = \, \exp\left( \sum_{x,x'} u_{x,x'}  f^\dagger_x c_{x'} \right) ,
\end{equation}
which takes a localized orbital $a^\dagger_x = \sum u^*_{x,x'} c^\dagger_{x'}$ and maps it to $f^\dagger_x$. $W$ is local in the sense that the coefficients $u_{x,x'}$ decay exponentially with distance; it also satisfies $W^\dagger W = 1$, so it is in fact a local unitary transformation. Because $W$ is local, it must satisfy the operator area law: it has a bounded OSEE. The evolution operator now takes the form
\begin{equation}
	({\rm Anderson \; localized}) \qquad e^{-i H t} \, = \,     W^\dagger \cdot   \left(  \bigotimes_{x} e^{- i t\, \varepsilon_x f^\dagger_x f_x }  \right) \cdot W .
\end{equation}
The point is that the operator in the middle, which is the only piece that is time-dependent in this expression, is a product operator, therefore
it cannot generate any entanglement. So, in an Anderson-localized system, the evolution operator itself satisfies the operator area law.

In many-body localized phases, the story is similar, except that the diagonalized evolution operator does not take the form of a tensor product anymore.
Instead, it is a sum of commuting terms that couple the localized orbitals (the $f^\dagger$'s) at arbitrary distances, with
exponentially decaying amplitudes,
\begin{equation}
	({\rm many-body \; localized}) \qquad e^{-i H t} \, = \,     W^\dagger \cdot  \exp \left( - i t \left[ \sum_x \varepsilon_x  f^\dagger_x f_x  +  \sum_{x,x'}  v_{x,x'}  f^\dagger_x f_x  f^\dagger_{x'} f_{x'}  + \dots \right]  \right) \cdot W .
\end{equation}
The latter are responsible for a dephasing mechanism, identified in Ref. \cite{serbyn2013universal}, which ultimately generates a logarithmic growth of the entanglement; these arguments can be repeated here to show that the OSEE must grow logarithmically\footnote{I am grateful to I. Cirac for a helpful discussion about this point.}.


\section{Local operators in Heisenberg picture: is it at all possible to get their OSEE from CFT?}

\begin{figure}[ht]
	\begin{center}
		\includegraphics[width=0.33\textwidth]{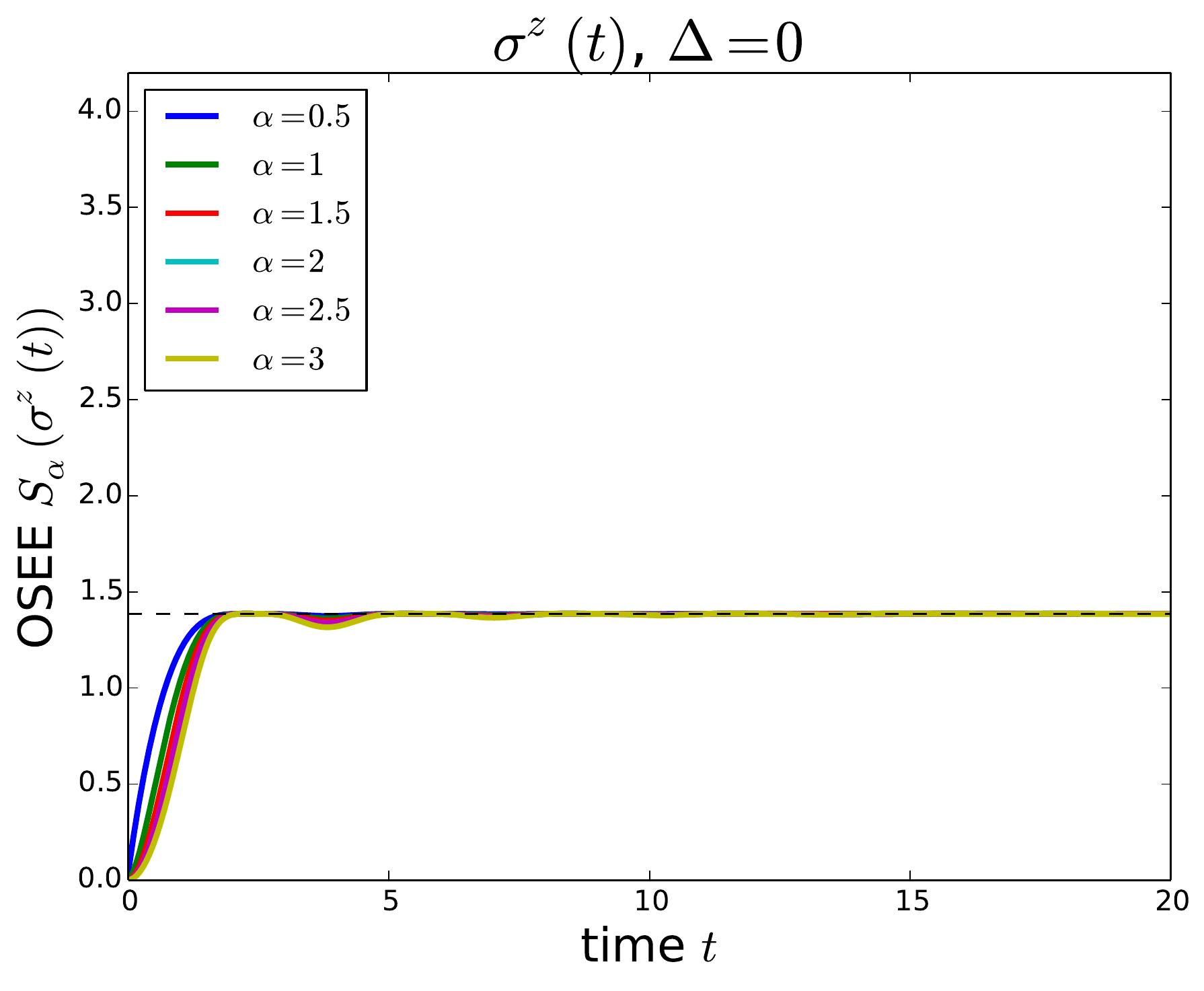}
		\hspace{-0.25cm} \includegraphics[width=0.33\textwidth]{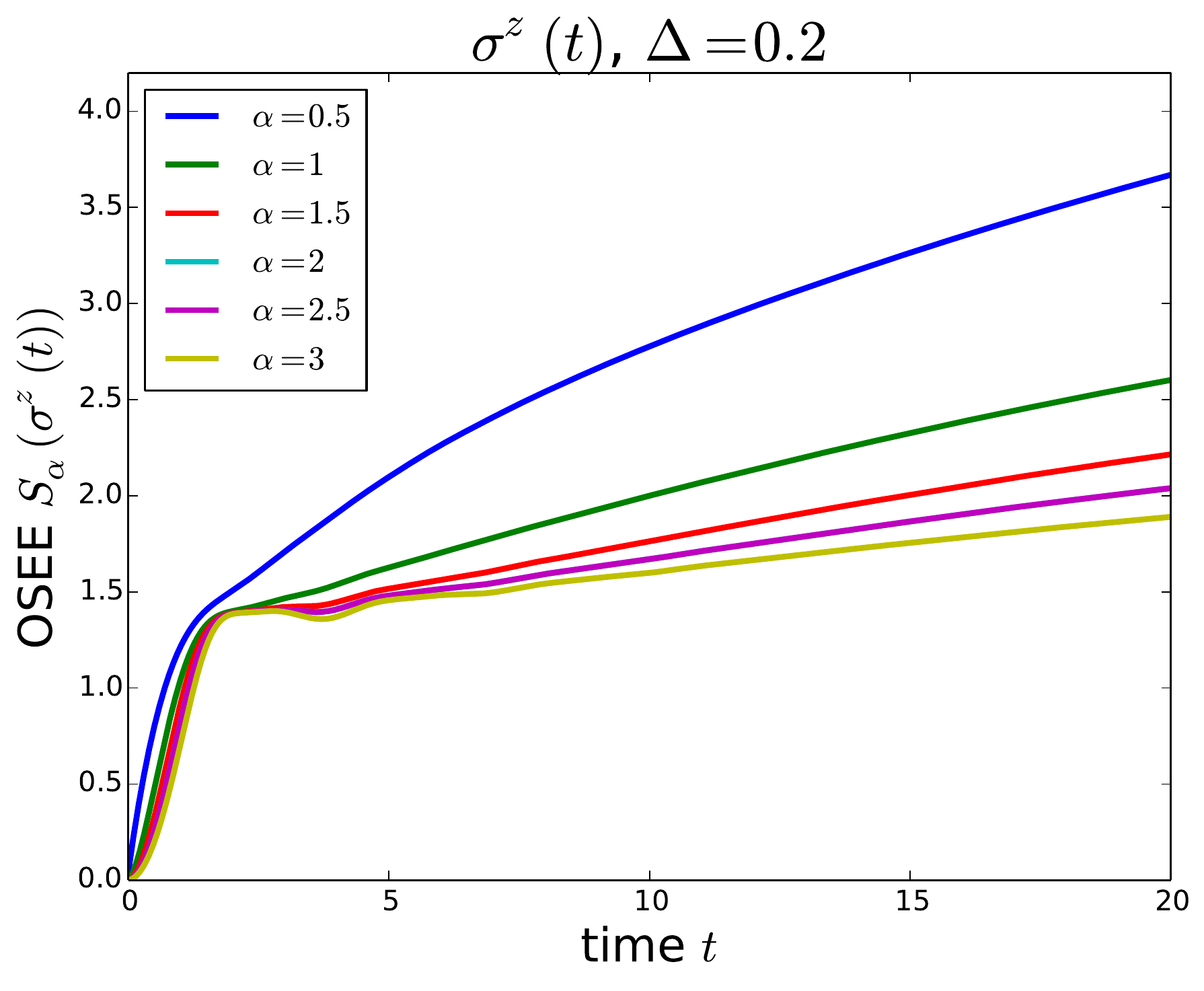}
		\hspace{-0.25cm} \includegraphics[width=0.33\textwidth]{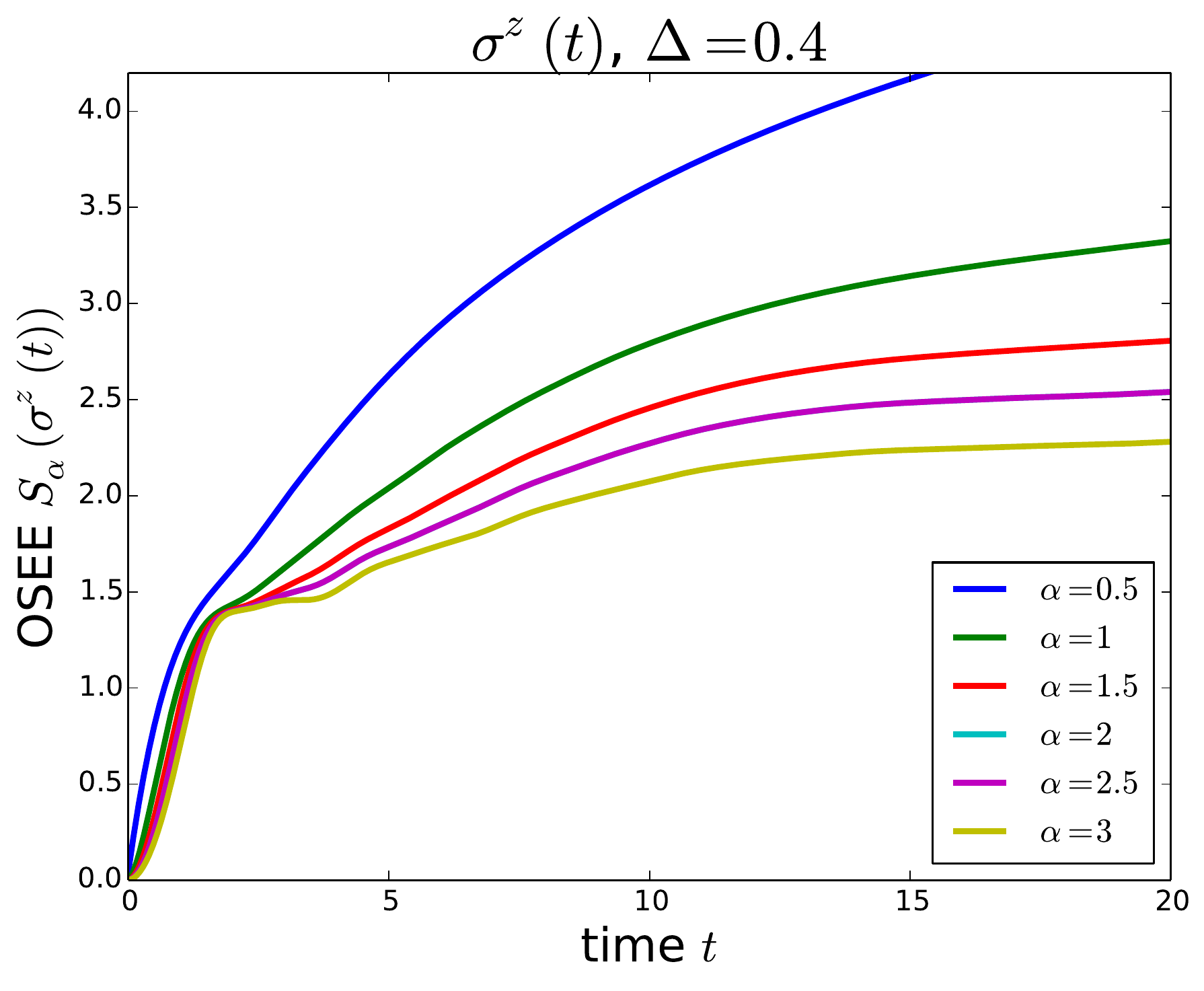} \\
		\includegraphics[width=0.33\textwidth]{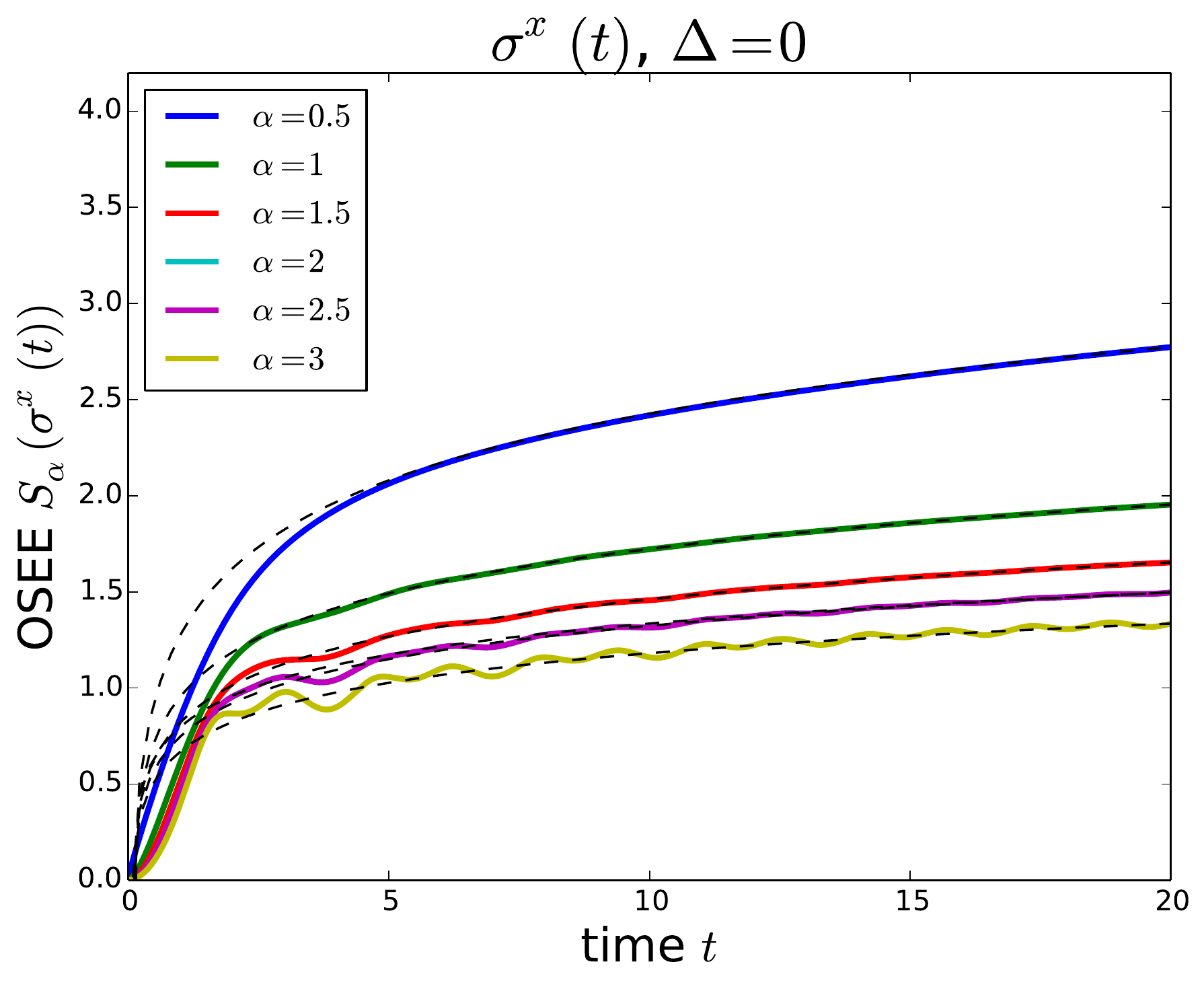}
		\hspace{-0.25cm} \includegraphics[width=0.33\textwidth]{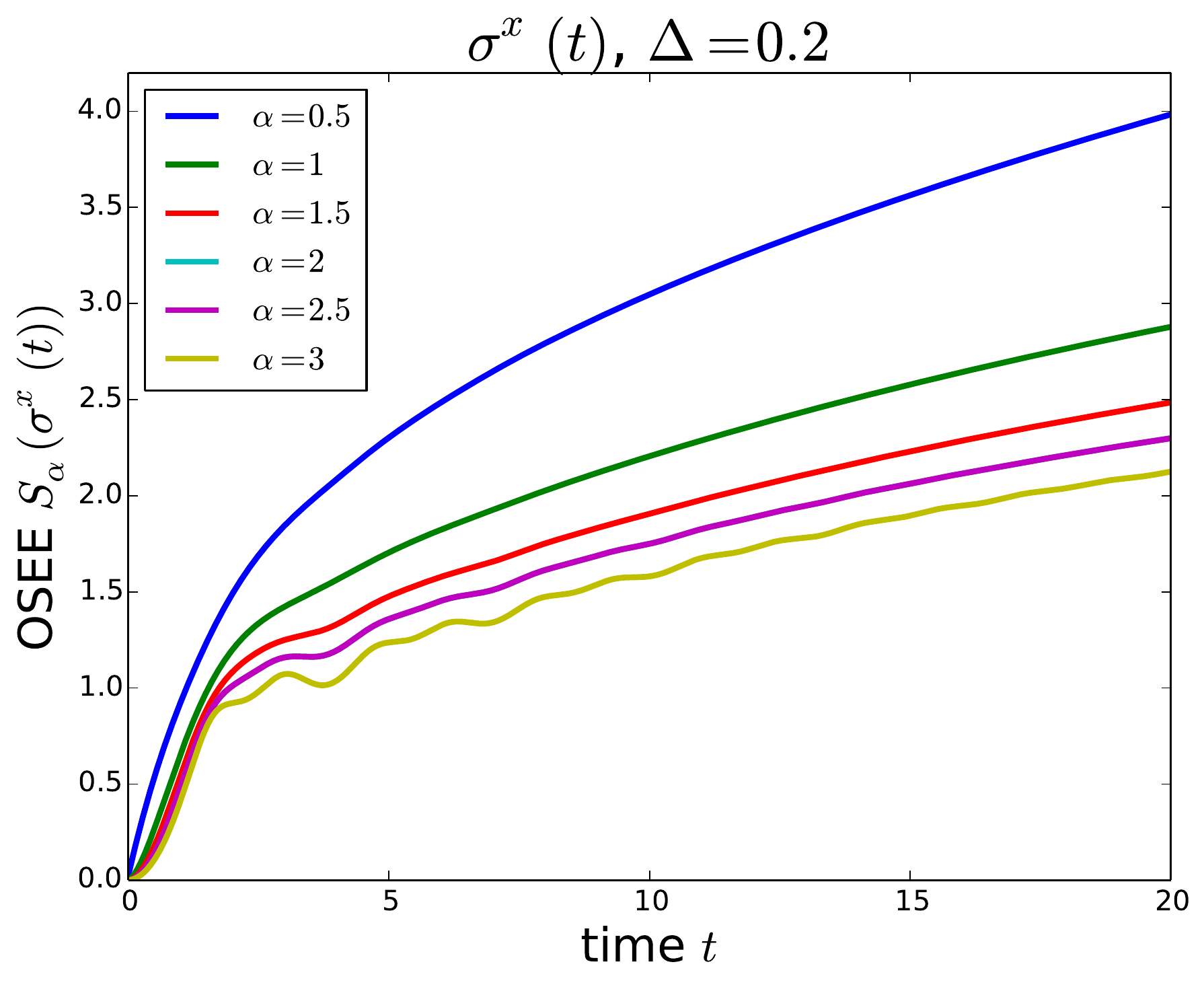}
		\hspace{-0.25cm} \includegraphics[width=0.33\textwidth]{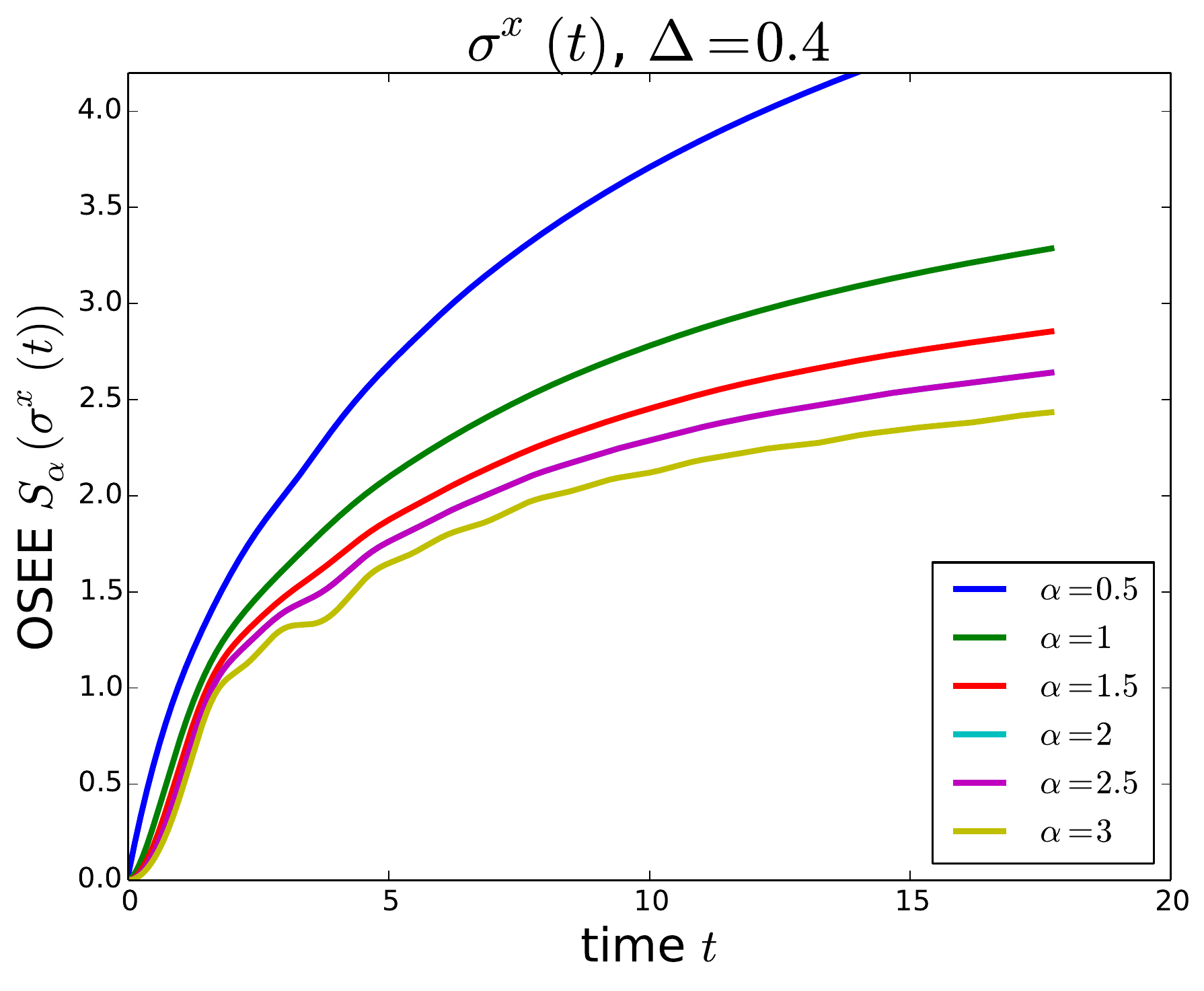} \\
	\end{center}
	\caption{OSEE of operators $\sigma^z_0(t)$ and $\sigma^x_0(t)$ in the XXZ chain, for a bipartition $[-L/2,0] \cup [1,L/2]$. The XXZ Hamiltonian is normalized as $H = -\frac{1}{2} \sum_x \left[  \sigma^+_x \sigma^-_{x+1}  +  \sigma^-_x \sigma^+_{x+1}  + \frac{\Delta}{2}  \sigma^z_x \sigma^z_{x+1}  \right]$. We use Time-Evolving Block Decimation (TEBD), on a system of $L=60$ sites. In the free fermion case (anisotropy parameter $\Delta =0$), the OSEE is bounded for $\sigma^z_0(t)$ (it quickly converges to $\log 4$), and it grows as $\frac{1}{6} (1+\frac{1}{\alpha}) \log t$ for $\sigma^x_0(t)$. When interactions are turned on (here $\Delta = 0.2$ and $\Delta = 0.4$), the growth of the OSEE is still sublinear.}
	\label{fig:OSEE_Heisenberg}
\end{figure}

Finally, we wish to revisit the original question tackled by Prosen and Pi\v{z}orn \cite{prosen2007operator}: how does the OSEE of a local operator in Heisenberg picture, $\phi(t) \, = \, e^{i H t} \phi e^{-i H t}$, evolve in time? They gave the following answer for free fermions (Ising chain \cite{prosen2007operator} and XY chain \cite{pivzorn2009operator}): depending on the local operator one is considering, it is either bounded, or it grows logarithmically (more details below). This observation, of course, is potentially very interesting for numerical purposes. Assuming that the OSEE would grow also at most logarithmically in {\it interacting} systems, then it would be efficient to represent all local operators in Heisenberg picture as MPOs, and this would allow to simulate the dynamics after quantum quenches at times much larger than the ones accessible in the standard Schr\"odinger picture. Similar observations were made in Ref. \cite{hartmann2009density}, which also concluded that Heisenberg-picture MPO algorithms should be dramatically more efficient than Schr\"odinger-picture MPS algorithms in a majority of dynamical problems. Thus, the question is:
\begin{center}
{\it (Q2) \qquad How fast does the OSEE of local operators in Heisenberg picture grow in generic interacting systems?} \vspace{0.1cm}
\end{center}
\noindent In Fig. \ref{fig:OSEE_Heisenberg}, we report some numerical results in the XXZ spin chain that clearly show that the OSEE is always
sublinear as a function of time. The results are compatible with a generic logarithmic growth of the OSEE, however the time window that is accessible to us
is quite small, and it would be interesting to have a more extensive numerical study that can confirm, or invalidate, the logarithmic growth.  \vspace{0.3cm}


\noindent This section reflects an attempt to attack this question with the analytical tools of 2d CFT. As already mentioned, the advantage of CFT is that it provides simple models of interacting many-body systems, with analytical calculations that usually remain easily tractable, thereby providing very helpful analytical insights for relatively minor effort. We shall see that the logarithmic growth of the OSEE can be obtained analytically in free fermion systems (we treat the example of the XXZ chain at $\Delta = 0$ explicitly), confirming the numerical observation of  Prosen and Pi\v{z}orn. However, the free fermion case happens to be peculiar. It follows from a rather unexpected connection with recent work on quantum quenches from strongly inhomogeneous initial states \cite{allegra2016inhomogeneous,dubail2016conformal}, and hints at a surprising breakdown of conformal invariance (see Fig. \ref{fig:regularization_scheme} below). We shall explain why it seems difficult to extend this calculation to interacting systems.   \vspace{0.3cm}

\noindent Overall, it will remain unclear at the end of this section whether CFT can be useful at all to understand the OSEE of operators in Heisenberg picture in interacting spin chains; we will leave this as an interesting open problem that deserves further investigation.  \vspace{0.3cm}

\subsection{The observations of Prosen and Pi\v{z}orn on the free fermion case}

Let us start by summarizing the observations of Ref. \cite{prosen2007operator}.
For simplicity, we consider only the case of a Hamiltonian of the form $H = \int \frac{dk}{2\pi} \varepsilon(k) c^\dagger(k) c(k)$, which conserves the number of fermions. But the statements also hold for Hamiltonians with pairing terms ({\it i.e.} $c^\dagger c^\dagger$ and $c c$).

\paragraph{The fermion creation/annihilation operator is an exact MPO.} The first observation \cite{prosen2007operator} is that
\begin{eqnarray*}
	c^\dagger_x (t) &=& \sum_{x' \in \mathbb{Z}} K(x-x',t) c^\dagger_{x'}
\end{eqnarray*}
for some kernel $K(x-x',t)$. For the XX chain, $K$ can be written with a Bessel function, but in general the precise form of $K(x-x',t)$ does not matter. What is important is
to observe that this expression of $c^\dagger_x (t)$ can be interpreted as an MPO with bond dimension $2$. Consequently, all finite products of fermion creation/annihilation operators, of the form $c_{x_1}^\dagger \dots c_{x_p}^\dagger  c_{x'_1} \dots c_{x'_q}$, are exact MPOs with finite bond dimension (the bond dimension is at most $2^{p+q}$). Of course, this implies that all such operators satisfy the operator area law, because their OSEE is obviously not larger than $(p+q) \log 2$.

\paragraph{The OSEE of a Jordan-Wigner string increases logarithmically in time.} The second observation \cite{prosen2007operator}, more interesting and intriguing, is
that the (endpoint of the) Jordan-Wigner (JW) string,
\begin{equation}
	{\rm JW}_x \, \equiv \, \prod_{x' \leq x} (-1)^{c_{x'}^\dagger c_{x'}}  \, = \, \prod_{x' \leq x} ( 1 - 2 c_{x'}^\dagger c_{x'}) ,
\end{equation}
when represented in Heisenberg picture, namely
\begin{equation}
	{\rm JW}_x(t) \, = \, \prod_{x' \leq x} [ 1 - 2 c_{x'}^\dagger (t) c_{x'}(t) ] ,
\end{equation}
has an OSEE that grows logarithmically at large time $t$. The prefactor of the logarithm of $t$ is always a rational number (for $\alpha=1$); Prosen and Pi\v{z}orn conjectured a formula for this prefactor in Ref. \cite{pivzorn2009operator}; we shall derive this prefactor for the XX chain below.  \vspace{0.4cm}

\paragraph{All local operators have an OSEE that increases at most logarithmically.} By multiplying JW strings with products of creation/annihilation operators $c_{x_1}^\dagger \dots c_{x_p}^\dagger  c_{x'_1} \dots c_{x'_q}$, one can represent all
local operators in any spin chain that can be mapped to fermions with a Jordan-Wigner transformation. So the two observations above are sufficient to conclude that {\it all local operators in Heisenberg picture have an OSEE that grows at most logarithmically, as long as the Hamiltonian is quadratic in the fermion creation/annihilation modes}. \vspace{0.3cm}

\subsection{OSEE of a Jordan-Wigner string: analytical derivation by mapping to a domain-wall initial state}

We now turn to the explicit calculation of the OSEE of a JW string in the XX chain. Without loss of generality, the endpoint of
the string can be placed at the origin, so we are interested in the OSEE of ${\rm JW}_0 (t)$. We proceed as follows (see also the appendix for more details about how to calculate the OSEE in free fermion systems). We start by turning the
operator into a state
\begin{equation}
	{\rm JW}_0(t) \, \longrightarrow \, \ket{{\rm JW}_0 (t)} \,\equiv\, \prod_{x \leq 0} [ 1 - 2 c_x^\dagger(t)  \tilde{c}^\dagger_x(t) ] \ket{0} ,
\end{equation}
namely we replace $c(t)$ by some new creation operator $\tilde{c}^\dagger (t)$, which anticommutes with all the $c$'s. $\ket{0}$ is the state that is annihilated
by all the $c$'s and all the $\tilde{c}$'s. As sketched in Fig. \ref{fig:cartoon_OSEE}.c, the main interest of rewriting operators as states in an enlarged Hilbert space (here a fermion Fock space) resides in the fact that the OSEE becomes the usual entanglement entropy; we will make use of this shortly. \vspace{0.3cm}

\noindent It is convenient to think of this state as coming from the time-evolution
\begin{equation}
	 \ket{ {\rm JW}_0(t)}  \, = \,e^{-i (H\otimes \tilde{1} - 1\otimes \tilde{H}) t} \ket{ {\rm JW}_0(0)} ,
\end{equation}
where $H\otimes \tilde{1} - 1\otimes \tilde{H}$ acts as
\begin{equation}
	\begin{array}{lll}
		\left[H\otimes \tilde{1} - 1\otimes \tilde{H} , c^\dagger(k) \right] \,=\, \varepsilon(k) c^\dagger(k)   & \qquad  \quad& \left[H\otimes \tilde{1} - 1\otimes \tilde{H}, \tilde{c}^\dagger(k) \right] \,=\, - \varepsilon(k) \tilde{c}^\dagger(k)     \\ \\
		\left[H\otimes \tilde{1} - 1\otimes \tilde{H}, c(k) \right] \,=\, -\varepsilon(k) c(k)   & \qquad  \quad& \left[H\otimes \tilde{1} - 1\otimes \tilde{H}, \tilde{c}(k) \right] \,=\,  \varepsilon(k) \tilde{c}(k)  .
	\end{array}
\end{equation}
Next, we perform the following Bogoliubov transformation,
\begin{equation}
	\label{eq:bogoliubov}
	\left( \begin{array}{c}  b^\dagger_x \\ d_x  \end{array} \right) \, \equiv  \,   \left( \begin{array}{cc}  \frac{2}{\sqrt{5}} & - \frac{1}{\sqrt{5}} \\  \frac{1}{\sqrt{5}} & \frac{2}{\sqrt{5}} \end{array}  \right) \left( \begin{array}{c}  c^\dagger_x \\ \tilde{c}_x \end{array} \right).
\end{equation}
This transformation does not do anything to the Hamiltonian $H\otimes \tilde{1} - 1\otimes \tilde{H}$, which already was (and still is) in diagonal form. However,
it simplifies the initial state:
\begin{equation}
	\ket{ {\rm JW}_0(0)} \propto  \prod_{x \leq 0}  d^\dagger_x b^\dagger_x \ket{0} \, .
\end{equation}
Notice that this is nothing but a product state, of the form $\ket{ {\rm DWIS}_b }   \otimes \ket{ {\rm DWIS}_d }   = \left(  \prod_{x \leq 0}  d^\dagger_x \ket{0_b} \right) \otimes \left(  \prod_{x \leq 0}  b^\dagger_x \ket{0_b} \right)$, where $\ket{0_{b,d}}$ is the vacuum for the $b$- and $d$-modes respectively. The two independent initial states for the $b$'s and the $d$'s are {\it domain-wall initial states} (DWIS), in the terminology of \cite{calabrese2008time, allegra2016inhomogeneous}.
At this point, the $b$- and $d$-modes are completely decoupled, and we are left with two independent time-evolutions,
\begin{equation}
	 \ket{ {\rm JW}_0(t)}  \, = \,e^{-i  H_b t} \ket{ {\rm DWIS}_b }   \otimes e^{-i  H_d t} \ket{ {\rm DWIS}_d }  
\end{equation}
where $H_b = \int \frac{dk}{2\pi} \varepsilon (k) b^\dagger(k) b(k)$ and $H_d = -\int \frac{dk}{2\pi} \varepsilon (k) d^\dagger(k) d(k)$. For the XX chain, the dispersion relation 
is $\varepsilon(k) = -\cos k$. \vspace{0.3cm}

\noindent The entanglement entropy of the XX chain evolving from a DWIS was calculated recently numerically in Ref. \cite{eisler2014surface}, and from CFT in Ref. \cite{dubail2016conformal}, relying on previous results \cite{allegra2016inhomogeneous}. The result is the following. For a bipartition $A\cup B$ with $A=(-\infty, x]$ and $B=(x,+\infty)$, the entanglement entropy
of $e^{- i H t} \ket{{\rm DWIS}}$ remains constant while $v t < |x|$, and then behaves as $S_\alpha ( e^{- i H t} \ket{{\rm DWIS}} ) = \frac{1}{12} (1+\frac{1}{\alpha}) \log \left[ t (1-(x/vt)^2)^{3/2} \right]$ when $vt > |x|$. We can import this result and apply it directly to the present situation. Since, here, we have two copies of the XX chain (one for the $b$'s and one for the
$d$'s), the OSEE of the JW string is exactly twice the entanglement entropy of the latter,
\begin{equation}
	S_\alpha ( {\rm JW}_0(t) ) \, = \, \frac{1}{6} \left(1+\frac{1}{\alpha}\right) \log \left[ t \left(1-(x/t)^2\right)^{\frac{3}{2}} \right] \, .
\end{equation}
We note that the recent paper \cite{SciPostPhys.1.2.014} studied a closely related problem about the growth of the (usual) entanglement entropy
in the Ising chain.

\paragraph{Breakdown of conformal invariance.} It is important to emphasize the following point. We clearly see that, even though the XX chain may be described by a CFT (in the sense that its Hamiltonian is gapless), and so we should be able to use some CFT techniques to analyze it, we have to deal with the DWIS, which breaks scale  invariance. Indeed, the Hamiltonian $H_b$ (resp. $H_d$) preserves the number of particles, so if the density is either zero or one, then no density fluctuation is
possible at all. Since $\ket{{\rm DWIS}_{b,d}}$ is precisely a state where the density is one or zero, then there can be no fluctuating degrees of freedom close to the boundaries. There is still a CFT hidden in this problem, but it is a CFT in a curved metric. It is very different from the traditional CFT setup \cite{calabrese2016quantum}, which we used in all previous sections. We refer to Refs. \cite{allegra2016inhomogeneous, dubail2016conformal} for more details about this. \vspace{0.3cm}

\subsection{Attempt (and failure) at a CFT calculation in the interacting case}

As we just emphasized, the free fermion case strongly suggests that it is not possible to get the generic behavior of the OSEE of operators in Heisenberg picture
as easily as, say, the generic linear growth of the OSEE of the evolution operator (section \ref{sec:evolution_op}). The traditional CFT
framework \cite{calabrese2016quantum}, which we used in sections $2$, $3$, $4$, apparently breaks down. In this last section, we briefly
discuss what happens if we try to attack this problem within this standard setup, and argue that it cannot be correct. Indeed, simple arguments lead to the
conclusion that the OSEE should either be bounded or increase linearly in time; both scenarios are clearly ruled out by the numerical results in Fig. \ref{fig:OSEE_Heisenberg}.

\subsubsection{OSEE of operator in Heisenberg picture as an out-of-time-ordered correlator}

\noindent We assume once again that we are dealing with an infinitely long one-dimensional quantum critical system, described by a CFT with gapless excitations that propagate at velocity $v$. We consider a local operator $\phi$ that acts at point $x=0$, and we want to calculate the OSEE of $\phi(t) = e^{i H t} \phi e^{-i Ht}$ for the bipartition $A \cup B$, $A = (-\infty, x)$, $B= [ x, +\infty )$. We use Eq. (\ref{eq:OSEE_replicas}) for integer $\alpha$,
\begin{eqnarray}
\nonumber	S_\alpha ( \phi(t) ) & = & \frac{1}{1-\alpha} \log \left(   \frac{ {\rm tr} [  \phi^\dagger(t)^{\otimes \alpha} \cdot   \mathcal{S}^A_{(1,\alpha,\alpha-1,\dots,2)}    \cdot \phi (t)^{\otimes \alpha} \cdot  \mathcal{S}^A_{(1,2,3,\dots,\alpha)}    ] }{   {\rm tr} [ \phi^\dagger(t)^{\otimes \alpha}   \cdot \phi (t)^{\otimes \alpha}] }   \right)   .
\end{eqnarray}
The numerator of the expression in the logarithm involves a correlator with operators $\phi$ and $\phi^\dagger$ acting at time $t$, while the two swap operators act at time $t=0$; we see, however, that they appear in the order $\Phi^\dagger(t) \mathcal{S}^A \Phi(t) \mathcal{S}^A$---with $\Phi^\dagger(t) \equiv \phi(t)^\alpha$---, as opposed to the time-ordered version which would look like $\Phi^\dagger(t) \Phi(t) \mathcal{S}^A  \mathcal{S}^A$. This type of correlator, dubbed {\it out-of-time-ordered correlator} or OTOC, has attracted a lot of attention recently in relation with quantum chaos and black holes and the AdS/CFT correspondence, following Refs.  \cite{shenker2013black,shenker2013multiple,kitaev2014hidden,kitaev2015simple}. \vspace{0.3cm}

\noindent As in sections $2$, $3$, $4$, one can rewrite the correlator appearing in Eq. (\ref{eq:OSEE_replicas}) in terms of twist operators, replacing the swap $\mathcal{S}^A$ by a pair of twist operators, $\mathcal{S}^A \rightarrow   \mathcal{T}_{(1,2,\dots, \alpha)} \mathcal{T}_{(1,\alpha,\dots,2)}$, with the two $\mathcal{T}$'s located at the two ends of the interval $A = (-\infty, x)$. It is convenient to think of the normalization of the twist operators as not being fixed at this point, so that $\mathcal{S}^A \propto \mathcal{T}(-\infty) \mathcal{T}(x)$, with some proportionality constant left undetermined. The two twist operators at $-\infty$, associated with $\mathcal{S}^A_{(1,2,\dots,\alpha)}$ and $\mathcal{S}^A_{(1,\alpha,\dots,2)}$ respectively, can be fused together to give the identity. So, in CFT, what we need to calculate is a ratio of the form
\begin{equation}
	\label{eq:OTOC_twists}
	  \frac{ {\rm tr} [  \phi^\dagger(t)^{\otimes \alpha} \cdot   \mathcal{T}_{(1,\alpha,\alpha-1,\dots,2)}  \cdot \phi (t)^{\otimes \alpha} \cdot  \mathcal{T}_{(1,2,3,\dots,\alpha)}   ] }{   {\rm tr} [ \phi^\dagger(t)^{\otimes \alpha}   \cdot \phi (t)^{\otimes \alpha}]  \;    {\rm tr} [  \mathcal{T}_{(1,\alpha,\alpha-1,\dots,2)}  \cdot \mathcal{T}_{(1,2,3,\dots,\alpha)} ]}
\end{equation}
where the two twist operators act at position $x$ and at time $0$. The new correlator of $\mathcal{T} \mathcal{T}$ in the denominator has been included in order to have a globally well-normalized expression, in the following sense. There are various operators that are located at the same position in Eq. (\ref{eq:OTOC_twists}), so there are several divergences that must be regulated in some way. But, whatever the regularization scheme, the ratio should always be one if $\phi$ is the identity operator; this is why the $\mathcal{T} \mathcal{T}$ factor in the denominator must be there. \vspace{0.3cm}

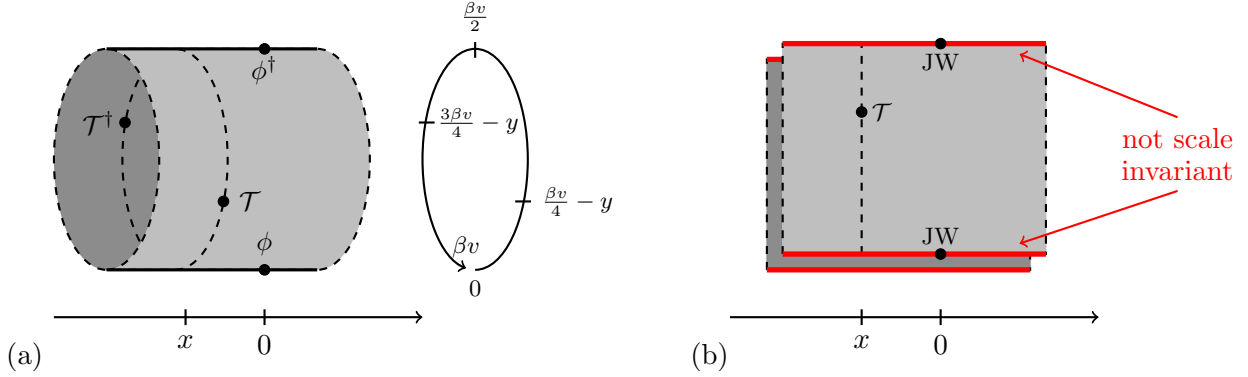
\begin{figure}[t]
	\begin{center}
	(a) \begin{tikzpicture}
	\begin{scope}[xshift=7.5cm,yshift=-5cm,scale=0.7]
		\draw[thick,->] (-1,-0.9) -- ++(7,0);
		\draw[thick] (3,-0.75) -- ++(0,-0.3) node[below]{$0$};
		\draw[thick] (1.5,-0.75) -- ++(0,-0.3) node[below]{$x$};
		\filldraw[very thick,draw=none,fill=gray!90] (0,0) arc (-90:270:1cm and 2.1cm);
		\draw[thick,dashed] (1.3,0) arc (-90:-270:1cm and 2.1cm);
		\filldraw[very thick,draw=none,fill=gray!50] (0,0) arc (-90:90:1cm and 2.1cm) -- ++ (4,0) arc (90:-90:1cm and 2.1cm) -- ++(-4,0);
		\draw[thick,dashed] (1.3,0) arc (-90:90:1cm and 2.1cm);
		\draw[thick,dashed] (0,0) arc (-90:90:1cm and 2.1cm) -- ++ (4,0) arc (90:-90:1cm and 2.1cm) -- ++(-4,0) arc (-90:-270:1cm and 2.1cm);
		\draw[very thick] (0,4.2) -- ++(4,0);
		\draw[very thick] (0,0) -- ++(4,0);
		\filldraw (0.35,2.8) circle (1mm) ++(-0.5,0) node{\small $\mathcal{T}^\dagger$};
		\filldraw (3,0) circle (1mm) ++(0,0.5) node{\small $\phi $};
		\filldraw (3,4.2) circle (1mm) ++(0,-0.4) node{\small $\phi^\dagger $};
		\filldraw (2.22,1.3) circle (1mm) ++(0.5,0) node{\small $\mathcal{T}$};
		\draw[thick,->] (7,0) node[below]{\footnotesize$0$} arc (-90:260:1cm and 2.1cm) node[above]{\footnotesize$\beta v$};
		\draw[thick] (7,4.2) ++(0,-0.17) -- ++(0,0.3) node[above]{\footnotesize$\frac{\beta v}{2}$};
		\draw[thick] (7.93,1.3) ++(-0.17,0) -- ++(0.3,0) node[right]{\footnotesize$\frac{\beta v}{4} - y$};
		\draw[thick] (6.1,2.8) ++(-0.17,0) -- ++(0.3,0) ++ (-0.15,0) node[right]{\footnotesize$\frac{3\beta v}{4} - y$};
	\end{scope}
	\end{tikzpicture}
	\qquad (b)\begin{tikzpicture}
	\begin{scope}[xshift=7.5cm,yshift=-5cm,scale=0.7]
		\draw[thick,->] (-1,-0.9) -- ++(7,0);
		\draw[thick] (1.5,-0.75) -- ++(0,-0.3) node[below]{$x$};
		\draw[thick] (3,-0.75) -- ++(0,-0.3) node[below]{$0$};
		\filldraw[gray!90] (-0.3,0) rectangle ++(5,4);	
		\draw[thick,dashed] (-0.3,0) rectangle ++(5,4);
		\draw[line width=2pt, red] (-0.3,4) -- ++(5,0);
		\filldraw[gray!50] (0,0.3) rectangle ++(5,4);
		\draw[thick,dashed] (0,0.3) rectangle ++(5,4);
		\draw[thick,dashed] (1.5,0.3) -- ++(0,4);
		\draw[line width=2pt, red] (0,0.3) -- ++(5,0);
		\filldraw[black] (3,0.3) circle (1mm) node[above]{\footnotesize${\rm JW}$};
		\draw[line width=2pt, red] (-0.3,0) -- ++(5,0);
		\draw[line width=2pt, red] (0,4.3) -- ++(5,0);
		\filldraw[black] (3,4.3) circle (1mm) node[below]{\footnotesize${\rm JW}$};
		\filldraw[black] (1.5,3) circle (1mm) node[right]{\footnotesize$\mathcal{T}$};
		\draw[thick, red,<-] (4.5,0.5) -- ++(3,1) node[above]{invariant};
		\draw[thick, red,<-] (4.5,4.1) -- ++(3,-1.2) node[below]{not scale};
	\end{scope}
	\end{tikzpicture}
	\end{center}
	\caption{(a) The regularization scheme (adapted from \cite{asplund2014holographic,roberts2014two}) in our tentative CFT calculation of the OSEE of some generic operator $\phi(t)$: we calculate a four-point function of the form $\left< \phi^\dagger \mathcal{T}^\dagger \phi \mathcal{T}\right>$ on the cylinder of circumference $\beta v$, and then take the analytic continuation $y \rightarrow i v t$, and $|x|, v t \gg \beta v$. (b) The calculation of the OSEE of a JW string: after the Bogoliubov transformation (\ref{eq:bogoliubov}) and other manipulations explained in the main text, it becomes apparent that---at least in that particular case---we are in fact dealing with two decoupled systems, with boundary conditions that break scale invariance.}
	\label{fig:regularization_scheme}
\end{figure}

\subsubsection{Tentative regularization scheme for the out-of-time-ordered correlator, and expected outcome}
\label{sec:discussion_OTOC}
In order to make sense of the expression (\ref{eq:OTOC_twists}) in CFT, we use a regularization scheme that is similar to the one in Refs. \cite{asplund2014holographic,roberts2014two}. Once again, we switch to imaginary time; the order of insertion of the operators is taken into account by
ensuring that they are inserted from right to left in increasing order of imaginary time. Thus, introducing some small parameter $\beta >0$, we see that one way of
getting the numerator of Eq. (\ref{eq:OTOC_twists}) is to first evaluate the expression 
\begin{equation*}
	 {\rm tr} [ ( e^{-(\frac{\beta}{4}-\frac{y}{v} ) H } \phi^\dagger  e^{- (\frac{\beta}{4}+\frac{y}{v} ) H }  )^{\otimes \alpha}  \cdot   \mathcal{T}_{(1,\alpha,\alpha-1,\dots,2)}  \cdot ( e^{-(\frac{\beta}{4}-\frac{y}{v} ) H } \phi  e^{-(\frac{\beta}{4}+\frac{y}{v} ) H }  )^{\otimes \alpha}   \cdot  \mathcal{T}_{(1,2,3,\dots,\alpha)}   ]
\end{equation*}
with $-\frac{\beta v}{4} <y < \frac{\beta v}{4}$, then do the analytic continuation $y \rightarrow i  v t$, and then finally take $\beta$ to zero. This expression (and similar ones for the denominator) is identical to an expectation value at finite inverse temperature $\beta$, involving operators at different imaginary times. So, for notational simplicity, we just write $\left< . \right>$ for such expectation values, and use the complex coordinate $z = x+i y$ to indicate the position of the different operators. Then the imaginary-time, regularized, version of the ratio (\ref{eq:OTOC_twists}) reads
\begin{equation}
	\label{eq:OTOC_im}
	  \frac{ \left<  \phi^{\dagger \otimes \alpha}( z_1, \overline{z}_1 ) \, \mathcal{T}_{(1,\alpha,\alpha-1,\dots,2)}  (z_2, \overline{z}_2 ) \, \phi^{\otimes \alpha} ( z_3, \overline{z}_3 ) \, \mathcal{T}_{(1,2,3,\dots,\alpha)}(  z_4, \overline{z}_4 )  \right> }{  \left<  \phi^{\dagger \otimes \alpha}( z_1, \overline{z}_1 )  \, \phi^{\otimes \alpha} ( z_3, \overline{z}_3 )  \right>    \left<  \mathcal{T}_{(1,\alpha,\alpha-1,\dots,2)}  (z_2, \overline{z}_2 ) \, \mathcal{T}_{(1,2,3,\dots,\alpha)}(  z_4, \overline{z}_4 )  \right> } ,
\end{equation}
with
\begin{equation}
	\label{eq:list_z}
	z_1 \, = \,i y+ i  \frac{\beta v}{4} , \qquad   z_2 \, = \,x   , \qquad   z_3 \, = \,  i y-i  \frac{\beta v}{4}  , \qquad   z_4 \, = \, x - i \frac{\beta v}{2}  .
\end{equation}
It is convenient to map the infinite cylinder of circumference $\beta v$, where the coordinates $z_j$ live, onto the plane, with the coordinates
\begin{equation}
	\label{eq:zeta_z}
	\zeta_j \, = \,  e^{ \frac{2\pi}{\beta v} z_j } .
\end{equation}
Now, what we need to analyze is the behavior of the four-point function $\frac{\left< \Phi_1^\dagger \mathcal{T}_2^\dagger  \Phi_3 \mathcal{T}_4 \right>}{ \left< \Phi_1^\dagger  \Phi_3\right> \left<  \mathcal{T}_2^\dagger \mathcal{T}_4 \right> }$ in the plane (we use the notations $\Phi  = \phi^{\otimes \alpha}$ and $\Phi_i = \Phi(z_i)$ for better readability), in the regime that corresponds to $y \rightarrow i v t$, $t \rightarrow \infty$. Carrying out this exercise explicitly is a little bit technical, so we defer it to the appendix. For our purposes, it is however sufficient to make a few elementary remarks, which already show that the present framework cannot lead to sensible
results. \vspace{0.3cm}

\noindent As any other four-point correlator in the plane, the one that is of interest to us now must be expressible as a sum of conformal blocks,
\begin{equation}
	\label{eq:block_G}
	\frac{\left<\Phi_1^\dagger \mathcal{T}_2^\dagger  \Phi_3 \mathcal{T}_4 \right>}{ \left< \Phi_1^\dagger  \Phi_3\right> \left<  \mathcal{T}_2^\dagger \mathcal{T}_4 \right> } \, = \, \sum_p \mathcal{G}_p (\eta) \overline{\mathcal{G}}_p (\overline{\eta}) , \qquad  \quad \eta \, =\, \frac{\zeta_{12}  \zeta_{34}}{ \zeta_{13}  \zeta_{24} }  , \; \overline{\eta} \, =\, \frac{\overline{\zeta}_{12}  \overline{\zeta}_{34}}{ \overline{\zeta}_{13}  \overline{\zeta}_{24} } ,
\end{equation}
where $\zeta_{ij} \equiv \zeta_i-\zeta_j$.
The sum runs over primary fields of the replicated CFT with $\alpha$ replicas, indexed by $p$. When $t \rightarrow \infty$, we see that $\eta \sim e^{ \frac{2 \pi t}{\beta}}$. In reasonable CFTs, such as for instance rational CFTs, the blocks behave algebraically in $\eta$ (cases where this algebraic behavior is violated correspond to so-called {\it logarithmic CFTs}---for an introduction, see the special volume of J. Phys. A \cite{gainutdinov2013logarithmic}---, where the behavior is logarithmic instead; however such cases appear in non-unitary theories which cannot arise as the low-energy description of one-dimensional quantum systems evolving unitarily, and are therefore of no interest to us here). In particular, if $h_p$ is the conformal dimension of the primary operator in the intermediate channel, then the block behaves as $\mathcal{G}_p(\eta) \sim \eta^{-h_p}$. Thus, there are only two possibilities:
\begin{itemize}
	\item the leading non-zero contribution comes from the identity block with $h_p=0$, such that $\frac{\left< \Phi^\dagger_1 \mathcal{T}^\dagger_2  \Phi_3 \mathcal{T}_4 \right>}{ \left< \Phi_1^\dagger  \Phi_3\right> \left<  \mathcal{T}^\dagger_2 \mathcal{T}_4 \right> }$ goes to a non-zero constant when $\eta \rightarrow \infty$. Then the OSEE at large time is just proportional to the logarithm of that constant. In other words, the OSEE of  $\phi(t)$ is bounded (at least for $\alpha$ integer and $\geq 2$)
	\item for some reason, the contribution from the identity block vanishes, and it is some other primary operator (with $h_p >0$) that gives the leading contribution $\frac{\left< \Phi^\dagger_1 \mathcal{T}^\dagger_2  \Phi_3 \mathcal{T}_4 \right>}{ \left< \Phi_1^\dagger  \Phi_3\right> \left<  \mathcal{T}^\dagger_2 \mathcal{T}_4 \right> } \sim \eta^{- h_p} \overline{\eta}^{- \overline{h}_p}$. Since $\eta \sim e^{\frac{2\pi t}{\beta}}$, we see that the OSEE of $\phi(t)$ increases linearly in time.
\end{itemize}
There is no room for other possibilities, such as {\it e.g.} logarithmic or power-law growths of the OSEE, based on this analysis and on this regularization scheme. Hence,
this appears to be in contradiction with our free fermion calculation above, and with the numerical results diplayed in Fig. \ref{fig:OSEE_Heisenberg}. \vspace{0.3cm}

\noindent To conclude this section: it is unclear whether or not it is possible to use CFT methods similar to the ones of section $2$, $3$, $4$ to tackle the
OSEE of local operators in Heisenberg picture for generic interacting systems. Certainly, the free fermion case cannot be tackled by such methods, because of the appearance of a
domain-wall initial state that breaks conformal invariance. But whether or not a similar
breakdown of conformal invariance appears also in generic interacting systems is unclear. We leave this as another open question, which deserves a more thorough investigation; it is a problem that is interesting
both for practical numerical purposes, and from a low-dimensional field theory perspective.


\section{Conclusion}

In these notes, we revisited the Operator Space Entanglement Entropy (OSEE) \cite{zanardi2001entanglement,prosen2007operator} through the prism of 2d CFT. 
We showed that the---now standard---tools developed by Cardy, Calabrese and others to tackle entanglement entropies and quantum quenches can be extended to
deal with the OSEE, in situations that are of primary interest for the development of numerical methods for 1+1d quantum dynamics. We provided derivations of
the following facts:
\begin{itemize}
	\item the thermal (Gibbs) density matrix $\rho_\beta$ at temperature $1/\beta > 0$ satisfies the operator area law: it has a bounded OSEE. As one approaches zero temperature,
	the OSEE increases as $S_\alpha (\rho_\beta)  \sim  \frac{c}{3} \left( 1 + \frac{1}{\alpha} \right) \log \beta$ (the logarithmic growth with $\beta$ had been observed previously in Ref. \cite{vznidarivc2008complexity}).
	\item after a global quench from an initial state with short-range correlations, the reduced density matrix has an OSEE that blows up linearly in time, before it decreases and saturates to a finite value. This implies that, although the reduced density matrix may be well approximated by an MPO with small dimension both at very short and at large times, it cannot be efficiently captured by an MPO in the transient regime.
	\item for systems described by CFT (and more generally for systems with ballistic propagation) the OSEE of the evolution operator $e^{- i H t}$ blows up linearly in time
	\item for free fermion Hamiltonians, the OSEE of a Jordan-Wigner string grows logarithmically in time. This was an early observation of Prosen and Pi\v{z}orn \cite{prosen2007operator}; here we provided an analytic derivation of this fact, relying on the results of \cite{allegra2016inhomogeneous, dubail2016conformal}.
\end{itemize}
On top of that, in section 3 and in section 5, we identified two questions that deserve to be further investigated:
\begin{center}
\begin{tabular}{llll}
{\it (Q1)}  & \qquad & {\it Does the GGE density matrix satisfy the operator are law?} \\
{\it (Q2)} & \qquad& {\it How fast does the OSEE of local operators in Heisenberg picture grow} \\
	&& {\it in generic interacting systems?}
\end{tabular}
\end{center}
In the simple free fermion example treated in section \ref{sec:GGE} (a quench from the XY to the XX chain), we provided the following answer to {\it (Q1)}: $\rho_{\rm GGE}$ satisfies the operator area law as a consequence of the occupation numbers $n_k$ being smooth as a function of $k$; this holds as long as the initial XY hamiltonian is gapped. It is natural to wonder whether or not the fact that the initial state possesses short-range correlations is a sufficient condition for $\rho_{\rm GGE}$ to obey an area law; it would be easy to investigate this question at least in other free fermion examples.
As for {\it (Q2)}, which is crucial in order to understand the performances of Heisenberg-picture DMRG \cite{hartmann2009density}, we were only able to conclude from numerics (Fig. \ref{fig:OSEE_Heisenberg}) that it is sublinear; it would be very interesting to have analytical results about this.

\paragraph{Note added.} A few days before this draft was submitted to the arXiv, a preprint by Zhou and Luitz appeared \cite{2016arXiv161207327Z}, that has a large overlap with our section 4.

\paragraph{Acknowledgements.} The ideas in these notes emerged from discussions with M. Collura, G. M\"oller, D. Karevski, V. Alba, and especially with F. Pollmann.
I also thank M.-C. Ba\~{n}uls and I. Cirac for very helpful and encouraging discussions at a later stage of the redaction, and J.-M. St\'ephan for carefully reading a first version of the manuscript and for suggesting improvements. I am grateful to V. Eisler for pointing out a mistake in the first arXiv version (namely, the fact that $H_{\rm GGE}$ is {\it not} local in the free fermion example in section 3.1, contrary to what was written previously).

I thank SISSA, Nordita (during the workshop "From quantum field theories to numerical methods"), Cambridge University, the MPQ Garching, and the Institut d'Etudes Scientifiques de Carg\`ese (during the school "Quantum integrable systems, CFTs and stochastic processes"), where parts of these notes were
written, for kind hospitality. I also acknowledge financial support from CNRS "D\'efi Inphyniti" and from the Conseil R\'egional and the Universit\'e de Lorraine.

\newpage

\appendix

\section{OSEE and free fermions}

Imagine we have a quadratic operator $O$, which can be put in diagonal form
\begin{equation}
	O \, = \, \prod_{a} \left[  \alpha_a  +  \beta_a  c_a^\dagger c_a \right]
\end{equation}
for some modes $c_a$, $c_a^\dagger$, $a=1, \dots, L$. Then the OSEE may be calculated as follows. We first turn $O$ into a
properly normalized state $\ket{O}$, as illustrated in Fig. \ref{fig:cartoon_OSEE}.(c)
\begin{equation}
	\ket{O} \, \equiv \, \prod_{a} \frac{\alpha_a  +  \beta_a  c_a^\dagger \tilde{c}^\dagger_a}{\sqrt{|\alpha_a|^2+|\beta_a|^2}} \ket{0}
\end{equation}
then we evaluate the correlations in that state,
\begin{equation}
	\bra{0} \left( \begin{array}{c}
		c^\dagger_a \\ \tilde{c}_a
	\end{array} \right)  \left( \begin{array}{cc}
		c_a & \tilde{c}^\dagger_a
	\end{array} \right) \ket{0} \, =\, \left( \begin{array}{cc}
		 \frac{ |\beta_a|^2 }{|\alpha_a|^2+|\beta_a|^2}   &    \frac{ \alpha_a \beta_a^* }{|\alpha_a|^2+|\beta_a|^2}   \\
		 \frac{ \alpha^*_a \beta_a }{|\alpha_a|^2+|\beta_a|^2}   &    \frac{ |\alpha_a|^2 }{|\alpha_a|^2+|\beta_a|^2} 
	\end{array} \right) .
\end{equation}
Thus, we obtain a hermitian $2L \times 2L$ matrix. Then we restrict this matrix to the subspace corresponding to the subsystem $A$, for a given bipartition of
the full system $A \cup B$. This gives a new hermitian $2L_A \times 2 L_A$ matrix, whose elements are the two-point correlators in the subsytem $A$.
Diagonalizing this matrix, we get $2 L_A$ real eigenvalues $n_i$, that are between $0$ and $1$. These are the occupation numbers of the modes that diagonalize
the correlation matrix for the subsystem $A$. As a result, the OSEE is given by
\begin{equation}
	S_\alpha ( O ) \, = \, \sum_{i=1}^{2L_A}  \frac{1}{1-\alpha}\log \left[  n_i^\alpha + (1-n_i )^\alpha  \right] .
\end{equation}
For more details, see the standard references \cite{peschel2003calculation,peschel2009reduced}.

\section{Twist operators for arbitrary permutations}

\paragraph{Replicas and the twist operator: some old results.}

Let $\alpha$ be a positive integer. Tensoring $\alpha$ replicas of a CFT---say in the plane $\mathbb{C}$---with central charge $c$, one gets a new CFT with
central charge $\alpha c$. It is well-known that one can consider twist operators (dating back\footnote{I thank Benjamin Doyon for pointing out these references to me.} to Refs. \cite{knizhnik1987analytic,dixon1987conformal}), such that, turning around the operator,
one goes from replica $j$ to replica $j+1$ (mod $\alpha$):
Let us call $\mathcal{T}(w,\overline{w})$ this twist operator. Any operator acting in a single copy $\phi_j(z) = 1 \otimes 1\otimes \phi_j (z) \otimes 1 \otimes\dots \otimes 1$ must have a monodromy
around $\mathcal{T}(w,\overline{w})$,
\begin{equation}
	\label{eq:twist}
	\phi_j(w + (z-w)e^{i \theta}) \mathcal{T}(w,\overline{w})  \, \rightarrow \,   \phi_{j+1}(z) \mathcal{T}(w,\overline{w})
\end{equation}
when $\theta$ goes from zero to $2\pi$. This may not be sufficient to characterize $\mathcal{T}(w,\overline{w})$ entirely; there may be many
operators that do that in the $\alpha^{\rm th}$-replicated theory, possibly with arbitrary high dimensions. We thus impose that $\mathcal{T}$ is, among the
many operators that satisfy (\ref{eq:twist}), the one with the smallest possible scaling dimension, which we write as $\Delta_\alpha = h_\alpha + \overline{h}_\alpha$. 
Notice that $\mathcal{T}$ treats right-moving and left-moving modes on the same footing, and so it must have spin zero: $h_\alpha = \overline{h}_\alpha$.
\vspace{0.4cm}

We then proceed to show that $\mathcal{T}(w,\overline{w})$ is a primary operator. To see that, notice that (\ref{eq:twist}) holds, 
in particular, for the stress-tensor on the $j^{\rm th}$ sheet, $T_j(z)$, such that the
total stress-tensor of the full theory, $T(z) = \sum_{j=1}^\alpha  T_j (z)$, has no monodromy. It follows that its OPE with $\mathcal{T}(w)$ is of the form
$T(z) \mathcal{T}(w) \, \simeq \,  \sum_{p\in \mathbb{Z}} (z-w)^{-2-p}  \mathcal{T}_p(w)$, for some operators $ \mathcal{T}_p$ with
conformal dimensions $\Delta_\alpha + p$. But, by definition, there cannot be such operators with a scaling dimension smaller than $\Delta_\alpha$. So all the terms with $p<0$ actually vanish. The OPE must then be of the form
\begin{equation}
	T(z) \mathcal{T}(w) \, \simeq \, \frac{h_\alpha}{(z-w)^2} \mathcal{T}(w) + \frac{1}{z-w} \partial  \mathcal{T}(w) + {\rm regular\, terms},
\end{equation}
as claimed. The conformal weight $h_\alpha$ is easily evaluated. One way of calculating it is to look at the full theory, with central charge $\alpha c$, on a long cylinder of circumference $\ell$, and of length $\Lambda \gg \ell$: the partition function behaves as
\begin{equation}
	Z \, \sim \, e^{ \Lambda \frac{\pi \alpha c}{6 \ell}} \, .
\end{equation}
Now insert the twist operator at both ends of the cylinder, at $\pm \Lambda/2$. This results in a new partition function
\begin{eqnarray}
\nonumber	Z_{\rm twist} & = & \left< \mathcal{T}(-\Lambda/2) \mathcal{T}(+\Lambda/2) \right> \times Z \\
	&\sim & e^{ - \Lambda \frac{2\pi}{\ell} \left(  \Delta_\alpha -\frac{\alpha c}{12}  \right)  } \, .
\end{eqnarray}
On the other hand, as one turns around the cylinder $\alpha$ times, it is clear that one goes through every replica exactly once, and so $Z_{\rm twist}$ is nothing
but the partition function of a CFT with central charge $c$, on a cylinder of circumference $\alpha \ell$. Therefore,
\begin{equation}
	Z_{\rm twist} \, \sim \, e^{ \Lambda \frac{\pi c}{6 \alpha\ell}} ,
\end{equation}
and, as a consequence,
\begin{equation}
	\Delta_\alpha \, = \, \frac{c}{12} \left(\alpha - \frac{1}{\alpha} \right)\, .
\end{equation}
This argument for fixing the dimension of twist operators seems to be very well-known, it appeared for instance\footnote{I thank Benoit Estienne for pointing out this reference.} in Ref. \cite{lunin2001correlation}.

\paragraph{Arbitrary elements of the permutation group $S_\alpha$.}

It is very natural to generalize this to arbitrary permutations in $\sigma \in S_\alpha$, namely to define the twist operator
$\mathcal{T}_{\sigma}$ as the operator with smallest possible scaling dimension such that
\begin{equation}
	\label{eq:sigma}
	\phi_j(w + (z-w)e^{i \theta}) \mathcal{T}(w,\overline{w})  \, \rightarrow \,   \phi_{\sigma(j)}(z) \mathcal{T}(w,\overline{w})
\end{equation}
when $\theta$ goes from zero to $2\pi$. It is clear, from the same argument as above, that this operator is primary. If $\sigma$ has cycles
of lengths $\alpha_1 +\dots +\alpha_c = \alpha$, then the dimension of $\mathcal{T}_\sigma$ is obviously
\begin{eqnarray}
\nonumber	\Delta_\sigma &  = & \Delta_{\alpha_1} + \dots + \Delta_{\alpha_c} \\
	&= & \frac{c}{12} \left(  \alpha - \frac{1}{\alpha_1} - \dots -\frac{1}{\alpha_c}  \right) .
\end{eqnarray}
Notice that $\Delta_\sigma$ is maximal for circular permutations. The OPE of these arbitrary twist operators must mimic the
group law of $S_\alpha$, which is a non-abelian group. So, to define the fusion of two twist operators, one need to a little bit
careful. One way to do this is to fix some curve $\Gamma \subset \mathbb{C}$, and to assume that the twist operators are
all lying on this curve; like beads on a string. Then the twist operators are naturally ordered: they appear in a given order as one
moves along $\Gamma$. This allows to write the non-abelian OPE satisfied by the twist operators,
\begin{equation}
	\mathcal{T}_{\sigma_1}(z) \mathcal{T}_{\sigma_2}(w) \, \simeq \;  \left|z-w\right|^{-\Delta_{\sigma_1} - \Delta_{\sigma_2} + \Delta_{\sigma_1 \circ \sigma_2} } \mathcal{T}_{\sigma_1 \circ \sigma_2}(w) + \dots
\end{equation}
where $z$ and $w$ lie on $\Gamma$. In general, calculating correlation functions of more than three twist operators is a difficult task; early attempts include Ref. \cite{lunin2001correlation}.

\section{More on the failed attempt at calculating the OSEE of operators in Heisenberg picture}

In this appendix we carry out an explicit calculation of the imaginary-time ratio (\ref{eq:OTOC_im}), in the case of two replicas only ($\alpha=2$), perform the Wick rotation,
and analyze the large-time behavior, following Ref. \cite{roberts2014two}.

\paragraph{Attempt at a direct calculation of $S_2( \phi(t) )$ within the regularization scheme of section $5.3.2$. The role of the entry $(1,1)$ of the $F$-matrix.}

For simplicity, let us assume that $\phi$ is a primary operator with conformal spin zero, and that $\phi = \phi^\dagger$. Our starting point is the four-point function of $\phi$ in the plane $\mathbb{C}$, expressed as a sum over conformal blocks
\begin{equation*}
	\left< \phi_1  \phi_2  \phi_3  \phi_4  \right>_{\mathbb{C}} \, = \, \frac{1}{| w_{13} |^{2 \Delta}  | w_{24} |^{2 \Delta}} \;  \sum_p  \mathcal{F}_p \left( \omega \right)  \overline{\mathcal{F}}_p \left( \overline{\omega} \right)  , \qquad  \quad   \omega = \frac{w_{12} w_{34}}{ w_{13} w_{24} } .
\end{equation*}
Here we are thinking of a diagonal theory, such as one of the minimal models of the $A_n$ series. As usual, there is some arbitrariness in the choice of the form of the prefactor $1/ (| z_{13} |^{2 \Delta}  | z_{24} |^{2 \Delta})$ and of the blocks themselves; here we have made our choice such that the conformal blocks transform conveniently under crossing, or '$F$-move',
\begin{equation*}
	\mathcal{F}_p (1-\omega)  \, = \, \sum_q  F_{p q} \, \mathcal{F}_q (\omega) 
\end{equation*}
with a matrix $F$ that has {\it constant} entries (they do not depend on $\omega$). Pictorially, this is of course the famous relation
$$
\begin{tikzpicture}
		\begin{scope}[xshift=-4cm,scale=0.7,rotate=90]
			\draw[thick] (-0.5,1) node[left]{$\phi_1$} -- (0,0) -- (-0.5,-1) node[right]{$\phi_4$};
			\draw[thick] (1.5,1) node[left]{$\phi_2$} -- (1,0) -- (1.5,-1) node[right]{$\phi_3$};
			\draw[thick] (0,0) -- (0.5,0) node[left]{$p$} -- (1,0);			
		\end{scope}
		\draw (-1.7,0.2) node{$\displaystyle = \quad \sum_q \; F_{p q}$};
		\begin{scope}[scale=0.7,yshift=0.35cm]
			\draw[thick] (-0.5,1) node[above]{$\phi_2$} -- (0,0) -- (-0.5,-1) node[below]{$\phi_1$};
			\draw[thick] (1.5,1) node[above]{$\phi_3$} -- (1,0) -- (1.5,-1) node[below]{$\phi_4$};
			\draw[thick] (0,0) -- (0.5,0) node[below]{$q$} -- (1,0);			
		\end{scope}
		\draw (1.6,0) node{.};
	\end{tikzpicture}
$$
As $\omega \rightarrow 0$, the conformal blocks behave as $\mathcal{F}_p (\omega )  \sim \frac{c_p}{ \omega^{2 h_\phi - h_p} }$ where $h_p \geq 0$ is the conformal dimension of the operator in the intermediate channel, and $c_p$ is a constant equal to a product of OPE coefficients. If we label the identity block by $p=1$, then $c_1 =1$ in the standard convention for normalization of the OPE coefficient $C^1_{\phi \phi} =1$ (or, equivalently, the standard normalization of the two-point function $\left<\phi \phi \right>$). Then we have
\begin{equation}
	\label{eq:limit0}
	\omega^{\Delta_\phi}  \mathcal{F}_p (\omega) \quad \underset{ \omega \rightarrow 0}{\longrightarrow} \quad  \left\{  \begin{array}{ll}  1 & {\rm if} \; p=1 \\ 0  & {\rm otherwise} .\end{array}  \right.
\end{equation}
Next, we use the conformal map $\zeta \mapsto w(\zeta) = \sqrt{\zeta}$ to write the two-point correlation function of the doubled operator $\Phi \equiv \phi \otimes \phi$ living on the two-sheeted plane with twist operators $\mathcal{T} \equiv \mathcal{T}_{(1,2)}$ inserted at $0$ and $\infty$. With $w_1 = \sqrt{\zeta_1}$, $w_2 = \sqrt{\zeta_3}$, $w_3 = -\sqrt{\zeta_1}$, $w_4 = -\sqrt{\zeta_3}$, we get
\begin{eqnarray}
\nonumber \left< \mathcal{T} (\infty)  \Phi (\zeta_1)  \Phi (\zeta_3) \mathcal{T} (0) \right> & = & \left| \frac{1}{16 \zeta_1 \zeta_3} \right|^{2\Delta_\phi} \sum_p  \mathcal{F}_p \left(  \frac{2 -  \frac{\sqrt{\zeta_1}}{\sqrt{\zeta_3}} - \frac{\sqrt{\zeta_3}}{\sqrt{\zeta_1}}  }{4} \right)  \overline{\mathcal{F}}_p \left(  \frac{2 -  \frac{\sqrt{\bar{\zeta}_1}}{\sqrt{\bar{\zeta}_3}} - \frac{\sqrt{\bar{\zeta}_3}}{\sqrt{\bar{\zeta}_1}}  }{4}   \right) .
\end{eqnarray}
With a special conformal transformations $\zeta \mapsto \frac{a \zeta+b}{c \zeta+d}$, we can get the same correlator with twists inserted at arbitrary positions $\zeta_2$ and $\zeta_4$. The result can be put in the form
\begin{equation*}
	\left< \Phi_1    \mathcal{T}_2  \Phi_3 \mathcal{T}_4 \right> \, = \, \frac{1}{ \left| \zeta_{13} \right|^{4 \Delta_\phi}  \left| \zeta_{24} \right|^{2 \Delta_2}  }  \sum_p  \mathcal{G}_p \left( \eta \right)  \overline{\mathcal{G}}_p \left( \overline{\eta}\right) , \qquad \eta = \frac{\zeta_{12} \zeta_{34}}{\zeta_{13} \zeta_{24}},
\end{equation*}
where $\Delta_2 = \frac{c}{8}$ is the scaling dimension of the twist operator $\mathcal{T}=\mathcal{T}_{(1,2)}$, and where we define the new conformal blocks in terms of the $\mathcal{F}$'s,
\begin{equation}
	\label{eq:defG}
	 \mathcal{G}_p (\eta) \, \equiv \,\left(  \frac{1}{16 \, \eta(1-\eta)}  \right)^{\Delta_\phi} \mathcal{F}_p \left(   \frac{2-\sqrt{\frac{\eta}{\eta-1}} - \sqrt{\frac{\eta-1}{\eta}} }{4} \right)  .
\end{equation}
Notice that, as a consequence of Eq. (\ref{eq:limit0}), we have
\begin{equation}
	\label{eq:limit1}
	 \mathcal{G}_p (\eta) \quad \underset{ \eta \rightarrow \infty}{\longrightarrow} \quad  \left\{  \begin{array}{ll}  1 & {\rm if} \; p=1 \\ 0  & {\rm otherwise} .\end{array}  \right.
\end{equation}
The quotient by the two-point functions is
\begin{equation*}
	\frac{ \left< \Phi_1 \mathcal{T}_2  \Phi_3  \mathcal{T}_4 \right> }{  \left< \Phi_1  \Phi_3 \right>  \left< \mathcal{T}_2 \mathcal{T}_4 \right>  }  \, = \, \sum_p  \mathcal{G}_p \left( \eta \right)  \overline{\mathcal{G}}_p \left( \overline{\eta}\right) ,
\end{equation*}
which is of the form (\ref{eq:block_G}) discussed in the main text. $\mathcal{G}_p$ has a branch cut between $0$ and $1$: indeed, as $\eta$ turns around $1$, intersecting the interval $(0,1)$ once, we see that we pick the other root in the argument of $\mathcal{F}$ in (\ref{eq:defG}). So, if we let $\mathcal{G}'_p ( \eta)$ be the analytic continuation of $\mathcal{G}_p ( \eta)$ when $\eta$ crosses the branch cut, then
\begin{eqnarray*}
\nonumber	\mathcal{G}'_p ( \eta) &=& \left(  \frac{ 1}{16 \, \eta (1-\eta)} \right)^{\Delta_\phi} \mathcal{F}_p \left(  \frac{2+\sqrt{\frac{\eta}{\eta-1}} + \sqrt{\frac{\eta-1}{\eta}} }{4} \right)  \\
\nonumber	 &=&  \left(  \frac{ 1}{16 \, \eta (1-\eta)} \right)^{\Delta_\phi}  \sum_q  F_{p q} \mathcal{F}_q \left(  \frac{2-\sqrt{\frac{\eta}{\eta-1}} - \sqrt{\frac{\eta-1}{\eta}} }{4} \right)  \\
	&=& \sum_q F_{pq}    \mathcal{G}_q ( \eta)  .
\end{eqnarray*}
Finally, we can come back to our problem: we set $\zeta_1 = e^{\frac{2\pi}{\beta v} (i y + i \frac{\beta v}{4}) }$, $\zeta_2 = e^{\frac{2\pi}{\beta v} x}$, $\zeta_3 = e^{\frac{2\pi}{\beta v} (i y-  i\frac{\beta v}{4}) }$ and $\zeta_4 = e^{\frac{2\pi}{\beta v} (x - i \frac{\beta v}{2})}$, see Eqs. (\ref{eq:list_z})-(\ref{eq:zeta_z}); then the cross-ratios become
\begin{equation*}
	\eta \, = \, \frac{1}{2} \left( e^{i \frac{\pi}{4}} e^{\frac{\pi}{\beta v} (x-i y)  }  +  e^{-i \frac{\pi}{4}} e^{-\frac{\pi}{\beta v} (x-i y)  }  \right)^2 \qquad \overline{\eta} \, = \, \frac{1}{2} \left( e^{-i \frac{\pi}{4}} e^{\frac{\pi}{\beta v} (x+i y)  }  +  e^{i \frac{\pi}{4}} e^{-\frac{\pi}{\beta v} (x+i y)  }  \right)^2 .
\end{equation*}
Taking the analytic continuation $y \rightarrow i v t$, this gives
\begin{equation*}
	\eta \, = \, \frac{1}{2} \left( e^{i \frac{\pi}{4}} e^{\frac{\pi}{\beta v} (x+ vt)  }  +  e^{-i \frac{\pi}{4}} e^{-\frac{\pi}{\beta v} (x+ v t)  }  \right)^2 \qquad \overline{\eta} \, = \, \frac{1}{2} \left( e^{-i \frac{\pi}{4}} e^{\frac{\pi}{\beta v} (x- v t)  }  +  e^{i \frac{\pi}{4}} e^{-\frac{\pi}{\beta v} (x- v t)  }  \right)^2 .
\end{equation*}
Notice that $\overline{\eta}$ is {\it not} the complex conjugate of $\eta$ after this operation. From now on, we focus on the case $x<0$, and we assume $\beta v \ll |x|$. Then as $t$ goes from zero to infinity, we see that $\overline{\eta}$ always stays close to $\infty$. On the contrary, $\eta$ starts from $\infty$ at $t=0$, then moves close to $1$, turns around $1$, crossing the segment $[0,1]$ once, and then goes back to $\infty$ as $t\rightarrow +\infty$. In other words, while $\overline{\mathcal{G}}_p (\overline{\eta})$ quickly converges to $\overline{\mathcal{G}}_p (\infty)$, the behavior of $\mathcal{G}_p (\eta)$ is slightly more subtle, because the branch cut between $0$ and $1$ is crossed once when $t$ goes from $0$ to $+\infty$. In consequence, we have $\mathcal{G}_p (\eta) \rightarrow \mathcal{G}'_p (\infty) =  \sum_q F_{pq}\mathcal{G}_q (\infty)$ when $t \rightarrow +\infty$. Putting everything together, we 
find
\begin{equation}
	\label{eq:OTOC_F11}
	\frac{ \left< \Phi_1 \mathcal{T}_2 \Phi_3   \mathcal{T}_4 \right> }{  \left< \Phi_1  \Phi_3 \right>  \left< \mathcal{T}_2 \mathcal{T}_4 \right>  }  \, \underset{t\rightarrow + \infty}{\longrightarrow} \, \sum_{p,q}  \overline{\mathcal{G}}_p  (\infty) F_{p q} \, \mathcal{G}_q  (\infty)  \, = \, F_{11}  ,
\end{equation}
where we used Eq. (\ref{eq:limit1}). Eq. (\ref{eq:OTOC_F11}) is the main result of our attempt at deriving the OSEE of $\phi(t)$: we see that, according to the discussion in  paragraph \ref{sec:discussion_OTOC}, the OSEE should either increase linearly or remain bounded, depending whether $F_{11}$ is zero or non-zero. At least, this is the result that we obtain within the setup of section $5.3$. As already pointed out in the main text, this is in contradiction both with the free fermion case, where an alternative analytical calculation is available (section 5.2) and leads to a logarithmic growth of the OSEE, and with the numerics of Fig. \ref{fig:OSEE_Heisenberg}. \vspace{0.3cm}

\paragraph{The example of $\left<\sigma^{\otimes 2}  \, \mathcal{T} \, \sigma^{\otimes 2} \, \mathcal{T} \right>$ in the Ising model.}
For concreteness, let us treat the example of the $\sigma$-field in the Ising model explicitly. $\sigma$ is a spinless primary operator of scaling dimension $\Delta_\sigma=1/8$. In the plane $\mathbb{C}$, we have
\begin{eqnarray*}
	\left< \sigma_1  \sigma_2  \sigma_3  \sigma_4 \right> &=&  \frac{1}{ |w_{13}|^{\frac{1}{4}} }   \frac{1}{ |w_{24}|^{\frac{1}{4}} }  \frac{ \left| 1+\sqrt{1-\omega} \right| + \left| 1-\sqrt{1-\omega} \right| }{2\,  |  \omega (1-\omega) |^{\frac{1}{4}}}, \qquad \quad \omega =  \frac{w_{12} w_{34}}{w_{13} w_{24}} \, .
\end{eqnarray*}
The two conformal blocks are
\begin{equation*}
	\mathcal{ F }_1 (\omega) = \frac{\sqrt{ 1+ \sqrt{1-\omega}  } }{\sqrt{2} \,  \omega^{\frac{1}{8}} (1-\omega)^{\frac{1}{8}}}   \qquad   \mathcal{F}_2 (\omega) = \frac{  \sqrt{ 1- \sqrt{1-\omega}  }  }{\sqrt{2}   \, \omega^{\frac{1}{8}}  (1-\omega)^{\frac{1}{8}}} ,
\end{equation*}
with the $F$-matrix
\begin{equation*}
	F \, = \, \frac{1}{\sqrt{2}} \left( \begin{array}{cc}  1 &  1 \\  1 & -1 \end{array} \right) .
\end{equation*}
Notice that Eq. (\ref{eq:limit0}) is satisfied. Thus, we find that the OTOC is, at large time,
\begin{equation*}
	\frac{ \left< \sigma^{\otimes 2}  \, \mathcal{T} \, \sigma^{\otimes 2} \, \mathcal{T} \right> }{  \left< \sigma^{\otimes 2}  \sigma^{\otimes 2} \right>  \left< \mathcal{T} \mathcal{T} \right>  }  \, \underset{t\rightarrow + \infty}{\longrightarrow} \, \frac{1}{\sqrt{2}} \, \neq \, 0.
\end{equation*}
This result predicts that the OSEE of $\sigma(t)$ should be bounded, which is in contradiction with the discussion of sections $5.1$-$5.2$: after the Jordan-Wigner transformation, the $\sigma$-field in the transverse Ising chain is an operator that is attached to a JW string, so its OSEE is known to grow logarithmically.

\bibliographystyle{ieeetr}
\bibliography{biblio}

\end{document}